\documentclass[preprint,a4paper,showabstract]{revtex4-2}
\setcitestyle{super}

\usepackage{graphicx}
\usepackage{epstopdf}
\usepackage{dcolumn}
\usepackage{bm}
\usepackage{hyperref}
\usepackage[utf8]{inputenc} 
\usepackage[T1]{fontenc}    
\usepackage{lmodern}
\usepackage{url}            
\usepackage{booktabs}       
\usepackage{amsfonts}       
\usepackage{nicefrac}       
\usepackage{microtype}      
\usepackage[version=3]{mhchem}
\usepackage{natbib}
\usepackage{amsmath}
\usepackage{amssymb}
\usepackage{xcolor}
\usepackage[range-units = single,range-phrase = \text{--},separate-uncertainty = true,detect-all]{siunitx}
	\DeclareSIUnit\linepair{lp}
	\DeclareSIUnit\pixels{px}
\usepackage{soul}
\setstcolor{red}
\usepackage{cancel}

\newcommand*{\al}[1]{\textcolor{black}{#1}}
\newcommand*{\alex}[1]{\textcolor{black}{#1}}
\newcommand*{\rev}[1]{\textcolor{black}{#1}}
\newcommand*{\ulysse}[1]{\textcolor{black}{#1}}
\renewcommand{\vec}[1]{\bm{\mathrm{#1}}}

\newcommand\diff{\mathrm{d}}
\newcommand\out{_\mathrm{out}}
\newcommand\rin{\ulysse{\vec{\rho}_\mathrm{in}}}
\newcommand\rout{\ulysse{\vec{\rho}\out}}

\newcommand\rp{\vec{r}_\mathrm{p}}

\begin{document}

\title{\al{Harnessing Forward Multiple Scattering} \\
for Optical Imaging Deep Inside an Opaque Medium}

\author{Ulysse Najar}
\affiliation{Institut Langevin, ESPCI Paris, PSL University, CNRS, 75005 Paris, France\\ \textnormal{{$^*$}Corresponding author (e-mail: alexandre.aubry@espci.fr)}}
\author{Victor Barolle}
\affiliation{Institut Langevin, ESPCI Paris, PSL University, CNRS, 75005 Paris, France\\ \textnormal{{$^*$}Corresponding author (e-mail: alexandre.aubry@espci.fr)}}
\author{Paul Balondrade}
\affiliation{Institut Langevin, ESPCI Paris, PSL University, CNRS, 75005 Paris, France\\ \textnormal{{$^*$}Corresponding author (e-mail: alexandre.aubry@espci.fr)}}
\author{Mathias~Fink}
\affiliation{Institut Langevin, ESPCI Paris, PSL University, CNRS, 75005 Paris, France\\ \textnormal{{$^*$}Corresponding author (e-mail: alexandre.aubry@espci.fr)}}
\author{Claude Boccara}
\affiliation{Institut Langevin, ESPCI Paris, PSL University, CNRS, 75005 Paris, France\\ \textnormal{{$^*$}Corresponding author (e-mail: alexandre.aubry@espci.fr)}}
\author{Alexandre Aubry$^*$}
\affiliation{Institut Langevin, ESPCI Paris, PSL University, CNRS, 75005 Paris, France\\ \textnormal{{$^*$}Corresponding author (e-mail: alexandre.aubry@espci.fr)}}

\date{\today}
\begin{abstract}
    \textbf{As light travels through a disordered medium such as biological tissues, it undergoes multiple scattering events. This phenomenon is detrimental to in-depth optical microscopy, as it causes a drastic degradation of contrast, resolution and brightness of the resulting image beyond a few scattering mean free paths. However, the information about the inner reflectivity of the sample is not lost; only scrambled. To recover this information, a matrix approach of optical imaging can be fruitful. Here, we report on a de-scanned measurement of a high-dimension reflection matrix $\mathbf{R}$ via low coherence interferometry. Then, we show how \al{a set of independent focusing laws can be extracted for each medium voxel} through an iterative multi-scale analysis of wave distortions contained in $\mathbf{R}$. It enables an optimal \al{and local} compensation of forward multiple scattering paths and provides a three-dimensional confocal image of the sample as the latter one had become digitally transparent. The proof-of-concept experiment is performed on a human opaque cornea and an extension of the penetration depth by a factor five is demonstrated compared to the state-of-the-art.}
\end{abstract}
\maketitle

\noindent {\large \textbf{Introduction}} 

Multiple scattering of waves concerns many domains of physics, ranging from optics or acoustics to solid-state physics, seismology, medical imaging, or telecommunications. In an inhomogeneous medium where the refractive index $n$ depends on the spatial coordinates $\mathbf{r}$, several physical parameters are relevant to characterize wave propagation: (\textit{i}) the scattering mean free path $\ell_s$, which is the average distance between two successive scattering events; (\textit{ii}) the transport mean free path $\ell_t$, which is the distance after which the wave has lost the memory of its initial direction. For a penetration depth $z$ smaller than $\ell_s$, ballistic light is predominant and standard focusing methods can be employed; for $z>\ell_s$, multiple scattering events result in a gradual randomization of the propagation direction before reaching the diffusive regime for $z>\ell_t$. Although it gives rise to fascinating interference phenomena such as perfect transmission~\cite{Gerardin2014,Horodynski2022} or Anderson localization~\cite{Billy2008,Hu2008}, multiple scattering still represents a major obstacle to deep imaging and focusing of light inside complex media~\cite{ntziachristos_going_2010,Bertolotti2022}.

\alex{To cope with the fundamental issue of multiple scattering, several approaches have been proposed to enhance the single scattering contribution drowned into a predominant diffuse background~\cite{Dunsby,ntziachristos_going_2010,badon_multiple_2017}. One solution is to perform a confocal discrimination and coherent time gating of singly-scattered photons by means of interferometry. This is the principle of optical coherence tomography~\cite{huang_optical_1991}, equivalent to ultrasound imaging for light. Nevertheless, a lot of photons associated with distorted trajectories are rejected by the confocal filter while they still contain a coherent information on the medium reflectivity. Originally developed in astronomy~\cite{babcock}, adaptive optics (AO) has been transposed to optical microscopy in order to address this issue~\cite{booth_adaptive_2014}. Nevertheless, it only compensates for low-order aberrations induced by long-scale fluctuations of the optical index and does not address high-order aberrations generated by forward multiple scattering events. To circumvent the latter problem, one has to go beyond a confocal scheme and investigate the cross-talk between the pixels of the image. This is the principle of matrix imaging in which the relation between input and output wave-fields is investigated under a matrix formalism.}


While a subsequent amount of work has considered the transmission matrix $\mathbf{T}$ for optimizing wave control and focusing through complex media~\cite{Tanter2001,Derode2003a,popoff_measuring_2010,Hsu2017,bouchet_maximum_2021,bender_depth-targeted_2022}, this configuration is not the most relevant for imaging purposes since only one side of the medium is accessible for most {in-vivo} applications. Moreover, in all the aforementioned works, the scattering medium is usually considered as a black box, while imaging requires to open it. To that aim, a {reflection matrix approach of wave imaging} (RMI) has been developed for the last few years~\cite{yoon_deep_2020,badon_distortion_2020,yoon_laser_2020,lambert_distortion_2020}. The objective is to determine, from the reflection matrix $\mathbf{R}$, the $\mathbf{T}$-matrix between sensors outside the medium and voxels mapping the sample~\cite{Gigan2022}. \alex{Proof-of-concept studies have reported penetration depths ranging from 7 $\ell_s$~\cite{kang_high-resolution_2017} to 10 $\ell_s$~\cite{badon_distortion_2020} but the object to image was a resolution target whose strong reflectivity artificially extends the penetration depth by several $\ell_s$ compared with direct tissue imaging~\cite{badon_multiple_2017}. Follow-up studies also considered the imaging of highly reflecting structures (e.g. myelin fibers) through an aberrating layer (e.g mouse skull)~\cite{yoon_laser_2020}, in a wavelength range that limits scattering and aberration from tissues~\cite{Kwon2023}. On the contrary, here, we want to address the extremely challenging case of three-dimensional imaging of biological tissues themselves (cells, collagen, extracellular matrix etc.) at large penetration depth ($z \sim 10\ell_s$), regime in which aberration and scattering effects are spatially-distributed over multiple length-scales.}

\alex{Inspired by previous works~\cite{badon_retrieving_2015,badon_spatio-temporal_2016}, full-field optical coherence tomography (FFOCT)~\cite{beaurepaire_full-field_1998,dubois_high-resolution_2002} will be used here to record the $\mathbf{R}-$matrix. In FFOCT, the incident wave-field is temporally- and spatially-incoherent. It enables, by means of low coherence interferometry, a parallel acquisition of a time-gated confocal image~\cite{barolle_manifestation_2021} at a much better signal-to-noise ratio than a traditional point scanning scheme for equal measurement time and power~\cite{mertz_introduction_2019}. By splitting the incident wave-field into two laterally-shifted components, a de-scanned measurement of $\mathbf{R}$ can be performed without a tedious raster scanning of the field-of-view~\cite{yoon_laser_2020}.} 

Another advantage of \alex{the de-scanned} basis is the direct access to the distortion matrix $\mathbf{D}$ through a Fourier transform. 
This matrix basically connects any focusing point with the distorted part of the associated reflected wavefront~\cite{badon_distortion_2020,lambert_distortion_2020}. 
A multi-scale analysis of $\mathbf{D}$ is here proposed to estimate the forward \alex{multiple} scattering component of the $\mathbf{T}$-matrix at an unprecedented spatial resolution ($\sim 6~\mu$m). Once the latter matrix is known, one can actually unscramble, in post-processing, all wave distortions and multiple scattering events undergone by the incident and reflected waves for each voxel. A three-dimensional confocal image of the medium can then be retrieved as if the medium had been made digitally transparent.

The experimental proof-of-concept presented in this paper is performed on a human {ex-vivo} cornea that we chose deliberately to be extremely opaque. Its overall thickness is of \alex{$10 \ell_s$}. FFOCT shows an imaging depth limit of \alex{$2\ell_s$} due to aberration and scattering. Strikingly, RMI enables to recover a full 3D image of the cornea at a resolution close to $\lambda/4$ ($\sim 230~$nm) and a penetration depth enhanced by, at least, a factor five.\\ 


\noindent {\large \textbf{Results}} \\

\noindent { \textbf{\alex{Measuring a} \ulysse{De-scan} Reflection Matrix}} \\

 \begin{figure}[h!]
    	\centering
    	\includegraphics[width=\linewidth]{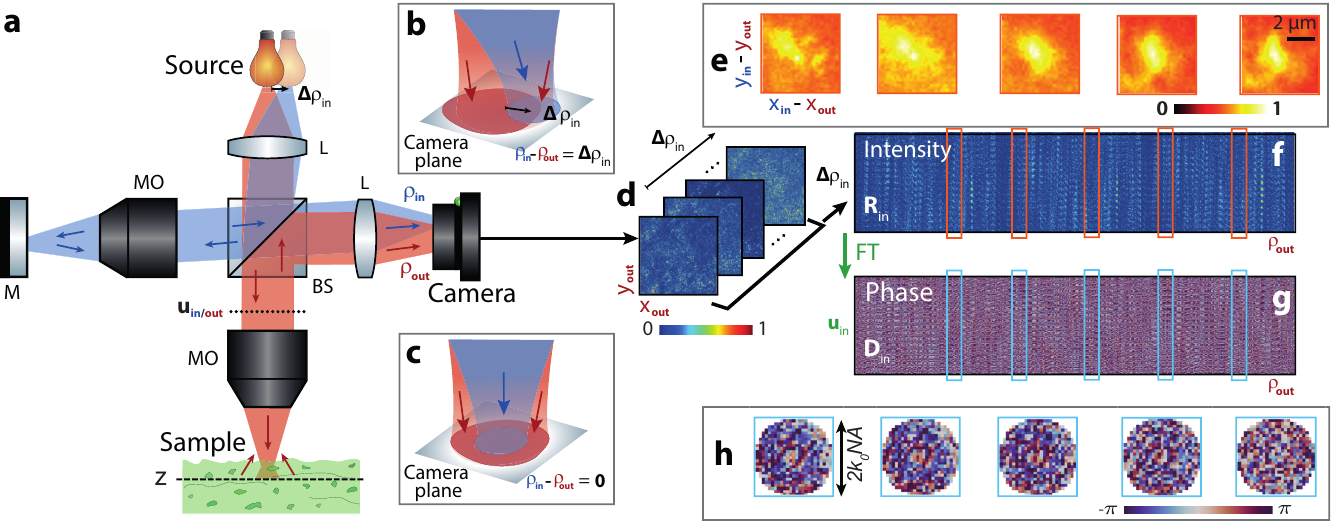}
    	\caption{\textbf{\ulysse{De-scanned} \alex{measurement} of the Reflection Matrix.} \textbf{a} Experimental setup (L: lenses, MO: microscope objectives, M: reference mirror, BS: beam splitter). Light from an incoherent source is split into two replica laterally shifted with respect to each other by a relative position $\Delta \bm{\rho}$ {(see Supplementary Section S1)}. By a game of polarization, each replica illuminates one arm of a Linnik interferometer. The sample beam (in red) illuminates the scattering sample through a microscope objective ($\mathrm{NA}=1.0$). The reference beam (in blue) is focused on a dielectric mirror through an identical microscope objective. Both reflected beams interfere on a CMOS camera whose surface is conjugated with focal planes of the MO. The amplitude and phase of the interference term are retrieved by phase-shifting interferometry. \textbf{b} Each pixel of the camera, depicted by its position $\bm{\rho}\out$, measures the reflection coefficient $R(\bm{\rho}_\textrm{in},\bm{\rho}\out,z)$ between de-scanned focusing points, \alex{$\bm{\rho}\out$ and $\bm{\rho}_\textrm{in}=\bm{\rho}\out+\Delta \bm{\rho}_\textrm{in}$}, 
    	at depth $z$ within the sample. \textbf{c} For $\Delta \bm{\rho}_\textrm{in}=\mathbf{0}$, the experimental set up is equivalent to a FFOCT apparatus and the interferogram directly provides a time-gated confocal image of the sample. \textbf{d} The set of interferograms are stored in the de-scanned reflection matrix \alex{$\mathbf{R}_{\textrm{in}}(z)=[R_\textrm{in}(\Delta \bm{\rho}_\textrm{in},\bm{\rho}\out,z)]$} displayed in panel \textbf{f}. \textbf{e} Each column of this matrix yields a {reflection point-spread function (RPSF)} associated with the focusing quality at point $\vec{\rho}\out$ (scale bar: $2~\mu$m). \textbf{g} The Fourier transform (FT) of each de-scanned wave-field provides the \alex{input} distortion matrix \alex{$\mathbf{D}_{\textrm{in}}(z)=[{D}_{\textrm{in}}( \mathbf{u}_\textrm{in},\bm{\rho}\out,z)]$.} \textbf{h} Each column of this matrix displays the distorted wave-front associated with each point $\vec{\rho}\out$ in the field-of-view. The optical data shown in panels d-h correspond to the acquisition performed at depth {$z=150~\mu$m}.
     }
    	\label{fig:setup}
    \end{figure}
    
\begin{figure}[h!]
            \centering
            \includegraphics[width=\linewidth]{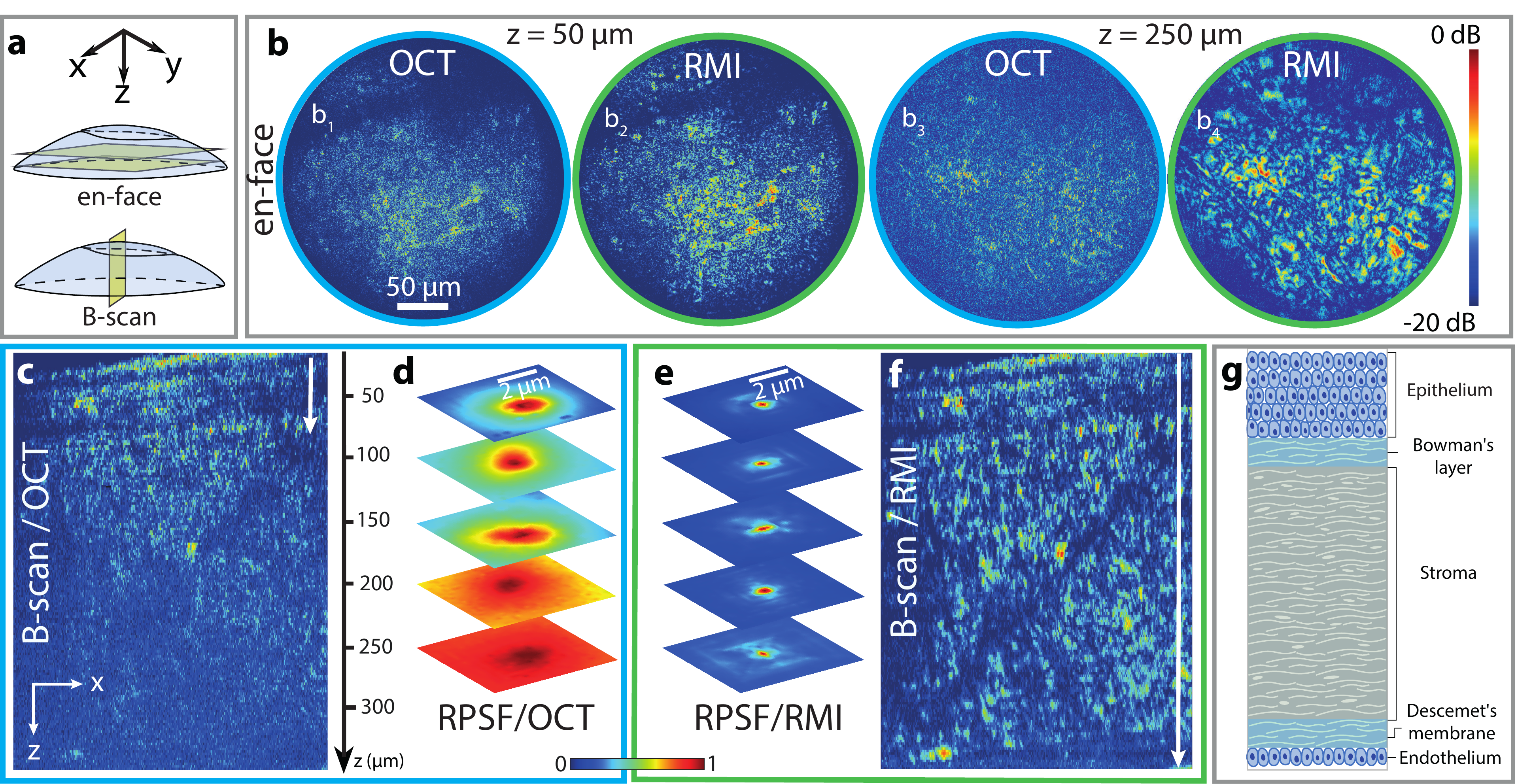}
            \caption{ \textbf{Volumetric matrix imaging of an opaque cornea.} \textbf{a.} Schematic of the imaging planes in the cornea. \textbf{b.} \emph{En-face} confocal images before (\textbf{b$_1$},\textbf{b$_3$}) and after (\textbf{b$_2$},\textbf{b$_4$}) the matrix imaging process for {$z=50~\mu$m} and $250~\mu$m, respectively (scale bar: $50~\mu$m). \textbf{c.} Longitudinal (\emph{x,z}) section of the initial confocal image. \textbf{d.} Original RSPFs from $z=50$ to $250~\mu$m (scale bar: $2~\mu$m). \textbf{e.}  Corresponding RPSFs after the matrix imaging process. \textbf{f.} Longitudinal (\emph{x,z}) section of the volumetric image at the end of the matrix imaging process. \textbf{g.} Schematic of a healthy human cornea. \alex{Each image is normalized at each depth by its averaged intensity.}
            }
            \label{fig2}\label{fig:Bscan}
        \end{figure}

\ulysse{
\alex{Our approach is based on a de-scanned measurement of the time-gated reflection matrix $\mathbf{R}$ from the scattering sample}.
Inspired by time-domain FFOCT~\cite{beaurepaire_full-field_1998,dubois_high-resolution_2002}, the corresponding set up is displayed in Fig.~\ref{fig:setup}a. It consists in a Michelson interferometer with microscope objectives in both arms (Fig.~\ref{fig:setup}a).}
In the first arm, a reference mirror is placed in the focal plane of a microscope objective (MO). The second arm contains the scattering sample to be imaged. 
Because of the broad spectrum of the incident light, interferences occur between the two arms provided that the optical path difference through the interferometer is close to zero. The length of the reference arm determines the slice of the sample (coherence volume) to be imaged and is adjusted in order to match with the focal plane of the MO in the sample arm. The backscattered light from each voxel of the coherence volume can only interfere with the light coming from the conjugated point of a reference mirror. The spatial incoherence of the light source actually acts as a physical confocal pinhole (Fig.~\ref{fig:setup}c). All these interference signals are recorded in parallel by the pixels of the camera in the imaging plane. \ulysse{Their amplitude and phase are retrieved by phase-stepping interferometry~\alex{\cite{dubois_high-resolution_2002}}.} The FFOCT signal is thus equivalent to a time-gated confocal image of the sample~\cite{barolle_manifestation_2021}.
Figures~\ref{fig2}b and c show en-face and axial FFOCT images of the opaque cornea at different depths. A dramatic loss in contrast is found beyond the epithelium ($z>70~\mu$m, see Fig.~\ref{fig:Bscan}g). It highlights the detrimental effect of multiple scattering for deep optical imaging. 

 \begin{figure}[h!]
    	\centering
    	\includegraphics[width=0.9\linewidth]{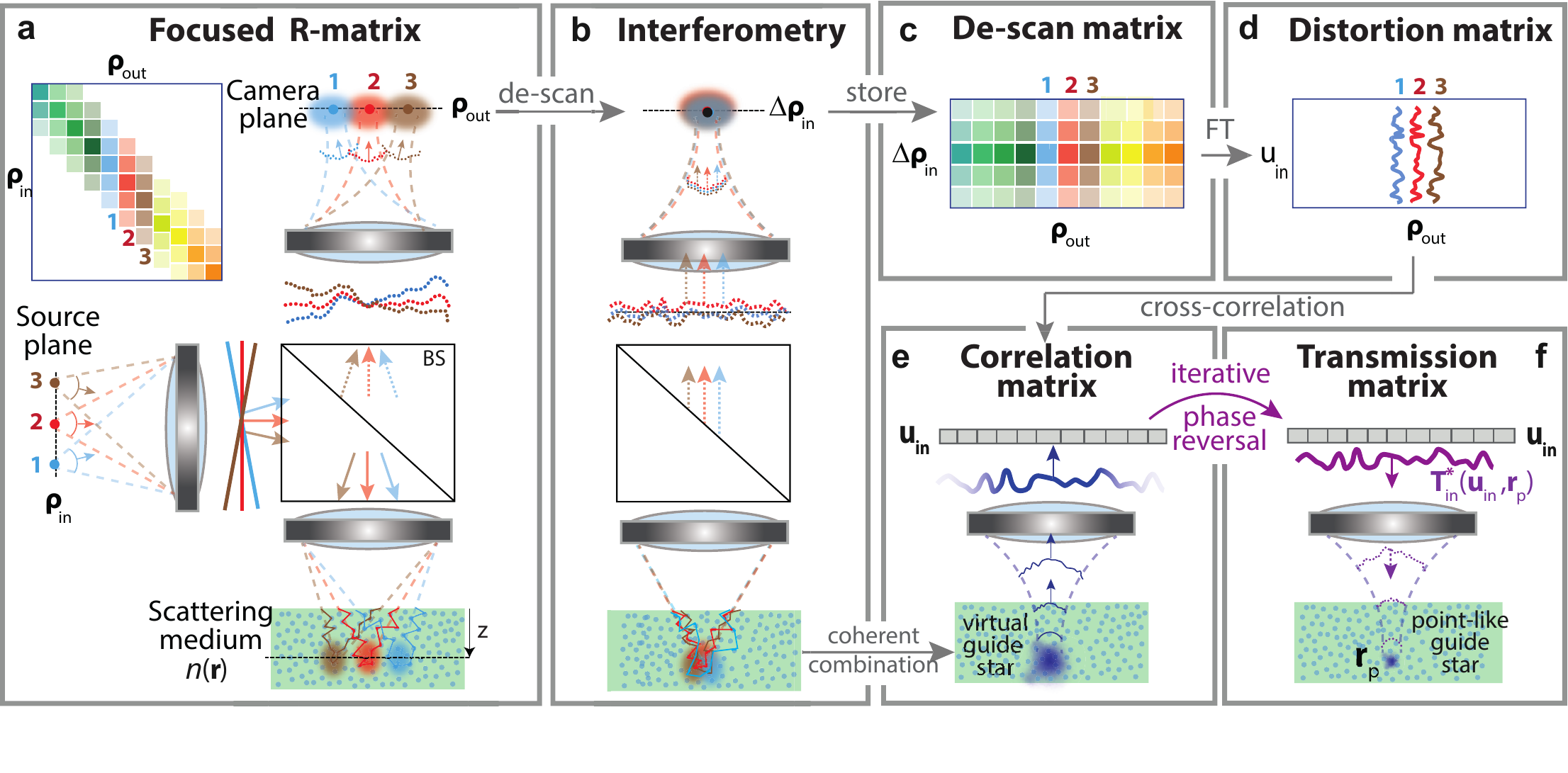}
    	\caption{\textbf{Different Stages of Matrix Imaging.} \alex{\textbf{a.} The focused $\mathbf{R}$-matrix contains the set of impulse responses $R(\bm{\rho}_\textrm{in},\bm{\rho}\out,z)$ between an array of point sources $\bm{\rho}_\textrm{in}$ and detectors $\bm{\rho}\out$ lying in planes conjugated with the focal plane of the microscope objective (BS: beam splitter).
     \textbf{b.} The interferometric set up displayed in Fig.~\ref{fig:setup} allows a \alex{de-scanned} measurement of $\mathbf{R}$ by scanning the relative position $\Delta \bm{\rho}_\textrm{in}=\bm{\rho}_\textrm{in}-\bm{\rho}_\textrm{out}$. \textbf{c.} \alex{Each column of the recorded matrix $\mathbf{R}_\textrm{in}$}(Eq.~\ref{descan}) corresponds to the RPSF measured by each camera pixel. \textbf{d.} A spatial Fourier transform (FT) over $\Delta \bm{\rho}_\textrm{in}$ provides the distortion matrix $\mathbf{D}_\textrm{in}$ (Eq.~\ref{FT}) linking each camera pixel with wave-front distortions seen from the input pupil plane ($\mathbf{u}_\textrm{in}$).  \textbf{e.} The correlation matrix \alex{$\mathbf{C}_\textrm{in}$} between those wave-fronts mimics the time-reversal operator associated with a virtual guide star that results from a {coherent} average of all the de-scanned focal spot{s} displayed in \textbf{b} (Supplementary Section \alex{\rev{S2}}). \textbf{f.}} IPR is then applied (Methods). The resulting wave-front compensates for aberrations and scattering inside the medium to produce a sharper guide star. \alex{It provides an estimation of one column} of $\mathbf{T}_{\textrm{in}}$ corresponding to the common mid-point $\rp$ of the input focal spots considered in panel \textbf{a}.
     }
    \label{fig:principle}
    \end{figure}
    
To overcome the multiple scattering phenomenon, \alex{one should go beyond a simple confocal image and record the cross-talk between the camera pixels. Experimentally, it consists in measuring the reflection matrix $\mathbf{R}$ associated with the sample (Fig.~\ref{fig:principle}a). Interestingly, this can be done by slightly modifying the illumination scheme of the FFOCT device, as displayed in Fig.~\ref{fig:setup}a.}
The incident wave-fields are still identical in each arm but are laterally shifted with respect to each other by a transverse position $\Delta \bm{\rho}_\textrm{in}$. Their spatial incoherence now acts as a de-scanned pinhole that gives access to the cross-talk between distinct focusing points (Fig.~\ref{fig:setup}b).
The interferogram recorded by the camera (Fig.~\ref{fig:setup}d) directly provides one line of the \alex{reflection matrix $\mathbf{R}_{\alex{\textrm{in}}}$ \alex{de-scanned at input} (Fig\alex{s}.~\ref{fig:principle}b \alex{and c}), such that
\begin{equation}
\label{descan}
   \alex{R_{\alex{\textrm{in}}}}(\Delta \bm{\rho}_\textrm{in},\bm{\rho}\out,z) = \alex{R}(\alex{\bm{\rho}_\textrm{out}}+\Delta \bm{\rho}_\textrm{in},\bm{\rho}\out,z),
\end{equation}
with $\mathbf{R}=[R(\bm{\rho}_\textrm{in},\bm{\rho}_\textrm{out},z)]$, the reflection matrix expressed in the canonical basis. Its coefficients $R(\bm{\rho}_\textrm{in},\bm{\rho}_\textrm{out},z)$ correspond to the response \alex{of the medium at depth $z$ between points $\bm{\rho}_\textrm{in}$ and $\bm{\rho}\out$ in the source and camera planes (Fig.~\ref{fig:principle}a).}    }
Scanning the relative position $\Delta \bm{\rho}_\textrm{in}$ is equivalent to recording the \alex{canonical} $\mathbf{R}$-matrix diagonal-by-diagonal \alex{(see Figs.~\ref{fig:principle}a and c)}. However, while a \alex{raster scan} (column-by-column acquisition) of $\mathbf{R}$ requires to illuminate the sample over a field-of-view $\Omega$ with $N=(\Omega/\delta_0)^2$ input wave-fronts~\alex{\cite{badon_smart_2016,kang_imaging_2015,yoon_laser_2020}}, the de-scanned basis \alex{allows} a much smaller number of \alex{field measurements}. 

This sparsity can be understood by expressing theoretically the \alex{de-scan matrix $  \mathbf{R}_{\textrm{in}}$} (Supplementary Section \rev{S2}):
\begin{equation}
\label{descan2}
     R_{\alex{\textrm{in}}}(\Delta \bm{\rho}_\textrm{in},\bm{\rho}\out,z) = \int_{\Omega} d\alex{\bm{\rho}_s} \,  H_\textrm{in} (\alex{\bm{\rho}_s}+\Delta \bm{\rho}_\textrm{in},\ulysse{\bm{\rho}_\textrm{in},z}) \gamma(\alex{\bm{\rho}_s}+\bm{\rho}\out,z) H\out (\alex{\bm{\rho}_s},\ulysse{\bm{\rho}\out,z})
\end{equation}
where $\gamma$ is the sample reflectivity.  $H_\textrm{in}(\alex{\bm{\rho}_s},\ulysse{\bm{\rho}_\textrm{in},z})$ and $H\out(\alex{\bm{\rho}_s},\ulysse{\bm{\rho}\out,z})$ are the local input and output point spread functions (PSFs) at points \ulysse{$(\bm{\rho}_\textrm{in},z)$} and \ulysse{$(\bm{\rho}\out,z)$} , respectively. This last equation confirms that the central line of \alex{$\mathbf{R}_{\alex{\textrm{in}}}$} ($\Delta \bm{\rho}_\textrm{in}=\mathbf{0}$), i.e. the FFOCT image, results from a convolution between the sample reflectivity $\gamma$ and the local confocal PSF, $H_\textrm{in} \times H\out$. 

The de-scanned elements allow us to go far beyond standard \alex{confocal} imaging. In particular, they will be exploited to unscramble the local input and output PSFs in the vicinity of each focal point. As a preliminary step, they can also be used to quantify the level of aberrations and multiple scattering. In average, the de-scanned intensity, $I(\Delta \bm{\rho}_\textrm{in}\alex{,\bm{\rho}\out,z})= |\alex{R_{\alex{\textrm{in}}}}(\Delta \bm{\rho},\bm{\rho}\out,z)|^2 $, can  actually be expressed as the convolution between the incoherent input and output PSFs~\cite{lambert_reflection_2020}:
\begin{equation}
\label{eq1}
 \langle I(\Delta \bm{\rho}_\textrm{in},\alex{\bm{\rho}\out,z}) \rangle \propto |H_\textrm{in}|^2 \stackrel{\Delta \bm{\rho}_\textrm{in}}{\circledast} |H\out|^2 (\Delta \bm{\rho}_\textrm{in},\alex{\bm{\rho}\out,z})
\end{equation}
where the symbol $\circledast$ stands for correlation product and $\langle \cdots \rangle $ for ensemble average. This quantity will be referred to as RPSF in the following (acronym for reflection PSF). Figure~\ref{fig:setup}e displays examples of RSPF extracted in depth of the opaque cornea. \alex{The} spatial extension $\delta_\textrm{R}$ of the RPSF indicates the focusing quality and dictates the number $M$ of central lines of $\alex{\mathbf{R}_\textrm{in}}(z)$ that contain the relevant information for imaging: 
\begin{equation}
  M\sim (\delta_R/\delta_0)^2  
\end{equation}
with $\delta_0\sim \lambda/(4NA)$, the confocal maximal resolution of the imaging system. For a field-of-view much larger than the spatial extension of the RPSF ($\Omega\gg\delta_R$), the de-scanned basis is thus particularly relevant for the acquisition of $\mathbf{R}$ ($M\ll N$).  \\

\clearpage
\noindent { \textbf{Quantifying the Focusing Quality}} \\

Figure \ref{fig:Bscan}{d} shows the depth evolution of the RPSF. It exhibits the following characteristic shape: a distorted and enlarged confocal spot due to aberrations on top of a \alex{diffuse} background~\cite{lambert_reflection_2020}. \rev{While the latter component is due to {multiple scattering}, the former component contains the contribution of singly-scattered photons but also a \rev{coherent backscattering} peak~\cite{lambert_reflection_2020} resulting from a constructive interference between multiple scattering paths~\cite{Albada1985,Wolf1985} (Supplementary Figure S13).}

Figure \ref{fig:Bscan}{d} clearly highlights two regimes. In the epithelium ($z<70~\mu$m), the \rev{single scattering} component is predominant and the image of the cornea is reliable although its resolution is affected by aberrations {(Fig.~\ref{fig2}b$_1$)}. Beyond this depth, the {multiple scattering} background is predominant and drastically blurs the image {(Fig.~\ref{fig2}b$_3$)}. The axial evolution of the \rev{single scattering rate} enables the measurement of the scattering mean free path $\ell_s$~\cite{Goicoechea2023} {(Supplementary Section \rev{S4})}. We find $\ell_s \sim  \alex{35}~\mu$m in the stroma {(Fig.~\ref{fig:Bscan}g)}, which confirms the strong opacity of the cornea. The penetration depth limit thus scales as $\ell_s$. {This value is modest compared with theoretical predictions~\cite{badon_multiple_2017} ($\sim 4\ell_s$) but is explained by the occurrence of strong aberrations at shallow depths, partially due to the index mismatch at the cornea surface (Fig.~\ref{fig2}d).}\\

\begin{figure}[h!]
            \centering
            \includegraphics[width=\linewidth]{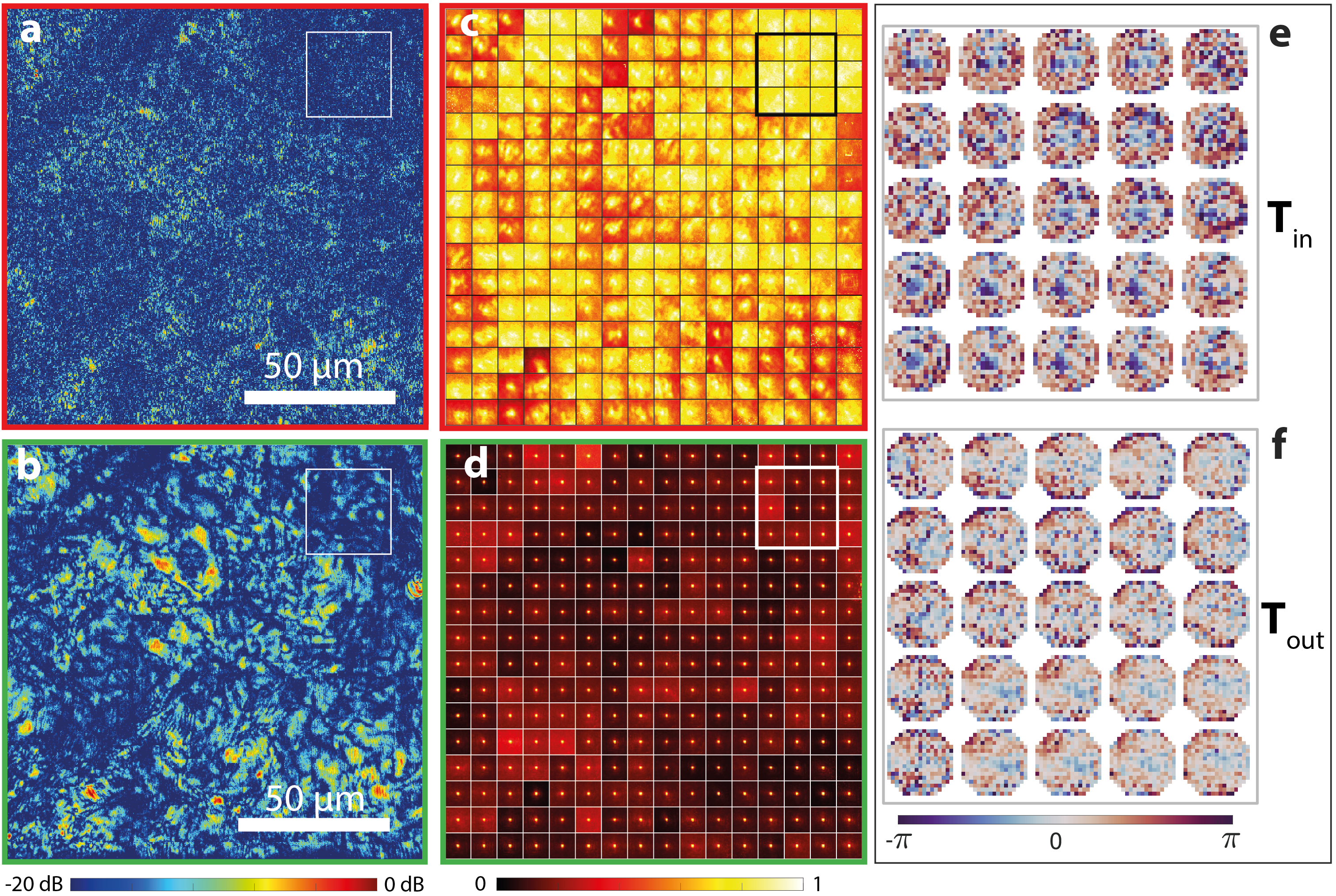}
            \caption{\textbf{\ulysse{Time-gated} Transmission Matrix for Local Compensation of Forward Multiple Scattering.} 
            \textbf{a,b.} Confocal field of view before and after the correction process at 200 $\mu$m-depth, respectively {(scale bar: $50~\mu$m)}. \textbf{c,d.} Maps of the local reflection point-spread functions (RPSFs) {(de-scan field-of-view: $7\times7~\mu$m$^2$)} over the field of view, before and after the correction process, respectively. \textbf{e,f.} {{Sub-part} of matrices, \al{$\bm{\mathcal{T}}_\mathrm{in}$} and \al{$\bm{\mathcal{T}}\out$}}, respectively, for the area delimited by the square box in panels \textbf{a}-\textbf{d}.}
            \label{fig:correction}
\end{figure}

The RSPF also fluctuates in the transverse direction. To that aim, a map of local RPSFs (Fig.~\ref{fig:correction}c) can be built by considering the back-scattered intensity over limited spatial windows (Methods). \alex{This map shows important fluctuations due to: (\textit{i}) the variations of the medium reflectivity that acts on the level of the confocal spot with respect to the diffuse background; (\textit{ii}) the lateral variations of the optical index upstream of the focal plane that induce distortions of the confocal peak}. Such complexity implies that any point in the medium will be associated with its own distinct focusing law. Nevertheless, spatial correlations subsist between RSPFs in adjacent windows (Fig.~\ref{fig:correction}c). Such correlations can be explained by a physical phenomenon often referred to as isoplanatism in AO~\cite{roddier_adaptive_1999} and that results in a locally-invariant PSF~\cite{Judkewitz2015}. We will now see how this local isoplanicity can be exploited for the estimation of the $\mathbf{T}$-matrices. 
\\

\clearpage

\noindent { \textbf{Iterative Phase Reversal of Wave Distortions}} \\

\alex{ To that aim, we will exploit and extend the distortion matrix concept introduced in a previous work~\cite{badon_distortion_2020}. 
Interestingly,} a Fourier transform over the coordinate $\Delta \bm{\rho}_\textrm{in}$ of each de-scanned wave-field, $\alex{R_{\alex{\textrm{in}}}}(\Delta \bm{\rho}_\textrm{in},\alex{\bm{\rho}}\out,\alex{z})$, \alex{actually} yields the wave distortions seen from the input pupil plane \alex{(Fig.~\ref{fig:principle}d)} :
\begin{equation}
\label{FT}
 \mathbf{D}_{\alex{\textrm{in}} }\alex{(z)} = \al{\mathbf{T}_0} \times \alex{\mathbf{R}_{\alex{\textrm{in}}}}\alex{(z)}
\end{equation}
where \al{$\mathbf{T}_0$} denotes the Fourier transform operator, $\al{T_0}(\mathbf{u},\Delta \bm{\rho})= \exp \left (-i 2\pi\mathbf{u}.\Delta \bm{\rho} / {\lambda f} \right )$, $\lambda$ the central wavelength and $f$ the MO focal length. $\ulysse{\mathbf{D}_{\alex{\textrm{in}}}(z)}=[{D}(\mathbf{u}_\textrm{in},\bm{\rho}_\textrm{out},z)]$ is the distortion matrix that connects any voxel (\ulysse{$\bm{\rho}\out,z$}) in the field-of-view to wave-distortions in the input pupil plane ($\mathbf{u}_\textrm{in}$). 

\alex{As expected in most of biological tissues,} this matrix exhibits local correlations that can be understood in light of the \alex{{shift-shift} memory effect~\cite{Judkewitz2015,osnabrugge_generalized_2017}: Waves produced by nearby points inside \alex{an anisotropic scattering} medium generate highly correlated random speckle patterns in the \alex{pupil plane}.} Figure \ref{fig:setup} illustrates this fact by displaying an example of distortion matrix (Fig.~\ref{fig:setup}g) and reshaped distorted wave-fields for different points \alex{$(\bm{\rho}\out,z)$} (Fig.~\ref{fig:setup}h). A strong similarity can be observed between distorted wave-fronts associated with neighboring points but this correlation tends to vanish when the two points are too far away. 

The next step is to extract and exploit \alex{this} local memory effect for imaging. To that aim, a set of correlation matrices $\mathbf{C}_\textrm{in} (\mathbf{r}_\textrm{p})$ shall be considered between distorted wave-fronts in the vicinity of each point $\mathbf{r}_\textrm{p}$ in the field-of-view (Methods). Under the hypothesis of local isoplanicity, each matrix $\mathbf{C}_\textrm{in} (\mathbf{r}_\textrm{p})$ is analogous to a $\mathbf{R}$-matrix associated with a virtual reflector synthesized from the set of output focal spots~\cite{lambert_distortion_2020} (see  Fig.~\ref{fig:principle}\alex{e} {and Supplementary Section \rev{S2}}). In this fictitious experimental configuration, an iterative phase-reversal \alex{(IPR)} process can be performed to converge towards the incident wave front that focuses perfectly through the heterogeneities of the medium onto this virtual \alex{guide star} (see Fig.~\ref{fig:principle}\alex{f} and Methods). 

\alex{IPR} repeated for each point $\rp$ yields \al{a set of pupil phase laws $\mathcal{T}_\textrm{in}(\mathbf{u},\mathbf{r}_p)$ forming the transmittance matrix $\bm{\mathcal{T}}_{\!\!\textrm{in}}$}. 
Its digital phase conjugation enables a local compensation of aberration and \alex{forward} multiple scattering. An updated de-scanned matrix can then be built:
\begin{equation}
\label{dopc}
     \alex{\vec{R}_{\alex{\textrm{in}}}}=    \al{\mathbf{T}_0^{\dag}}  \times \left [ {\al{\bm{\mathcal{T}}}_{\!\!\textrm{in}}^* \circ \mathbf{D}}_{\alex{\textrm{in}}} \right ]
\end{equation}
where the symbol $\dag$ stands for transpose conjugate and $\circ$ for the Hadamard product. The same process can be repeated by exchanging input and output to estimate the output 
\al{transmittance} matrix $\bm{\mathcal{T}}_{\!\!\textrm{out}}$ \alex{(Methods)}. \al{The element wise product between the free space transmission matrix $\mathbf{T}_0$ and the transmittance matrix $\bm{\mathcal{T}}$ constitutes an estimator of the time-gated transmission matrix $\mathbf{T}$. The latter matrix contains the impulse responses $T(\mathbf{u},\mathbf{r})$ between the pupil plane $\mathbf{u}$ and each voxel $\mathbf{r}$ inside the medium around the ballistic time $\tau_B$. Note that this matrix not only contains a ballistic (possibly aberrated) component but also grasps forward multiple scattering paths which display a time-of-flight in the same coherence time as ballistic photons. In the following, we show how these complex trajectories can be harnessed thanks to RMI}.\\

\noindent { \textbf{Multi-Scale Analysis of the Distortion Matrix}} \\

\al{To that aim, a critical aspect} is the choice of the spatial window over which wave distortions shall be analyzed. On the one hand, the isoplanatic assumption is valid for low-order aberrations that are associated with extended isoplanatic patches. On the other hand, forward multiple scattering gives rise to high-order aberrations that exhibit a coherence length that decreases with depth until reaching the size of a speckle grain beyond $\ell_t$~\cite{Judkewitz2015}. However, each spatial window should be large enough to encompass a sufficient number of independent realizations of disorder~\cite{lambert_ultrasound_2022}. Indeed, the bias of our $\mathbf{T}-$matrix estimator scales as follows (see Supplementary Section \rev{S3}):
\begin{equation}
\label{bias}
|\delta \al{\mathcal{T}}(\mathbf{u},\rp)|^2 \sim {1}/({\mathcal{C}^2 N_W})
\end{equation}
with $N_W$ the number of resolution cells in each spatial window. $C$ is a coherence factor that is a direct indicator of the focusing quality~\cite{mallart_adaptive_1994}.

\begin{figure}[h!]
    \centering  
    \includegraphics[width=1\linewidth]{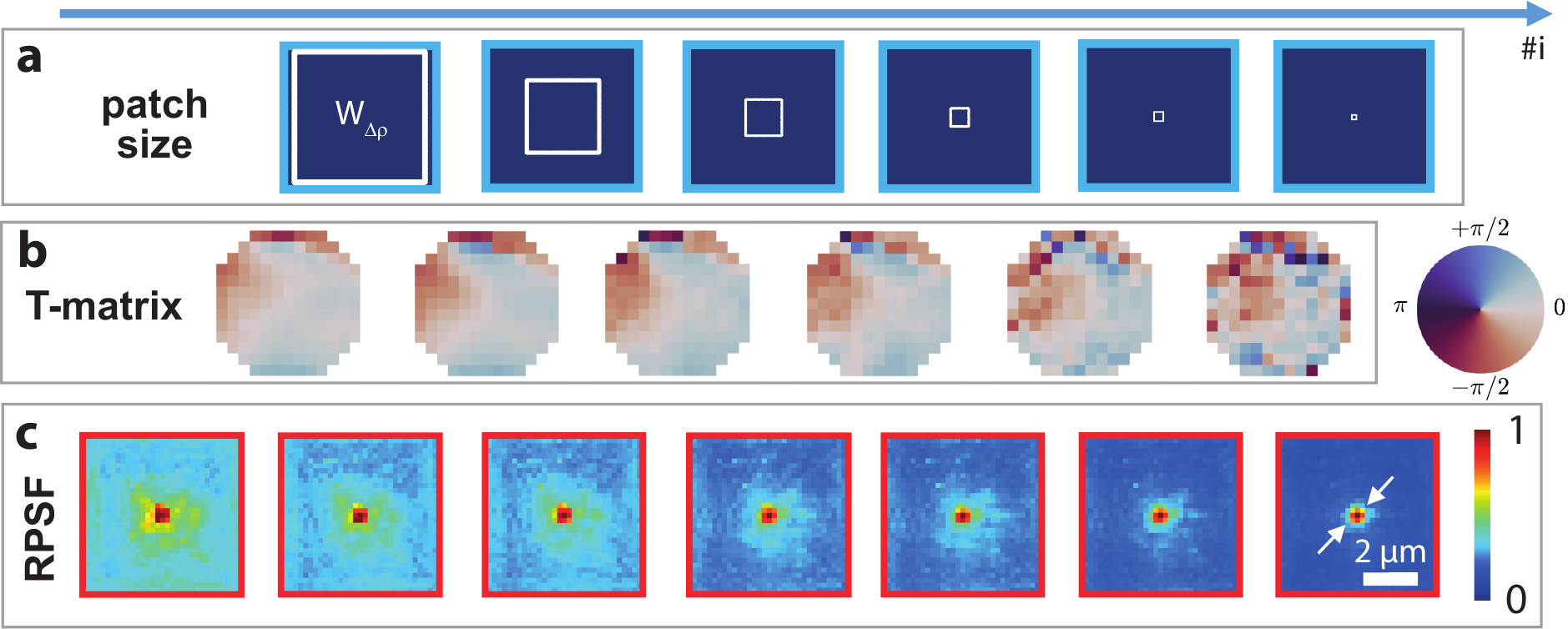}
    \caption{
    \textbf{Multi-scale analysis of wave distortions.}
   \textbf{a.} {The entire field-of-view is $138\times138~\mu$m$^2$. At each step, it is divided into a set of spatial windows whose dimension gradually decreases: from 138, 100, 50, 25, 13 to 6 $\mu$m} 
   \textbf{b.} Evolution of the \al{pupil transmittance  $\al{\mathcal{T}}(\mathbf{u}\out,\rp)$} for one point $\rp$ of the field-of-view at each iteration step. \textbf{c.} Corresponding local RPSF {at $\rp$ before and after} compensation of aberration and scattering using digital phase-conjugation of the optical transfer function displayed in panel \textbf{b} (scale bar: $2~\mu$m).
   Data are from the cross-section at $200~\mu$m depth within the sample.}
   \label{fig5}
\end{figure}

\alex{To limit this bias while addressing the scattering component of \al{$\bm{\mathcal{T}}$}, an iterative multi-scale analysis of $\mathbf{D}$ is proposed (Methods). It consists in gradually reducing the size of the virtual guide star by: (\textit{i}) alternating the correction at input and output (Supplementary Section \rev{S3}); (\textit{ii}) dividing by two the size of overlapping spatial windows at each iterative step (Fig.~\ref{fig5}a). Thereby the RPSF extension is gradually narrowed (Fig.~\ref{fig5}b) and the coherence factor $\mathcal{C}$ increased. The spatial window can thus be reduced accordingly at the next step while maintaining an acceptable bias (Eq.~\ref{bias}). It enables the capture of finer angular and spatial details of the $\al{\bm{\mathcal{T}}}-$matrix at each step (Fig.~\ref{fig5}c) while ensuring the convergence of IPR. As discussed further, the end of the process is monitored by the memory effect that shall exhibit the \al{${\bm{\mathcal{T}}}-$}matrix (Supplementary Section \rev{S3}). \rev{The whole process is validated by a reference imaging experiment on a resolution target placed behind an opaque tissue layer (Supplementary Figure S8).}}\\ 

\noindent { \textbf{\rev{Transmittance} Matrix and Memory Effect}} \\

Figures~\ref{fig:correction}e and f show a sub-part of the \al{$\bm{\mathcal{T}}-$}matrices measured at depth $z=200~\mu$m \alex{for final patches of {$6\times 6~\mu$m$^2$}}. Spatial reciprocity should imply equivalent input and output aberration phase laws. This property is not checked by our estimators. Indeed, the input aberration phase law accumulates not only the input aberrations of the sample-arm but also those of the reference arm \alex{(Supplementary Section \rev{S4}).} Therefore, the sample-induced aberrations can be investigated independently from the imperfections of the experimental set up by considering the output matrix \al{$\bm{\mathcal{T}}_{\! \! \textrm{out}}$}.

\begin{figure}[h!]
    \centering
    \includegraphics[width=\linewidth]{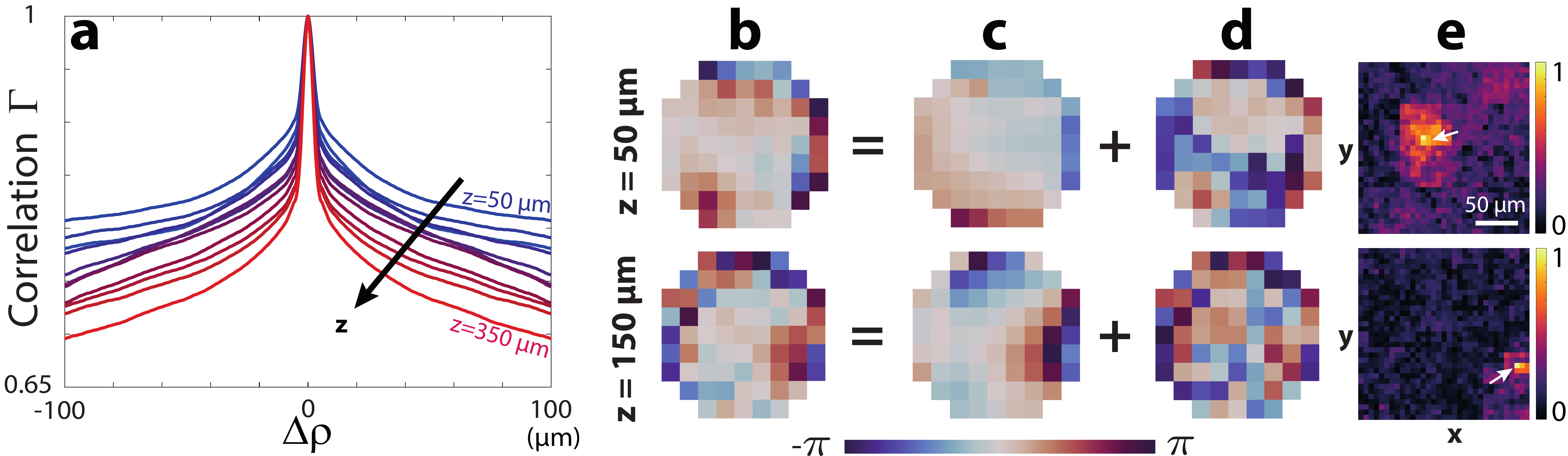}
    \caption{\textbf{Revealing the memory effect exhibited by the $\al{\bm{\mathcal{T}}}-$matrix.} \textbf{a.} Transverse evolution of the mean correlation function of the transmitted wave-field from shallow (blue) to large (red) depths. \textbf{b.} The phase of each transmitted wave-field is the sum of: \textbf{c.} a spatially-invariant aberration phase function;  \textbf{d.} a complex scattering law exhibiting high spatial frequencies. \textbf{e.} The spatial correlation of the latter component with the $\al{\bm{\mathcal{T}}}-$matrix provides a map of the corresponding isoplanatic patch (scale bar: $50~\mu$m).}
    \label{fig:CorrelationsLoisAberrations}
\end{figure}

An analysis of its spatial correlations~\cite{lambert_ultrasound_2022} (Methods) \rev{and its angular decomposition (Supplementary Figure \alex{S12})} shows that wave distortions induced by the cornea are made of two contributions : (\textit{i}) {an almost spatially-invariant aberrated component } (Fig.~\ref{fig:CorrelationsLoisAberrations}a) associated with long-scale fluctuations of the refractive index (Fig.~\ref{fig:CorrelationsLoisAberrations}c)
; (\textit{ii}) a forward multiple scattering component (Fig.~\ref{fig:CorrelationsLoisAberrations}d) 
\rev{giving rise to an angular dispersion of photons between the cornea surface and the focal plane. The latter component is} associated with a short-range memory effect whose extension drastically decreases in depth (Figs.~\ref{fig:CorrelationsLoisAberrations}a,e). \alex{The access to \rev{this} contribution fundamentally differentiates RMI from conventional AO that only provides an access to the irrotational component of wave distortions~\cite{wu2023singleshot} (Supplementary Section \rev{S4}).} 

\alex{The memory effect is also a powerful tool to monitor the convergence of the IPR process. When the spatial window is too small (3$\times$3 $\mu$m$^2$), IPR provides a spatially-incoherent \al{${\bm{\mathcal{T}}}-$matrix} and leads to a bucket-like image (Supplementary Figure \rev{S7}). This observable thus indicates when the convergence towards \al{${\bm{\mathcal{T}}}$} is fulfilled or when the algorithm shall be stopped.} \\

\clearpage

\noindent { \textbf{Deep Volumetric Imaging}} \\

Eventually, the \al{$\bm{\mathcal{T}}$}-matrix can be used to compensate for local aberrations over the whole field-of-view. \alex{To that aim, a} digital phase conjugation \alex{is} performed at input and output (Eq.~\ref{dopc}). The comparison between the initial and resulting images (Figs.~\ref{fig:correction}a,b) demonstrates the benefit of a local compensation of aberration and scattering. The drastic gain in resolution and contrast provided by RMI enables to reveal a rich arrangement of biological structures (cells, striae, \textit{etc.}) that were completely blurred by scattering in the initial image. For instance, a stromal stria, indicator of keratoconus~\cite{Grieve2017}, is clearly revealed on the RMI B-scan (Fig.~\ref{fig2}f) while it was hidden by the multiple scattering fog on the initial image (Fig.~\ref{fig2}c). The B-scan shows that RMI  provides a full image of the cornea with the recovery of its different layers throughout its thickness ($350~\mu$m $\sim \alex{10}\ell_s$, {see also Supplementary Movies}).

The gain in contrast and resolution can be quantified by investigating the RSPF {after RMI}. A close-to-ideal confocal resolution {($ 230~$nm vs. $\delta_0 \sim215~$nm)} is reached throughout the cornea thickness  (Fig.~\ref{fig2}e). The confocal-to-diffuse ratio is increased by a factor up to 15~dB in depth {(Supplementary Section \rev{S4})}. Furthermore, the map of local RPSFs displayed in Fig.~\ref{fig:correction}d shows the efficiency of RMI for addressing extremely small isoplanatic patches.\\

\noindent {\large  \textbf{Discussion}} \\

\alex{In this experimental proof-of-concept, we demonstrated the capacity of RMI to exploit forward multiple scattering for deep imaging of biological tissues. This work introduces several crucial elements, thereby leading to a better imaging performance than previous studies.}

\alex{First, the proposed IPR algorithm outperforms iterative time reversal processing~\cite{badon_distortion_2020} for local compensation of aberrations in scattering media because it can evaluate the focusing laws over a larger angular domain (Supplementary Figure \rev{S2}). Second, the bias of our $\al{\mathbf{{T}}}$-matrix estimator has been expressed analytically (Eq. \ref{bias}) as a function of a coherence factor that grasps the blurring effect of aberrations and multiple scattering. This led us to define a multi-scale strategy for matrix imaging with a fine monitoring of its convergence based on the memory effect. The latter observable is a real asset as it provides an objective criterion to: (\textit{i}) optimize the resolution of our $\mathbf{T}-$matrix estimator (Supplementary Section \rev{S3}); (\textit{ii}) compare our approach with alternative methods such as the CLASS algorithm~\cite{kang_high-resolution_2017,yoon_laser_2020,Kwon2023} (Supplementary Section \rev{S5}). Our multi-scale process enables us to target isoplanatic areas more than four times smaller than CLASS. Interestingly, those two approaches are based on the maximization of different physical quantities: the confocal intensity for CLASS; the coherence of the wave-field induced by a virtual guide star for IPR. Hence they are, in principle, perfectly complementary and could be advantageously combined in the future.}

Although this experimental proof-of-concept is promising for deep optical imaging of biological tissues, it also suffers from several limitations that need to be addressed in future works. First, FFOCT is not very convenient for 3D in-vivo imaging since it requires an axial scan of the sample. Another possibility would be to move the reference arm and measure $\mathbf{R}$ as a function of the time-of-flight. An access to the time (or spectral) dependence of the $\mathbf{R}-$matrix is actually critical to reach a larger penetration depth. 
Indeed, the focusing law extracted from a time-gated $\mathbf{R}-$matrix is equivalent in the time domain to a simple application of time delays between each angular component of the wave-field. Yet, the diffusive regime requires to address independently each frequency component of the wave-field to make multiple scattering paths of different lengths constructively interfere on any focusing point in depth. \alex{On the one hand, the exploitation of the {chromato-axial} memory effect~\cite{Zhu2020} will be decisive to ensure the convergence of IPR over isoplanatic volumes~\cite{Bureau2023}. On the other hand, the {tilt-tilt} memory effect~\cite{osnabrugge_generalized_2017} can also be leveraged by investigating the distortion matrix, not only in the pupil plane, but in any plane lying between the medium surface and the focal plane, thereby mimicking a multi-conjugate AO scheme~\cite{Kang2023}.}

Beyond the diffusive regime, another blind spot of this study is the medium movement during the experiment~\cite{Jang2014,Scholler2020}. In that respect, the matrix formalism shall be developed to include the medium dynamics. Moving speckle can actually be an opportunity since it can give access to a large number of speckle realizations for each voxel. A high resolution $\mathbf{T}-$matrix could be, in principle, extracted without relying on any isoplanatic assumption~\cite{Osmanski2012}. 

To conclude, this study is a striking illustration of a pluri-disciplinary approach in wave physics. A passive measurement of the $\mathbf{R}-$matrix is indeed an original idea coming from seismology~\cite{Campillo2003}. The $\mathbf{D}-$matrix is inspired by stellar speckle interferometry in astronomy~\cite{Labeyrie1970}. The $\mathbf{T}-$matrix is a concept that has emerged both from fundamental studies in condensed matter physics~\cite{rotter_light_2017} and more applied fields such as MIMO communications~\cite{Foschini1998} and ultrasound therapy~\cite{Tanter2001}. The emergence of high-speed cameras and the rapid growth of computational capabilities now makes matrix imaging mature for deep in-vivo optical microscopy.\\

\noindent {\large \textbf{Methods}} \\

\noindent {\textbf{Experimental set up}} 

The full experimental setup is displayed in {Supplementary Figure S1}. It is made of two parts: (\textit{i}) a polarized Michelson interferometer illuminated by a broadband LED source (Thorlabs M850LP1, $\lambda_\circ=850~$nm, $\Delta\lambda=35~$nm) in a pseudo-Kohler configuration, thereby providing at its output two identical spatially-incoherent and broadband wave-fields of orthogonal polarization, the reference one being shifted by a lateral position $\Delta \bm{\rho}_\textrm{in}$ by tilting the mirror in the corresponding arm; (\textit{ii}) a polarized Linnik interferometer with microscope objectives (Nikon N60X-NIR, $\mathrm{M}=60\times$, $\mathrm{NA}=1.0$) in the two arms and a CMOS camera (Adimec Quartz 2A-750, 2Mpx) at its output. The de-scanned beam at the output the first interferometer illuminates the reference arm of the second interferometer and is reflected by the reference mirror placed in the focal plane of the MO.  The other beam at the output of the first interferometer illuminates the sample placed in the focal plane of the other MO. The CMOS camera, conjugated with the focal planes of the MO, records the interferogram between the beams reflected by each arm of the Linnik interferometer. The spatial sampling of each recorded image is {$\delta_0=230~$nm} and the field-of-view is $275\times 275~\mu$m$^2$.\\ 

\noindent {\textbf{Cornea}} 

The human cornea under study is a pathological surgical specimen that was provided by the Quinze-Vingts National Eye Hospital operating room at the time of keratoplasty. The use of such specimens was approved by the Institutional Review Board (Patient Protection Committee, Ile-de-France V) and adhered to the tenets of the Declaration of Helsinki as well as to international ethical requirements for human tissues. The ethics committee waived the requirement for informed written consent of patient; however, the patient provided informed oral consent to have their specimen used in research.\\

\noindent {\textbf{Experimental procedure}} 

The experiment consists in the acquisition of the de-scanned reflection matrix $\mathbf{R}_{\textrm{in}}$. To that aim, an axial scan of the sample is performed over the cornea thickness ($350~\mu$m) with a sampling of $2~\mu$m (i.e 185 axial positions). For each depth, a transverse scan of the de-scanned position $\Delta \bm{\rho}_\textrm{in}$ is performed over a {$2.9\times2.9$ $\mu$m$^2$} area with a spatial sampling $\delta_0=230~$nm (that is to say 169 input wave-fronts instead of 10$^6$ input wave-fronts in a canonical basis). For each scan position $(\Delta \rho,z)$, a complex-reflected wave field is extracted by phase shifting interferometry from four intensity measurements. This measured field is averaged over 5 successive realisations (for denoising). The integration time of the camera is set to 5 ms. Each wave-field is stored in the de-scanned reflection matrix $\alex{\vec{R}_\textrm{in}}=[\alex{R_\textrm{in}}(\Delta \bm{\rho}_\textrm{in},\bm{\rho}\out)]$ (Fig.~\ref{fig:setup}). The duration time for the recording of $\alex{\vec{R}_\textrm{in}}$ is of $\sim 30$ s at each depth. The post-processing of the reflection matrix (IPR and multi-scale analysis) to get the final image took only a few minutes on Matlab. The experimental results displayed in Fig.~\ref{fig:correction} and \ref{fig5} at a single depth $z=200~\mu$m have been obtained by performing a de-scan over a {$7\times 7~\mu$m$^2$ area with a spatial sampling $\delta_0=230~$nm (961 input wave-fronts).}\\ 

\noindent {\textbf{Local RPSF}} 

To probe the local RPSF, the field-of-view is divided into regions that are defined by their central midpoint \alex{$\rp=(\bm{\rho}_\textrm{p},z)$} and their \alex{lateral} extension $L$. A local average of the back-scattered intensity can then be performed in each region:
\begin{equation}
   I(\Delta \bm{\rho}_\textrm{in},\rp)=\langle |\alex{R_\textrm{in}}(\Delta \bm{\rho}_\textrm{in},\alex{\bm{\rho}\out,z})|^2 W_{L}(\alex{\bm{\rho}\out - \bm{\rho}_\textrm{p}}) \rangle_{\alex{\bm{\rho}\out}}
\end{equation}
where \alex{$W_{L}(\bm{\rho}\out - \bm{\rho}_\textrm{p}) = 1$ for $|x\out - x_\mathrm{p}|<L$ and $|y\out - y_\mathrm{p}|<L$} , and zero otherwise. \\

\noindent \alex{\textbf{Multi-scale compensation of wave-distortions}}

\alex{The multi-scale process consists in an iterative compensation of aberration and scattering phenomena at input and output of the reflection matrix. To that aim, wave distortions are analyzed over spatial windows $W_L$ that are gradually reduced at each step $q$ of the procedure, such that: 
\begin{equation}
L=FOV/2^q 
\end{equation}
where $FOV$ denotes the initial field-of-view.}

\alex{The whole procedure is summarized in Supplementary Figure \rev{S4}. At each stage of this iterative process, the starting point is the de-scanned reflection matrix $\mathbf{R}^{(q-1)}_\textrm{in}$, obtained at the previous step, $\mathbf{R}^{(0)}_\textrm{in}$ being the reflection matrix recorded by our experimental set up (Fig.~\ref{fig:setup}). An input distortion matrix $\mathbf{D}_\textrm{in}^{(q)}$ is deduced from $\mathbf{R}_\textrm{in}^{(q)}$ via a numerical Fourier transform (Eq.~\ref{FT}). A local correlation matrix of wave distortions is then built around each point $\rp$ of the field-of-view:
\begin{equation}
\label{corr_in}
    C_\textrm{in} (\mathbf{u}_\textrm{in}, \mathbf{u}'_\textrm{in},\rp)= \left \langle {D}_\textrm{in}^{(q)}(\mathbf{u}_\textrm{in},\alex{\bm{\rho}\out,z}){D}_\textrm{in}^{(q)*}(\mathbf{u}'_\textrm{in},\alex{\bm{\rho}\out,z}) W_{{L}}(\alex{\bm{\rho}\out - \bm{\rho}_\textrm{p}})\right \rangle _{\alex{\bm{\rho}\out}}
\end{equation}
}

\alex{IPR is then applied to each correlation matrix $\mathbf{C}_\textrm{in}(\rp)$ (see further and Supplementary Section \rev{S3}). The resulting input phase laws, $\hat{\phi}_\textrm{in}(\rp)$, are used to compensate for the wave distortions undergone by the incident wave-fronts:
\begin{equation}
    {\mathbf{R}}'_\textrm{in}= \al{\mathbf{T}_0^{\dag}} \times \left [\exp \left ( -i \hat{\bm{\phi }}_\textrm{in} \right) \circ \mathbf{D}_\textrm{in}^{(q-1)} \right ]
\end{equation}
The corrected matrix ${\mathbf{R}}'_\textrm{in}$ is only intermediate since phase distortions undergone by the reflected wave-fronts remain to be corrected.}

\alex{To that aim, an output de-scanned matrix  ${\mathbf{R}'_{\textrm{out} }}(z)$} is deduced from the input de-scanned matrix  \alex{${\mathbf{R}'_{\textrm{in}}}(z)$} using the following change of variable (Supplementary Figure \rev{S5}):
\begin{equation}
\label{change}
   \alex{ R'_\textrm{out}}(\bm{\rho}_\textrm{in},\Delta \bm{\rho}_\textrm{out},z)=\alex{R'_\textrm{in}}(-\Delta \bm{\rho}_\textrm{out},\bm{\rho}_\textrm{in}+\Delta \bm{\rho}_\textrm{out},z)
\end{equation}
with $\Delta \bm{\rho}_\textrm{out}=\bm{\rho}\out-\bm{\rho}_\textrm{in}=-\Delta \bm{\rho}_\textrm{in}$. An output distortion matrix is then built by applying a Fourier transform over the de-scanned coordinate:
\begin{equation}
 \alex{ \mathbf{D}'_{\textrm{out}}=  \vec{R}'_{\textrm{out}}}\times \al{\mathbf{T}_0^T }
\end{equation}
where the superscript $^T$ stands for matrix transpose. From \alex{${\mathbf{D}}'_{\textrm{out}}$}, one can build a correlation matrix $\mathbf{C}\out$ for each point $\rp$:
\begin{equation}
\label{corr_out}
    C\out (\mathbf{u}\out, \mathbf{u}'\out,\rp)= \left \langle \alex{{D}'_\textrm{out}}(\alex{\bm{\rho}_\textrm{in} },\mathbf{u}\out,\alex{z_p})\alex{{D}_\textrm{out}^{'*}}(\alex{\bm{\rho}}_{\textrm{in} },\mathbf{u}'\out,\alex{z_p}) W_{L}(\alex{\bm{\rho}_\textrm{in} - \bm{\rho}_\textrm{p}})\right \rangle _{\alex{\bm{\rho}_\textrm{in}}}
\end{equation}
The IPR algorithm described further is then applied to each matrix $\mathbf{C}\out(\rp)$. \alex{The resulting output phase laws, $\hat{\bm{\phi}}_\textrm{out}(\rp)$, are leveraged to compensate for the residual wave distortions undergone by the reflected wave-fronts:
\begin{equation}
    {\mathbf{R}}_\textrm{out}^{(q)}= \left [  \mathbf{D}'_\textrm{out} \al{\circ} \exp \left ( -i \bm{\phi }_\textrm{out} \right) \right ] \times \al{\mathbf{T}}_0^{*}
\end{equation}
The RPSFs displayed in Fig.~\ref{fig5}c are extracted from the matrices ${\mathbf{R}}_\textrm{out}^{(q)}  $ obtained at the end of each iteration of the multi-scale process. An input de-scanned matrix, combining the input and output corrections, is finally obtained by performing the following change of variables:}\alex{
\begin{equation}
\label{change2}
   \alex{ R^{(q)}_\textrm{in}}(\Delta \bm{\rho}_\textrm{in},\bm{\rho}_\textrm{out},z)=\alex{R^{(q)}_\textrm{out}}(\bm{\rho}_\textrm{out}-\Delta \bm{\rho}_\textrm{in},-\Delta \bm{\rho}_\textrm{in},z)
\end{equation}
This matrix $\mathbf{R}^{(q)}_\textrm{in}$ is the starting point of the next stage of the multi-scale process, and so on.}

\alex{\al{The $\bm{\mathcal{T}}$}-matrices correspond to the cumulative function of the aberration phase laws:
\begin{equation}
    \al{\bm{\mathcal{T}}}^{(q)}_{\!\textrm{in/out}}=\al{\bm{\mathcal{T}}}^{(q-1)}_\textrm{in/out} \circ \exp \left ( i \bm{\phi }_\textrm{in/out}^{(q)} \right) = \prod_{k=1}^q \exp \left ( i \bm{\phi }_{\!\textrm{in/out}}^{(k)} \right)
\end{equation} 
Figure~\ref{fig5}b shows the evolution of one line of the \al{transmittance} matrix $\al{\bm{\mathcal{T}}}^{(q)}_\textrm{out}$ throughout the RMI process. The iterative procedure is stopped by investigating the correlation properties of this estimator (see further and Supplementary Section \rev{S3}).}\\

\noindent {\textbf{Iterative phase reversal algorithm.}} 

The IPR algorithm is a computational process that provides an estimator of the \al{pupil transmittance matrix}, \alex{$\al{\mathcal{T}}(\mathbf{u},\rp)=\exp \left [ i \phi(\mathbf{u},\rp) \right ] $}, that links each point $\mathbf{u}$ of the pupil plane with each voxel $\rp$ of the cornea volume. To that aim, the correlation matrix $\mathbf{C}$ computed over the spatial window $W_{L}$ centered around each point $\rp$ is considered \alex{(Eqs.~\ref{corr_in} and ~\ref{corr_out})}.  Mathematically, the algorithm is based on the following recursive relation:
\begin{equation}
\alex{\hat{\bm{\phi}}^{(n)} (\rp) = \mathrm{arg}\left\{  \mathbf{C}(\rp) \times \exp \left [ i \hat{\bm{\phi}}^{(n-1)}(\rp) \right ]  \right\}}
\end{equation}
where \alex{$\hat{\bm{\phi}}^{(n)}$} is the estimator of \alex{${\bm{\phi}}$} at the $n^\textrm{th}$ iteration of the phase reversal process. \alex{$ \hat{\phi}^{(0)}$} is an arbitrary wave-front that initiates the process (typically a flat phase law) and $\alex{\hat{\bm{\phi}}}= \lim_{n\to\infty} \alex{\hat{\bm{\phi}}}^{(n)}$ is the result of IPR. \\

\noindent {\textbf{Aberration and Scattering Components of the $\mathbf{T}$-matrix.}}

The spatial correlation of transmitted wave-fields are investigated at each depth $z$ by computing the correlation matrix of $\al{\bm{{\mathcal{T}}}}\out$: $\mathbf{C}_{\al{\bm{\mathcal{T}}}}=\al{\bm{{\mathcal{T}}}}\out \times \al{\bm{{\mathcal{T}}}}\out^{\dagger}$. A mean correlation function $\Gamma$ can be computed by performing the following average:
\begin{equation}
    \Gamma (\Delta \bm{\rho},z)= \left \langle C_{\al{\mathcal{T}}}(\bm{\rho}_\textrm{in},\bm{\rho}_\textrm{in}+\Delta \bm{\rho},z) \right \rangle_{\bm{\rho}_\textrm{in}}
\end{equation}
The correlation function $\Gamma$ displayed in Fig.~\ref{fig:CorrelationsLoisAberrations}a shows that the matrix $\al{\bm{{\mathcal{T}}}}\out$ can be decomposed as a spatially-invariant component \al{$\bm{\mathcal{A}}\out$} and a short-range correlated component \al{$\bm{\mathcal{S}}\out$}. Each component can be separated by performing a singular value decomposition of $\al{\bm{\mathcal{T}}}\out$, such that
\begin{equation}
   \al{\bm{{\mathcal{T}}}}\out= \sum_{p=1}^N s_p \mathbf{U}_p \mathbf{V}_p^\dagger
\end{equation}
where $s_p$ are the positive and real singular values of $\al{\bm{{\mathcal{T}}}}\out$ sorted in decreasing order, $\mathbf{U}_p $ and $ \mathbf{V}_p$ are unitary matrices whose columns correspond to the singular vectors of $ \al{\bm{{\mathcal{T}}}}\out $ in the pupil and focal planes, respectively. The first eigenspace of $\al{\bm{{\mathcal{T}}}}\out$ provides its spatially-invariant aberrated component:
 $   \al{\bm{\mathcal{A}}\out} = s_1 \mathbf{U}_1 \mathbf{V}_1^\dagger.$
The higher rank eigenstates provide the \alex{forward} multiple scattering component \al{$\bm{\mathcal{S}}\out$}. Lines or columns of the associated correlation matrix $\alex{\mathbf{C}_{\al{\mathcal{S}}}=\al{\bm{\mathcal{S}}}\out \times \al{\bm{\mathcal{S}}}^\dag \out}$ provides the isoplanatic patches displayed in Fig.~\ref{fig:CorrelationsLoisAberrations}e.\\

\clearpage

\noindent\textbf{Data availability.} The optical data generated in this study are available at Zenodo~\cite{Najar_2023} (\href{https://zenodo.org/record/7665117}{https://zenodo.org/record/7665117}).\\

\noindent\textbf{Code availability.}
Codes used to post-process the optical data within this paper are available from the corresponding author. \\

\noindent\textbf{Acknowledgments.}
The authors wish to thank A. Badon for initial discussions about the experimental set up, K. Irsch for providing the corneal sample and A. Le Ber for providing the iterative phase reversal algorithm. \\The authors are grateful for the funding provided by the European Research Council (ERC) under the European Union's Horizon 2020 research and innovation program (grant agreement nos. 610110 and 819261, HELMHOLTZ* and REMINISCENCE projects, MF and AA, respectively). This project has also received funding from Labex WIFI (Laboratory of Excellence within the French Program Investments for the Future; ANR-10-LABX-24 and ANR-10-IDEX-0001-02 PSL*, MF).\\

\noindent\textbf{Author Contributions Statement.}
A.A. initiated and supervised the project. C.B., V.B. and A.A. designed the experimental setup. U.N., V.B. and P.B. built the experimental set up.  U.N. and V.B. developed the post-processing tools. U.N. performed the corneal imaging experiment. U.N. and A.A. analyzed the experimental results. V.B.and A.A. performed the theoretical study. A.A. and U.N. prepared the manuscript. U.N., V.B., P.B., M.F., C.B., and A.A. discussed the results and contributed to finalizing the manuscript. \\

\noindent\textbf{Competing interests.}
A.A., M.F., C.B. and V.B. are inventors on a patent related to this work held by CNRS (no. US11408723B2, published August 2022). All authors declare that they have no other competing interests. 


\clearpage 

\clearpage

\renewcommand{\thetable}{S\arabic{table}}
\renewcommand{\thefigure}{S\arabic{figure}}
\renewcommand{\theequation}{S\arabic{equation}}
\renewcommand{\thesection}{S\arabic{section}}

\setcounter{equation}{0}
\setcounter{figure}{0}
\setcounter{section}{0}

\begin{center}
\Large{\bf{Supplementary Information}}
\end{center}
\normalsize
This document provides further information on: (i) the experimental set up; (ii) the theoretical expression of the de-scanned matrix; (iii) the measurement of the scattering mean free path; (iv)  the theoretical expression of the correlation matrix; (v) the estimation of the transmission matrix; (vi) the contrast enhancement provided by reflection matrix imaging.

\clearpage 
 
\section{Detailed experimental set-up}

\begin{figure}[h!]
    \centering
    \includegraphics[width=\linewidth]{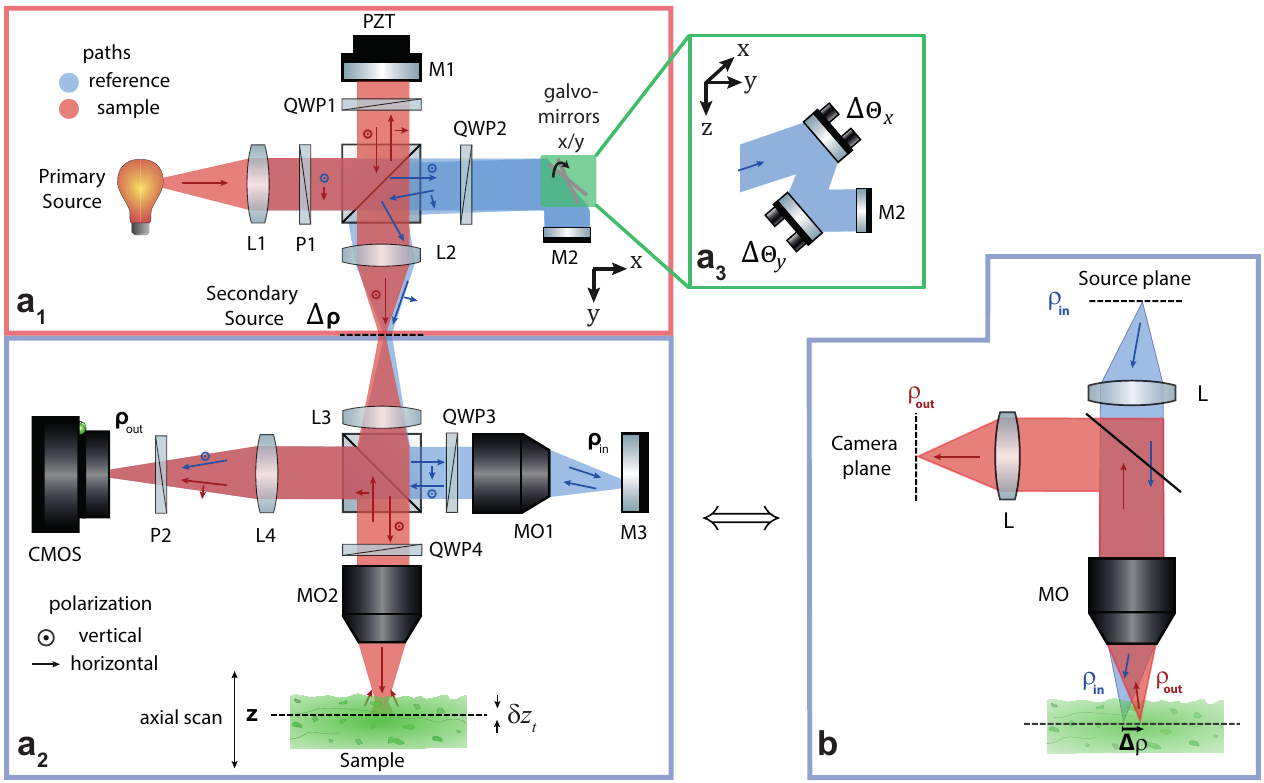}
    \caption{\textbf{De-scanned measurement of the reflection matrix.}  \textbf{a.} Experimental set up. P: polarizer, L: lens, QWP: quarter-wave plate, M: mirror, PZT: piezo-electric actuator, PBS: polarisation beam splitter, MO: microscope objective. The apparatus is made up of two parts (\textbf{a}$_1$,\textbf{a}$_2$). 
    \textbf{a}$_1$. Michelson interferometer illuminated by an incoherent light source at its input and generating two twin incoherent beams of orthogonal polarization and laterally shifted from each other at its output. The polarised beam splitter (PBS1) separates the impinging light into a reference path (in blue) and a sample path (in red). 
    The scan mirrors (\textbf{a}$_3$) tilt the reference beam by angles $\Delta\Theta_x$ and $\Delta\Theta_y$ in both transverse directions $x$ and $y$. This allows the scan of the point-spread function along the de-scanned coordinates $\Delta_x$ and $\Delta_y$ in a plane conjugate to the reference mirror plane.
    \textbf{a}$_2$. Michelson interferometer with microscope objectives (MO) in both arms (Linnik configuration). Both beams have orthogonal polarizations and each interferometer arm includes a quarter-wave plate (QWP). The output beams are collected by the L4 lens and interfere on the camera after having been projected on a $45^\circ$-rotated polarizer (P2). 
    \textbf{b.} Equivalent layout in the case of a coherent measurement. The source plane, the focal plane, and the camera planes are conjugated. Displacing a point source $\bm{\rho}_\textrm{in}$ in the source plane discretely scans the focal plane inside the sample. The illuminated area is imaged in the camera plane; in an epi-detection configuration.}
    \label{fig:setup}
\end{figure}

The full experimental set up is displayed in Supplementary Fig.~\ref{fig:setup}. The setup is divided into two building blocks, labelled (a) and (b). The first component is a Michelson interferometer [Supplementary Fig.~\ref{fig:setup}a]. The light source is a broadband LED (Thorlabs M850LP1, $\lambda_\circ=850~$nm, $\Delta\lambda=35~$nm), is placed in a plan conjugated with the focal plane within the sample, so as to illuminate the focal plane with the image of the source. The source's illumination pattern is not uniform and is smaller than the maximal extension of the field of view allowed by the microscope objectives' numerical aperture.
As a result, the sample's intensity in the focal plane is modulated by the image of the source. In order to get a uniform illumination of the whole field of view, we set up a pseudo-Koehler illumination apparatus: 
An aspheric lens and a diaphragm are placed right in front of the source, such that the incident beam is collimated in the diaphragm plane. This plane is considered the source plane, and is conjugated to the sample plane by (L1). This way, the image of the source is defocused in the sample plane. This ensures an incoherent~\cite{mertz_introduction_2019}, yet uniform, illumination of the field of view. 

The incident light is collimated using a converging lens (L1) with a focal length $f_1~=~150~$mm. 
The beam transmitted through this lens (L1) is linearly polarized at 45$^\circ$ by a polarizer (P1) so that it is then equally reflected (sample arm) and transmitted (reference arm) by the polarized beam splitter (PBS1). 

The sample beam reflected by (PBS1) is horizontally polarized. It propagates through a quarter-wave plate (QWP1), is reflected by a plane mirror (M1), whose normal axis lies along the optical axis and that is mounted on a piezoelectric actuator (PZT). The reflected beam passes again through the quarter-wave plate (QWP1). This sequence induces a polarization rotation by 90$^{\circ}$ of the reflected beam with respect to the incident beam in the sample arm. The reflected wave can be then transmitted through the beam splitter (PBS1) with a vertical polarization and finally focused in a secondary source plane conjugated with the source plane by means of the lens (L2) of focal length $f_2~=~125~$mm.

The reference beam, vertically polarized at the exit of the polarizer (P1), is transmitted by the beam splitter (PBS1), propagates through a quarter-wave plate (QWP2), is reflected by a set of two galvanometric scan mirrors, and then by the reference mirror (M2).
The set of scan mirrors enables a 2D rotation of the incident wave-field by angles $\bm{\Theta}=(\Theta_x,\Theta_y)$ with respect to the optical axis. 
After reflection on the reference mirror (M2) and on the scan mirrors back again, the reflected beam propagates again through (QWP2). This round trip through (QWP2) enables a 90$^{\circ}$ rotation of the polarization: Whereas the incident light is V-polarized when it enters the reference arm, it is H-polarized when exiting it (see Fig.~\ref{fig:setup}a).
Therefore, the reference beam is reflected by the beam splitter (PBS1) before being focused by the lens (L2) in the secondary source plane. 

Finally, in the secondary source plane, the wave-field is made of two images of the incident light orthogonally polarized and translated with respect to each other by a relative position $\Delta\bm{\rho}=(\Delta x , \Delta y )$. This lateral shift is dictated by the tilt $\bm{\Theta}=( \Theta_x,\Theta_y)$ of the reference beam: $\Delta x = 2 f_2 \tan \Theta_x$ and $\Delta y = 2 f_2 \tan \Theta_y$. 
The factor 2 results from the double reflection on each scan mirror, due to the reflection on the (M2) mirror.
Note also that the optical path difference between the two arms is set to zero by equalizing the length of sample and reference arms for $\Delta\bm{\theta}=\text{0}$.

After the Michelson interferometer, the two orthogonally polarized twin beams enter a Michelson interferometer with two identical microscope objectives in both arms (a configuration known as a Linnik interferometer) [Supplementary Fig.~\ref{fig:setup}b]. 
They are again collimated by a lens (L3) of focal length $f_3$=200~mm. 
The two lenses (L2) and (L3) thus constitute a 4$f$ system which compensates the effects of diffraction between the two interferometers. 

The vertically polarized light (sample beam) is transmitted by a polarized beam splitter cube (PBS2), propagates through a quarter-wave plate (QWP4) before being focused in the focal plane of an immersion microscope objective (MO2, Nikon, 60$\times$, NA=1.0). The light reflected by the sample is then collected by (MO2) and propagates again through the quarter-wave plate (QWP4). Because single scattering tends to preserve polarization, the corresponding wave-field undergoes a 90$^{\circ}$ polarization rotation and gets reflected by the beam splitter (PBS2) before being focused in the plane of the camera using the converging lens (L4) of focal length $f_4=$200~mm. The combination of this lens (L4) with the microscope objective (MO1) entails a magnification $M_4$ of 60.

Regarding the horizontally-polarized beam at the exit of the lens (L3), it is reflected by the beam splitter (PBS2), passes through the quarter-wave plate (QWP3) before being focused by the microscope objective (MO1) identical to (MO2). The light is then reflected by the reference mirror (M3) placed in the focal plane of (MO2) before being collected again by the same microscope objective (MO2). The reflected light comes through the quarter-wave plate (QWP3). As in the other arm, the polarization of the  reflected beam exhibits a 90$^{\circ}$ rotation of its polarization. The beam is now vertically polarized and transmitted by the beam splitter (PBS2), before being focused on the camera with the lens (L4). 

{The detection scheme consists in recording the interferogram between the sample and reference beams based on their orthogonal polarisations projected through a 45-degree polarizer (P2). This experimental configuration allows an enhancement of single scattering and forward multiple scattering  with respect to diffuse light in the sample arm. Indeed, the former components roughly exhibit the same time-of-flight and polarization as reference light while the latter one is characterized by a fully randomized polarization and a longer time-of-flight distribution. This filtering of diffuse light is deliberate since the post-processing method described in the accompanying paper addresses the forward multiple scattering contribution and not the randomly-scattered diffuse light.}

The CMOS camera (Adimec Quartz 2A-750, 2Mpx) records the interferogram {between sample and reference beams} with a spatial sampling equal to $\delta_0=$230~nm given the magnification $M_4$. 
The volume of the sample from which photons can interfere with the reference beam is called the ``coherence volume". Its position is dictated by the optical path difference between the reference and sample arms. Its thickness is inversely proportional to the light spectrum bandwidth~\cite{Fercher1996}:
\begin{equation}
    \delta z_t= \frac{2 \ln2}{n \pi} \left ( \frac{\lambda_0^2}{\Delta \lambda}\right )
\end{equation}
with $\lambda_0$ the central wavelength of the light source and $\Delta \lambda$ its spectral bandwidth. In the present case, $\delta z_t \sim 10 $ $\mu$m. A critical tuning of the experimental set up consists in adjusting the coherence volume with the focal plane of the microscope objective. In a volumetric sample, whose refractive index differs from that of water, the coherence volume no longer coincides with the focusing plane. This focusing defect accumulates with the transverse aberrations generated by the heterogeneities of the medium. However, it is possible to compensate for it by a fine tuning of the length of the reference arm. 

The experimental procedure then consists in recording the de-scanned reflection matrix ${\mathbf{R}_{\textrm{in}}}(z)$ at each depth $z$ of the sample. This latter parameter is swept by means of a motorized axial displacement of the sample carrier. The scan of the relative position $\Delta \bm{\rho}$ between the incident wave-fields in the sample and reference arms is controlled by the tilt imposed by the galvanometer (M2). For each couple $(\Delta \bm{\rho},z)$, the CCD camera conjugated with the MO focal plane records the output intensity:
\begin{equation}
I_{\alpha}(\Delta \bm{\rho},\bm{\rho}_\textrm{out},z) =  \int_{0}^T |e^{i\alpha}E_\textrm{out}(\bm{\rho}_\textrm{out},t)+E_\textrm{out}^{\textrm{(ref)}*}(\Delta \bm{\rho},\bm{\rho}_\textrm{out},t)|^2  \mathrm{d}t
\label{intensity}
\end{equation}
with $t$ the absolute time, $\mathbf{r}$ the position vector on the CCD screen, $E_\textrm{out}(\mathbf{r},\tau)$ the scattered wave field associated with the sample arm, $E_\textrm{out}^{\textrm{(ref)}}(\mathbf{r},\tau)$ the reference wave field; $T$ the integration time of the CCD camera, and $\alpha$ an additional phase term controlled with a piezoelectric actuator placed on mirror (M1) of the first interferometer [Supplementary Fig.~\ref{fig:setup}a]. 
The interference term between the sample and reference beams is extracted from the four intensity patterns (Eq.~\ref{intensity}) recorded at $\alpha=0,$ $\pi/2$, $3\pi/2$ and $\pi$ (phase-stepping interferometry) {and provides the de-scanned wave-field}:
\begin{equation}
\label{cor}
{R_\textrm{in}}(\Delta \bm{\rho},\bm{\rho}_\textrm{out},z)  = \frac{1}{T} \int_{0}^T E_\textrm{out}(\bm{\rho}_\textrm{out},t)E_\textrm{out}^{\textrm{(ref)}*}(\Delta \bm{\rho},\bm{\rho}_\textrm{out},t)  \mathrm{d} t 
\end{equation}

Note that previous studies~\cite{badon_retrieving_2015,badon_spatio-temporal_2016} reported on the passive measurement of the de-scanned reflection matrix at the surface of a scattering sample. Although the experimental set up presented in those studies shares some similarities with the current set up displayed in Supplementary Fig.~\ref{fig:setup} (low-coherence interferometry), there are also several major differences. First, the two arms are illuminated by the light reflected by the sample in Ref.~\onlinecite{badon_retrieving_2015}; hence there is no reference arm.  Second, the reflection matrix is measured at the surface of the sample as a function of time-of-flight, while the current set up measures a time-gated reflection matrix as a function of depth inside the sample. At last, a much larger integration time is required to record the reflection matrix in Ref.~\cite{badon_retrieving_2015} because of the absence of reference arm.

\clearpage

\section{Theoretical analysis}

\subsection{Theoretical expression of the de-scanned matrix}

In this section, we express theoretically  the de-scanned matrix recorded by the experimental set up in Supplementary Figs.~\ref{fig:setup}a,b. To that aim, we will rely on the simple Fourier optics model proposed in a recent paper~\cite{barolle_manifestation_2021} to describe the manifestation of aberrations in FFOCT. For the sake of simplicity, this model is scalar. The large numerical aperture imposes that the recorded wave-field is associated with single scattering events taking place in the focal plane of the MO. 

The wave field $E_{\textrm{out}}(\bm{\rho}_\textrm{out},z)$ reflected by the sample arm in the camera plane can then be expressed as follows~\cite{barolle_manifestation_2021}:
\begin{equation}
\label{Esample}
    E_{\textrm{out}}(\bm{\rho}_\textrm{out},z,\omega) =  \int_{\Sigma_{0}  }\iint_{  \Sigma_{\rho}}  H(\bm{\rho}_\textrm{out},\bm{\rho}_s,z) \gamma(\bm{\rho}_s,z)  H(\bm{\rho}_s,\bm{\rho}_0,z) E_{0}(\bm{\rho}_0,\omega) \diff\bm{\rho}_s \diff\bm{\rho}_0.
\end{equation}
$E_{0}(\bm{\rho}_0,\omega)$ is the incident wave-field in the secondary source plane $\Sigma_{0}$ at frequency $\omega$. Light propagation between $\Sigma_{0}$ and the focal plane $\Sigma_{\rho}$ is described by the impulse response $H(\bm{\rho}_0,\bm{\rho}_s,z)$ between a point in the secondary source plane at transverse coordinate $\bm{\rho}_0$ and a point at transverse coordinate $\bm{\rho}_s$ in the focal plane and at depth $z$ inside the sample. It accounts for sample-induced aberrations. $\gamma(\bm{\rho}_s,z)$ represents the sample reflectivity at depth $z$. By spatial reciprocity, the propagation of the reflected wave-field from the sample to the detector plane is also modelled by the impulse response  $H(\bm{\rho}_s,\bm{\rho}_\textrm{out})$. The relatively narrow bandwidth ($\Delta \lambda \ll \lambda$) of the light source and the use of achromatic optical elements (lens, beam splitter, quarter wave plate) allow us to neglect the dependence of $H$ on frequency $\omega$.

Replacing $\gamma(\bm{\rho}_s,z)$ by a uniform reflectivity in Eq.~\ref{Esample} and taking into account the lateral shift of the reference wave-field induced by the galvanometer M2 [Supplementary Fig.~\ref{fig:setup}] leads to the following previous expression for $E^{(\textrm{ref})}_{\textrm{out}}(\bm{\rho}_\textrm{out},z)$~\cite{barolle_manifestation_2021}: 
\begin{equation}
\label{Eref}
    E^{(\textrm{ref})}_{\textrm{out}}(\bm{\rho}_\textrm{out},\Delta \bm{\rho},z) =  \iint_{\Sigma_{0}  } H_{\textrm{ref}}(\bm{\rho}_\textrm{out}-\bm{\rho}_0)  E_0(\bm{\rho}_0+\Delta \bm{\rho}) \diff\bm{\rho}_0.
\end{equation}
where $H_{\textrm{ref}}$ is the impulse response associated with the reference arm (way and return path) that we assume as spatially-invariant [$H_{\textrm{ref}}(\bm{\rho}_\textrm{out},\bm{\rho}_0)=H_{\textrm{ref}}(\bm{\rho}_\textrm{out}-\bm{\rho}_0)$]. 

The de-scanned wave-field is obtained by extracting the interference term between the reflected wave-fields coming from the sample and reference arms:
\begin{equation}
{R_\textrm{in}}(\bm{\rho}_\textrm{out},\Delta \bm{\rho},z)  =\langle E_\textrm{out}(\bm{\rho}_\textrm{out},\omega) E^{\textrm{(ref)}*}_{\textrm{out}}(\bm{\rho}_\textrm{out},\omega) \rangle
\end{equation}
Assuming a spatially-incoherent incident wave-field [$\langle E_0(\bm{\rho}_0) E_0^*(\bm{\rho}'_0)\rangle=  I_0 \delta (\bm{\rho}_0 - \bm{\rho}'_0) $] and injecting Eqs.~\ref{Esample} and \ref{Eref} into the last equation leads to the following expression for {the} coefficients of ${\mathbf{R}_\textrm{in}}$:
\begin{eqnarray}
{R_\textrm{in}}(\bm{\rho}_\textrm{out},\Delta \bm{\rho},z) &= & I_0 \iint H_\textrm{out}(\bm{\rho}_\textrm{out},\bm{\rho},z) \gamma(\bm{\rho}_s,z) H_\textrm{in} (\bm{\rho}_s,\bm{\rho}_\textrm{out}+\Delta \bm{\rho},z)  
 \diff\bm{\rho} 
    \label{terme_interference}
\end{eqnarray}
with 
\begin{equation}
\label{SR}
H_\textrm{out}\equiv H \mbox{   and   } H_\textrm{in}\equiv H \odot  H^*_\textrm{ref}.
\end{equation}
The symbol $\odot$ stands for the convolution product over the variable $\bm{\rho}_\textrm{in}$. The last equation means that: (\textit{i}) the output focusing matrix, $\mathbf{H}_\textrm{out}$, and the associated $\mathbf{T}-$matrix, $\mathbf{T}_\textrm{out}=\rev{\mathbf{T}_0}\times \mathbf{H}_\textrm{out}$, only grasp the sample-induced aberrations; (\textit{ii}) the input focusing matrix, $\mathbf{H}_\textrm{in}$, and the associated $\mathbf{T}-$matrix, $\mathbf{T}_\textrm{in}=\rev{\mathbf{T}_0}\times \mathbf{H}_\textrm{in}$, also contain the aberrations undergone by the incident and reflected reference beams (Supplementary Section \ref{sectionSR}).  

For $\Delta \bm{\rho}=\mathbf{0}$ (conventional FFOCT set up), the recorded wave-field (Eq.~\ref{terme_interference}) is equivalent to a time-gated confocal image~\cite{barolle_manifestation_2021}. It can actually be expressed as the convolution between the sample reflectivity $\gamma$ and the confocal PSF $H_\textrm{in} \times H_\textrm{out}$:
\begin{equation}
{R_\textrm{in}}(\bm{\rho}_\textrm{out},\Delta \bm{\rho}=\mathbf{0},z) = I_0 \iint H_\textrm{out}(\bm{\rho}_\textrm{out},\bm{\rho}_s,z) H_\textrm{in} (\bm{\rho}_s,\bm{\rho}_{out},z) \gamma(\bm{\rho}_s,z) \diff \bm{\rho}_s.   
\end{equation}
On the one hand, the confocal nature of the recorded wave-field implies a transverse resolution $\delta \rho_0 \sim \lambda/4\mathrm{NA}$. On the other hand, the axial resolution is either controlled by the thickness $\delta z_t$ of the coherence volume or the depth-of-field $\delta z_0$ of the microscope objective: $\delta z_0= n \lambda/\mathrm{NA}^2$. In the present case, $\delta z_0 \sim 1 $ $\mu$m$<\delta z_t\sim10$  $\mu$m. The axial resolution is thus given by the depth-of-field. $\delta \rho_0$ and $\delta z_0$ thus dictate the values of the transverse and axial sampling of the de-scanned matrix $\mathbf{R}_\textrm{in}$ in our experiment.

\clearpage

\subsection{Relation between the de-scanned matrix and the focused reflection matrix}

In this section, we investigate to which extent the de-scanned matrix recorded by the experimental set up in Supplementary Figs.~\ref{fig:setup}a,b can be considered equivalent to the focused reflection matrix that would be recorded by the fictitious coherent set up displayed in Supplementary Fig.~\ref{fig:setup}c. 

The coefficients of a focused reflection matrix recorded by the fictitious coherent set up displayed in Fig.~\ref{fig:setup} can be expressed as:
\begin{eqnarray}
{R}(\bm{\rho}_\textrm{out},\bm{\rho}_\textrm{in},z) &= & I_0 \iiint H(\bm{\rho}_{out},\bm{\rho}_s,z) \gamma(\bm{\rho}_s,z) H (\bm{\rho}_s,\bm{\rho}_{in},z)  
 \diff\bm{\rho} 
    \label{terme_interference2}
\end{eqnarray}
A strict equality between Eqs.~\ref{terme_interference} and \ref{terme_interference2} is only obtained if ${H}_\textrm{in}\equiv H$. This condition is fulfilled only for a perfect reference arm: $\rev{{\mathcal{T}}}_{\! \!\textrm{ref}}\equiv \mathbf{1}_{k<NA}$ and $H \odot  H^*_\textrm{ref} \equiv H $. In theory, the incoherent set up of Fig.~\ref{fig:setup}a is thus equivalent to the fictitious coherent set up of Fig.~\ref{fig:setup}b.
\begin{equation}
R_\textrm{in}(\Delta \bm{\rho},\bm{\rho}_\textrm{out})=R(\bm{\rho}_\textrm{out}+\Delta \bm{\rho},\bm{\rho}_\textrm{out},z)
\end{equation}
In reality, the reference arm always exhibits aberrations such as a slight defocus of the reference mirror M3 in Fig.~\ref{fig:setup}b or a slight defocus of the reference beam in the secondary source plane at the output of first interferometer. 

\subsection{{The distortion matrix}}\label{sec:SM_distortion}

The distortion matrix is related to the de-scanned matrix by a simple Fourier transform:
\begin{equation}
    \vec{D}_{\textrm{in}}(z) = \rev{\mathbf{T}_0} \times \mathbf{R}_{\textrm{in}}(z),
\end{equation}
or in terms of matrix coefficients,
\begin{equation}
    D(\mathbf{u}_\textrm{in},\bm{\rho}_{\textrm{out}},z) =\sum_{\Delta \bm{\rho}} R_\textrm{in}(\Delta \bm{\rho}, \bm{\rho}_\textrm{out},z)  \exp \left ( -i \frac{2\pi}{\lambda f} \mathbf{u}_\textrm{in} \cdot \Delta \bm{\rho} \right ).
\end{equation}
Injecting Eq.~\ref{terme_interference} into the last equation yields
\begin{equation}
\label{D}
    D(\mathbf{u}_\textrm{in},\bm{\rho}_{\textrm{out}},z) =\sum_{\Delta \bm{\rho}}I_0 \iint H_\textrm{out}(\bm{\rho}_\textrm{out},\bm{\rho}_s,z) \gamma(\bm{\rho}_s,z) H_\textrm{in} (\bm{\rho}_s,\bm{\rho}_{out}+\Delta \bm{\rho},z) \exp \left ( -i \frac{2\pi}{\lambda f} \mathbf{u}_\textrm{in} \cdot \Delta \bm{\rho} \right )  
 \diff\bm{\rho}_s .
\end{equation}
In a previous paper~\cite{badon_distortion_2020}, we showed that a singular value decomposition of $\mathbf{D}$ enables to decompose the field-of-view into isoplanatic modes and extract the associated aberration phase laws. However, this demonstration was based on the condition that the $\mathbf{D}-$matrix is dominated by its correlations in the focal plane. This is the case for a specular reflector such as a resolution target or a medium of continuous reflectivity but no longer valid for a random distribution of heterogeneities like in the opaque cornea under study. In the accompanying paper, we propose a more general solution to overcome aberrations and scattering in optical microscopy: An iterative multi-scale analysis of wave distortions.

To that aim, the field-of-detection should be subdivided into overlapping regions that are defined by their central midpoint $\mathbf{r}_p=(\bm{\rho}_p,z_p)$ and their spatial extension $L$. All of the distorted components associated with focusing points $\bm{\rho}_\textrm{out}$ located within each region are extracted and stored in a local distortion matrix $\mathbf{D}'_\textrm{in}(\mathbf{r}_p)$:
\begin{equation}
\label{eq:window2}
    D'(\mathbf{u}_\textrm{in},\bm{\rho}_\textrm{out}, \mathbf{r}_p) = D(\mathbf{u}_\textrm{in},\bm{\rho}_\textrm{out},z_p) ~ W_L(\bm{\rho}_\textrm{out} - \bm{\rho}_p),
\end{equation}
where $W_{L}(x,y) = 1$ for $|x|<L$ and $|y|<L$, and zero otherwise.

At this stage, a local isoplanatic assumption shall be made over each region of size $L$. This hypothesis implies that the PSFs ${H}_\textrm{in/out}$ are invariant by translation in each region. This leads us to define local spatially-invariant PSFs ${H^{(l)}_\textrm{in/out}}$ around each central midpoint $\rp$ such that: 
\begin{equation}
\label{local_iso}
H_\textrm{in/out}(\bm{\rho}_s,\bm{\rho}_\textrm{in/out},z_p) = H^{(l)}_\textrm{in/out}(\bm{\rho}_s-\bm{\rho}_\textrm{in/out},\mathbf{r}_p).
\end{equation}
Under this assumption, Eq.~\ref{D} can be rewritten as follows:
\begin{equation}
{D'}(\mathbf{u}_\textrm{in},\bm{\rho}_\textrm{out}, \rp) = 
  \underbrace{\rev{\mathcal{T}}_{\!\!\textrm{in}}(\mathbf{u}_\textrm{in},\rp)}_{\mbox{transmittance}}
  \underbrace{\iint \gamma(\bm{\rho}_s+\bm{\rho}_\textrm{out},z) H_\textrm{out}^{(l)}(\bm{\rho}_s,\mathbf{r}_p)  \exp \left ( - i \frac{2\pi}{\lambda f}  \mathbf{u}_\textrm{in} \cdot \bm{\rho} \right ) \diff \bm{\rho}_s}_{\mbox{virtual source}}.
  \label{Dalex}
\end{equation}
Around each point $\rp$, the aberrations can be modelled by a \rev{local} transmittance $ \rev{\mathcal{T}}_{\!\!\textrm{in}}(\mathbf{u}_\textrm{in},\rp)$. This transmittance is 
the Fourier transform of the input PSF ${H^{(l)}_\textrm{in}}(\bm{\rho}_s,\rp)$: 
\begin{equation}
  \rev{\mathcal{T}}_{\!\!\textrm{in}}(\mathbf{u}_\textrm{in},\rp)=\iint {H^{(l)}_\textrm{in}}(\bm{\rho}_s,\rp) \exp \left ( \frac{2\pi}{\lambda f} \mathbf{u}_\textrm{in} \cdot \bm{\rho}_s \right )  \diff {\bm{\rho}_s} 
\end{equation}
The physical meaning of this last equation is the following: Each distorted wave-field corresponds to the diffraction of a virtual source synthesized inside the medium modulated by the pupil transmittance $ \rev{\mathcal{T}}_{\!\!\textrm{in}}\rev{(\mathbf{u}_\textrm{in},\rp)}$ \rev{of the sample seen from point $\rp$}. Each virtual source is spatially incoherent due to the random reflectivity of the medium, and its size is governed by the spatial extension of the output focal spot. The idea is now to smartly combine each virtual source to generate a coherent guide star and estimate the local transmittance $ \rev{\mathcal{T}}_{\!\!\textrm{in}}$ independently from the sample reflectivity.

\subsection{{Covariance Matrix of Wave Distortions}}

To do so, the correlation matrix $\mathbf{C}_\textrm{in}=\mathbf{D}_\textrm{in}\mathbf{D}_\textrm{in}^\dag$ is an excellent tool. Its coefficients write as follows 
\begin{equation}
\label{corr_in}
    C_\textrm{in} (\mathbf{u}_\textrm{in}, \mathbf{u}'_\textrm{in},\rp)= N_\mathcal{W}^{-1} \sum_{\bm{\rho}_\textrm{out}} {D}'(\mathbf{u}_\textrm{in},\bm{\rho}_\textrm{out},\rp){D}'^*(\mathbf{u}'_\textrm{in},\bm{\rho}_\textrm{out},\rp)
\end{equation}
The matrix $\mathbf{C}_\textrm{in}(\rp)$ can be decomposed as the sum of its ensemble average, the covariance matrix $\left \langle \mathbf{C}_\textrm{in}\right \rangle (\mathbf{r}_p)$, and a perturbation term $\delta \mathbf{C}_\textrm{in}(\mathbf{r}_p) $:
\begin{equation}
\label{C}
   \mathbf{C}_\textrm{in}(\mathbf{r}_p)= \left \langle \mathbf{C}_\textrm{in}\right \rangle (\mathbf{r}_p) +  \delta \mathbf{C}_\textrm{in}(\mathbf{r}_p).
\end{equation}
The intensity of the perturbation term scales as the inverse of the number $N_\textrm{W}=(L/\delta \rho_0)^2$ of resolution cells in each sub-region~\cite{robert_greens_2008,lambert_ultrasound_2022}:
\begin{equation}
\label{perturbation}
    \left \langle \left |\delta C_\textrm{in}(\mathbf{u},\mathbf{u}',\mathbf{r}_p)\right |^2 \right \rangle = \frac{ \left \langle \left | C_\textrm{in}(\mathbf{u},\mathbf{u},\mathbf{r}_p)\right |^2 \right \rangle}{N_\mathcal{W}}
\end{equation}
This perturbation term can thus be reduced by increasing the size $L$ of the spatial window $W_L$, but at the cost of a resolution loss.

Under assumptions of local isoplanicity (Eqs.~\ref{local_iso} and \ref{Dalex}) and random reflectivity, 
\begin{equation}
\label{random}
\langle \gamma(\bm{\rho}_s,z)\gamma^*(\bm{\rho}'_s,z) \rangle= \langle | \gamma |^2 \rangle \delta (\bm{\rho}_s - \bm{\rho}'_s),
\end{equation}
with $\delta$, the Dirac distribution, 
the covariance matrix can be expressed as follows~\cite{lambert_distortion_2020}:
\begin{equation}
\label{rhoaveeq}
\left \langle \mathbf{C}_\textrm{in} \right \rangle (\rp) =  \left [\rev{\bm{\mathcal{T}}_{\!\!\textrm{in}}} (\rp) \circ \rev{\mathbf{T}_0} \right ] \times  \mathbf{C}_H (\rp) \times \left [\rev{\mathbf{T}_0} \circ \rev{\bm{\mathcal{T}}_{\!\!\textrm{in}}} (\rp) \right ]^{\dag} ,
\end{equation}
or, in terms of matrix coefficients,
\begin{eqnarray}
\label{rhoaveeq2}
\left \langle \mathbf{C}_\textrm{in} \right \rangle (\mathbf{u}_\textrm{in},\mathbf{u}'_\textrm{in},\rp) &=&  \rev{\mathcal{T}}_{\!\!\textrm{in}} (\mathbf{u}_\textrm{in},\rp)  \rev{\mathcal{T}}^*_{\!\!\textrm{in}} (\mathbf{u}'_\textrm{in},\rp)  \iint \diff\bm{\rho}_s \left |H_\textrm{out}^{(l)}(\bm{\rho}_s,\rp) \right |^2 \exp \left [-i\frac{2\pi}{\lambda f} (\mathbf{u}_\textrm{in}-\mathbf{u}'_\textrm{in})\cdot\bm{\rho}_s \right ] \nonumber \\ 
&=& \rev{\mathcal{T}}_{\!\!\textrm{in}} (\mathbf{u}_\textrm{in},\rp) \rev{\mathcal{T}}^*_{\!\!\textrm{in}} (\mathbf{u}'_\textrm{in},\rp)  \underbrace{ \left [\rev{\mathcal{T}}_{\!\!\textrm{out}} \stackrel{\mathbf{u}}{\circledast} \rev{\mathcal{T}}_{\!\!\textrm{out}}\right ] (\mathbf{u}_\textrm{in}-\mathbf{u}'_\textrm{in},\rp)}_{=C_H(\mathbf{u}_\textrm{in},\mathbf{u}'_\textrm{in},\rp)} .
\end{eqnarray}
$\mathbf{C}_H$ is a reference correlation matrix that would be measured in an homogeneous cornea for a virtual reflector whose scattering distribution corresponds to the output focal spot intensity $|H_\textrm{out}^{(l)}(\bm{\rho}_s,\rp)|^2$.
The covariance matrix $\left \langle \mathbf{C}_\textrm{in} \right \rangle (\rp)$ thus corresponds to the same experimental situation but for a virtual reflector embedded into the heterogeneous cornea under study. 

\section{Estimation of the transmittance matrix}

\subsection{{Iterative Time Reversal}}
\label{IPC}
For such an experimental configuration, it has been shown that an iterative time reversal (ITR) process converges towards a wavefront that focuses perfectly through the heterogeneous medium onto this scatterer~\cite{prada_eigenmodes_1994,prada_decomposition_1996}. 
Hence, let us consider the following fictitious experiment that consists in a phase conjugating mirror placed in the pupil plane of the microscope objective and the virtual reflector placed in its focal plane (see Supplementary Fig.~\ref{S2_IPRvsSVD}b).  It gives rise to a stationary wave-field, $\bm {\psi}=\bm{\psi}^{+} + \bm{\psi}^{-}$, made of down-going and up-going wave-fields, $\bm{\psi}^-$ and $\bm{\psi}^+$. Both wave-fields check the following relationships in the pupil plane:
\begin{equation}
\label{cin2}
\bm{\psi}^+_\textrm{u}=\mathbf{C}_\textrm{in} \times \bm{\psi}^-_\textrm{u}
\end{equation}
and
\begin{equation}
\psi^-_\mathbf{u}=\epsilon \psi^{+*}_\mathbf{u}.
\end{equation}
with $\epsilon$ the reflectivity of the phase conjugating mirror. 
Combining the two previous equations leads to the following eigenequation:
\begin{equation}
\label{eigen}
\bm{\psi}^-_\mathbf{u}=|\epsilon|^2\mathbf{C}_\textrm{in}^*\mathbf{C}_\textrm{in} \times \bm{\psi}^-_\mathbf{u}.
\end{equation}
The ITR process has thus eigenmodes which can be determined by the diagonalization of the time reversal operator $\mathbf{C}_\textrm{in}^*\mathbf{C}_\textrm{in} $. In particular, the first eigenvector $\mathbf{U}_\textrm{in}$ of $\mathbf{C}_\textrm{in}^*\mathbf{C}_\textrm{in}$, which is also the first singular vector of $\mathbf{D}_\textrm{in}$, corresponds to the wave-front that optimizes the energy backscattered by the virtual reflector.

If the virtual reflector was point-like, this wave-front would be a perfect estimator of $\rev{\mathcal{T}}_{\!\!\textrm{in}}$. Its phase conjugate would perfectly compensate for aberrations and focuses through the heterogeneous medium onto the point-like target~\cite{prada_eigenmodes_1994,prada_decomposition_1996}.
However, here the virtual guide star is enlarged compared to the diffraction limit. This wave-front is of finite angular support $\delta {u}_c$ and tends to focus on the virtual reflector but with a resolution width $\delta \rho_c \sim \lambda f/\delta u_c$ larger than the diffraction limit~\cite{lambert_ultrasound_2022} (see Supplementary Fig.~\ref{S2_IPRvsSVD}). Its phase is thus a good estimator of $\rev{\mathcal{T}}_{\!\!\textrm{in}}$ over the angular domain $\delta u_c$ but absolutely not elsewhere. 

This assertion is illustrated by Supplementary Figs.~\ref{S2_IPRvsSVD}b and c that show the modulus and phase of the first singular vector $\mathbf{U}_\textrm{in}$ of $\mathbf{D}_\textrm{in}$ at depth $z=200$ $\mu$m in the cornea. As anticipated, the modulus of $\mathbf{U}_\textrm{in}$ exhibits a main central lobe at small spatial frequencies (delimited by a circle white line in Supplementary Fig.~\ref{S2_IPRvsSVD}b) but is extremely low at high angles of incidence. This means that the phase of $\mathbf{U}_\textrm{in}$ is a good estimator for $|\mathbf{u}_\textrm{in}|<\delta u_c$ but is not reliable beyond $u_c$ (Supplementary Fig.~\ref{S2_IPRvsSVD}c).
\begin{figure}[ht!]
\includegraphics[width=0.8\textwidth]{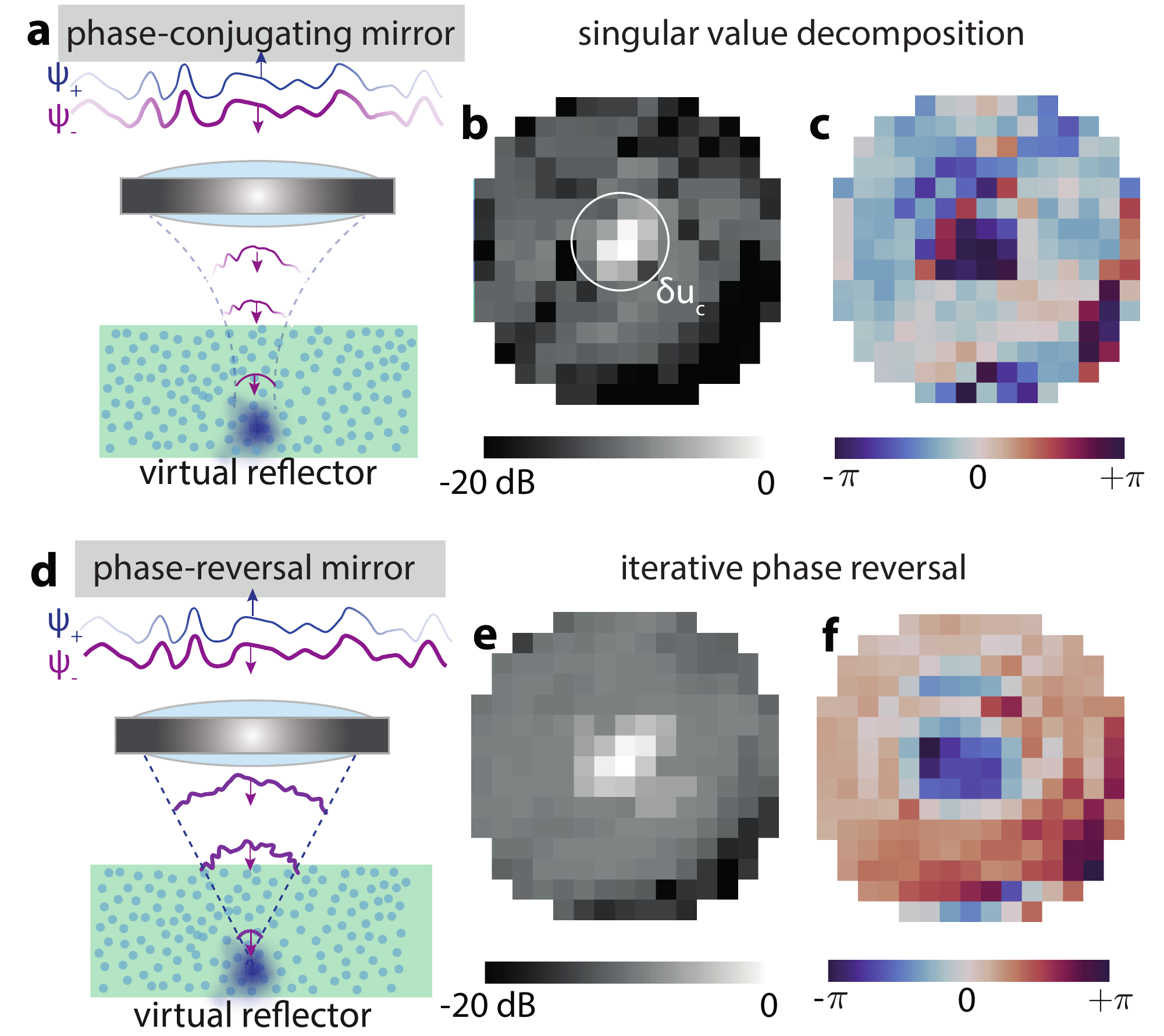}
  \caption{\textbf{Iterative Time Reversal vs. Iterative Phase Reversal.} \textbf{a.} The first eigenstate of $\mathbf{C}_\textrm{in}$ corresponds to the eigenmode that would arise between a phase conjugating mirror in the pupil plane and the virtual reflector. \textbf{b}-\textbf{c.} Absolute value and phase of $\mathbf{U}_1$ the first eigenvector of $\mathbf{C}_\textrm{in}$ at the first iteration of the multi-scale analysis ($z=200$ $\mu$m). \textbf{d.} The IPR process converges towards the wave-front $\rev{\hat{\mathcal{T}}}_{\!\!\textrm{in}}$ that would be obtained if an iterative phase reversal mirror was used to focus on the virtual reflector. \textbf{e}-\textbf{f.} Absolute value of $\mathbf{C}_\textrm{in}\times \rev{\hat{\mathcal{T}}}_{\!\!\textrm{in}}$ and phase of $\rev{\hat{\mathcal{T}}}_{\!\!\textrm{in}}$ at the first iteration of the multi-scale analysis ($z=200$ $\mu$m).}
  \label{S2_IPRvsSVD}
\end{figure}

\clearpage

\subsection{{Iterative Phase Reversal}}

To circumvent that issue, the iterative phase reversal (IPR) algorithm has been developed. It consists in replacing the virtual phase conjugating mirror of Supplementary Fig.~\ref{S2_IPRvsSVD}a by a phase reversal mirror (Supplementary Fig.~\ref{S2_IPRvsSVD}b). As a phase conjugating mirror, the latter mirror reverses the phase of the incident wave-field but back-emits a wave of constant amplitude, such that:
\begin{equation}
\psi^-_\mathbf{u}=\exp\left[i \, \mathrm{arg}\left \lbrace   \psi^{+*}_\mathbf{u} \right \rbrace \right ]. 
\end{equation}
Combined with Eq.~\ref{cin2}, the latter equation yields the following
relation for the down-going wave-field:
\begin{equation}
\psi^-_\mathbf{u}=\exp\left[i \, \mathrm{arg}\left \lbrace   \mathbf{C}_\textrm{in} \times \bm{\psi}^-_\textrm{u} \right \rbrace \right ]. 
\end{equation}
Unlike Eq.~\ref{eigen}, this is not an eigenequation but it can be solved iteratively [see Eq.~18 of the accompanying manuscript]. By definition, the resulting wave-front $\rev{\hat{\mathcal{T}}}_{\!\!\textrm{in}}$  is of constant modulus over the pupil. To see the angular domain addressed by $\rev{\hat{\mathcal{T}}}_{\!\!\textrm{in}}$ , one can investigate the modulus of $\mathbf{C}_\textrm{in} \times \rev{\hat{\mathcal{T}}}_{\!\!\textrm{in}}$ (see Supplementary Fig.~\ref{S2_IPRvsSVD}e). Comparison with Supplementary Fig.~\ref{S2_IPRvsSVD}b shows that the IPR process addresses each angular component of the imaging process, leading to a more reliable estimation of the $\mathbf{T}-$matrix over the whole pupil (Supplementary Fig.~\ref{S2_IPRvsSVD}f). While ITR is guided by a maximization of the energy back-scattered by the virtual reflector, IPR optimizes the coherence of the wave-front over the whole pupil aperture, thereby leading, in principle, to a diffraction-limited focal spot onto the virtual scatterer (Supplementary Fig.~\ref{S2_IPRvsSVD}c).

In Supplementary Section~\ref{IPRvsSVD}, the IPR and ITR approaches will be compared quantitatively when incorporated in a multi-scale process. Prior to that, the bias of the $\mathbf{T}-$matrix estimator provided by IPR is established theoretically to justify this strategy.

\subsection{{Bias of the $\mathcal{T}$-matrix estimator}}\label{bias}

The IPR process assumes the convergence of the correlation matrix $\mathbf{C}_\textrm{in}$ (Eq.~10) towards its ensemble average $\left \langle \mathbf{C}_\textrm{in} \right \rangle $, the covariance matrix~\cite{lambert_distortion_2020,lambert_ultrasound_2022}. In fact, this convergence is never fully realized and $\mathbf{C}_\textrm{in}$ should be decomposed as the sum of this covariance matrix $\left \langle \mathbf{C}_\textrm{in}\right \rangle $ and the perturbation term $\delta \mathbf{C}_\textrm{in} $ (Eq.~\ref{C}). 
In the following, we express theoretically the bias induced by this perturbation term on the estimation of $\rev{\hat{\bm{\mathcal{T}}}}_{\!\!\textrm{in}}$. In particular, we will show how it scales with the parameter $N_\mathcal{W}$ and the focusing quality. We consider here the input correlation matrix $\mathbf{C}_\textrm{in}$ but a similar demonstration can be performed at output. For sake of lighter notation, the dependence over $\mathbf{r}_p$ is omitted in the following. 

To understand the parameters controlling the error $\delta \rev{{\bm{\mathcal{T}}}}_{\!\!\textrm{in}} $ between $\rev{\hat{\bm{\mathcal{T}}}}_{\!\!\textrm{in}}$ and $\rev{{\bm{\mathcal{T}}}}_{\!\!\textrm{in}}$, one can express $\rev{\hat{\bm{\mathcal{T}}}}_{\!\!\textrm{in}}$ as follows:
\begin{equation}
\rev{\hat{\bm{\mathcal{T}}}}_{\!\!\textrm{in}}=\exp \left ( j \mbox{arg} \left \lbrace  \mathbf{C}_\textrm{in} \times \rev{\hat{\bm{\mathcal{T}}}}_{\!\!\textrm{in}} \right \rbrace  \right )   =  \frac{ \mathbf{C}_\textrm{in} \times \rev{\hat{\bm{\mathcal{T}}}}_{\!\!\textrm{in}} }{|| \mathbf{C}_\textrm{in} \times \rev{\hat{\bm{\mathcal{T}}}}_{\!\!\textrm{in}} ||}
\end{equation}
By injecting Eq.~\ref{C} into the last expression, $\rev{\hat{\bm{\mathcal{T}}}}_{\!\!\textrm{in}}$ can be expressed, at first order, as the sum of its expected value $\rev{{\bm{\mathcal{T}}}}_{\!\!\textrm{in}}$ and a perturbation term $\delta \rev{{\bm{\mathcal{T}}}}_{\!\!\textrm{in}} $:
\begin{equation}
  \rev{\hat{\bm{\mathcal{T}}}}_{\!\!\textrm{in}}=\underbrace{\frac{ \langle \mathbf{C}_\textrm{in}\rangle \times \rev{{\bm{\mathcal{T}}}}_{\!\!\textrm{in}}  }{|| \langle \mathbf{C}_\textrm{in} \rangle \times \rev{{\bm{\mathcal{T}}}}_{\!\!\textrm{in}} ||}}_{= \rev{{\bm{\mathcal{T}}}}_{\!\!\textrm{in}}} + \underbrace{\frac{  \delta \mathbf{C}_\textrm{in} \times \rev{{\bm{\mathcal{T}}}}_{\!\!\textrm{in}}}{|| \langle \mathbf{C}_\textrm{in} \rangle \times \rev{{\bm{\mathcal{T}}}}_{\!\!\textrm{in}}  ||}}_{\simeq \delta \rev{{\bm{\mathcal{T}}}}_{\!\!\textrm{in}}}.
\end{equation}
The bias intensity can be expressed as follows:
\begin{equation}
\label{bias1}
    | \delta \rev{{{\mathcal{T}}}}_{\!\!\textrm{in}} |^2=\frac{ \rev{{\bm{\mathcal{T}}}}_{\!\!\textrm{in}}^{\dag} \times   \delta \mathbf{C}_\textrm{in}^{\dag}\times \delta \mathbf{C}_\textrm{in} \times   \rev{{\bm{\mathcal{T}}}}_{\!\!\textrm{in}} }{ \rev{{\bm{\mathcal{T}}}}_{\!\!\textrm{in}}^{\dag}\times \langle \mathbf{C}_\textrm{in}\rangle^{\dag} \times \langle \mathbf{C}_\textrm{in} \rangle \times   \rev{{\bm{\mathcal{T}}}}_{\!\!\textrm{in}}}
\end{equation}
Using Eq.~\ref{perturbation}, the numerator of the previous equation can be expressed as follows:
 \begin{equation}
  \rev{{\bm{\mathcal{T}}}}_{\!\!\textrm{in}}^{\dag} \times   \delta \mathbf{C}_\textrm{in}^{\dag}\times \delta \mathbf{C}_\textrm{in} \times   \rev{{\bm{\mathcal{T}}}}_{\!\!\textrm{in}}= M^2  \langle | {\delta C}_\textrm{in} (\mathbf{u},\mathbf{u}')|^2\rangle = M^2 \langle | \langle {C}_\textrm{in} \rangle (\mathbf{u},\mathbf{u})|^2\rangle /N_\mathcal{W}  .
 \end{equation}
 Injecting Eq.~\ref{rhoaveeq2} into the last equation leads to the following expression for the numerator of Eq.~\ref{bias1}:
  \begin{equation}
 \rev{{\bm{\mathcal{T}}}}_{\!\!\textrm{in}}^{\dag} \times   \delta \mathbf{C}_\textrm{in}^{\dag}\times \delta \mathbf{C}_\textrm{in} \times   \rev{{\bm{\mathcal{T}}}}_{\!\!\textrm{in}}= M^2    \left | \rev{\mathcal{T}}_{ \! \textrm{out}} \stackrel{\mathbf{u}}{\circledast} \rev{\mathcal{T}}_{\! \textrm{out}} (\mathbf{0}) \right |^2  /N_\mathcal{W} .
 \end{equation}
 
The denominator of Eq.~\ref{bias1} can be expressed as follows:
\begin{equation}
\rev{{\bm{\mathcal{T}}}}_{\!\!\textrm{in}}^{\dag}\times \langle \mathbf{C}_\textrm{in}\rangle^{\dag} \times \langle \mathbf{C}_\textrm{in} \rangle \times   \rev{{\bm{\mathcal{T}}}}_{\!\!\textrm{in}}= M^2 \left | \sum_{\mathbf{u}}  \rev{\mathcal{T}}_{\! \textrm{out}} \stackrel{\mathbf{u}}{\circledast} \rev{\mathcal{T}}_{\! \textrm{out}} (\mathbf{u})  \right |^2
\end{equation}
The bias intensity is thus given by:
\begin{equation}
   | \delta \rev{\mathcal{T}}_{\! \! \textrm{in}} |^2 = \frac{\left | \rev{\mathcal{T}}_{\! \textrm{out}} \stackrel{\mathbf{u}_\textrm{out}}{\circledast} \rev{\mathcal{T}}_{\! \textrm{out}} (\mathbf{0}) \right |^2}{N_\mathcal{W}\left | \sum_{\mathbf{u}}  \rev{\mathcal{T}}_{\! \textrm{out}} \stackrel{\mathbf{u}}{\circledast}\rev{\mathcal{T}}_{\! \textrm{out}} (\mathbf{u})  \right |^2}\end{equation}
{In the last expression, we recognize the ratio between the coherent intensity (energy deposited exactly at focus) and the mean incoherent intensity. This quantity is known as the coherence factor in ultrasound imaging~\cite{mallart_adaptive_1994,robert_greens_2008}:
\begin{equation}
    \mathcal{C}_\textrm{out}= \frac{  \sum_{\mathbf{u}}  \rev{\mathcal{T}}_{\! \textrm{out}}\stackrel{\mathbf{u}}{\circledast} \rev{\mathcal{T}}_{\! \textrm{out}} (\mathbf{u}) }{  \rev{\mathcal{T}}_{\! \textrm{out}} \stackrel{\mathbf{u}}{\circledast} \rev{\mathcal{T}}_{\! \textrm{out}} (\mathbf{0})  } =\frac{|H^{(l)}_\textrm{out} (\bm{\rho}=\mathbf{0})|^2}{   \delta_R^{-2}\int d\bm{\rho} |H^{(l)}_\textrm{out} (\bm{\rho})|^2} 
\end{equation}
In the speckle regime (Eq.~\ref{random}) and for 3D imaging, the coherence factor $\mathcal{C}$ ranges from 0,
for strong aberrations and/or multiple scattering background, to $4/9$ in the ideal case~\cite{Silverstein2001}. The bias intensity can thus be rewritten as:
\begin{equation}
\label{bias2}
   {| \delta \rev{\mathcal{T}}_{\!\! \textrm{in}} (\mathbf{u}) |^2} = \frac{1}{\mathcal{C}^2_\textrm{out} N_{\mathcal{W}}} 
\end{equation}
This last expression justifies the multi-scale analysis proposed in the accompanying paper. A gradual increase of the focusing quality, quantified by $\mathcal{C}$, is required to address smaller spatial windows that scale as $N_{\mathcal{W}}$. Following this scheme, the bias made of our $\mathbf{T}-$matrix estimator can be minimized and the iterative phase reversal algorithm converges towards a satisfying estimator.}

\subsection{{Numerical validation of the iterative phase reversal process}}

The IPR algorithm is now validated by means of a numerical simulation. The numerical simulation emulates an imaging experiment in an epi-detection configuration, as depicted in Supplementary Fig.~\ref{fig:corr_in_out_rpsf_aberr_SIMUS}a. The experimental conditions (numerical aperture, focal length, `\textit{etc.}) are identical to our experiment. The field-of-view contains $N=61 \times 61$ independent resolution cells. For sake of simplicity, a plane object of random complex reflectivity $\gamma(\bm{\rho})$ is considered in the focal plane of the microscope objective and the isoplanatic assumption is also made. Under these assumptions, the coefficients of the reflection matrix can be expressed in the pupil basis as follows:
\begin{equation}
R(\mathbf{u}_\textrm{out},\mathbf{u}_\textrm{in})=\rev{\mathcal{T}}(\mathbf{u}_\textrm{out}) \tilde{\gamma}(\mathbf{u}_\textrm{out}+\mathbf{u}_\textrm{in})\rev{\mathcal{T}}(\mathbf{u}_\textrm{in})
\end{equation}
where $ \tilde{\gamma}(\mathbf{u})=\int d\bm{\rho}_s \gamma(\mathbf{\rho}_s) \exp \left( 2 \pi \mathbf{u} \cdot \bm{\rho}_s/(\lambda f) \right)$, the Fourier transform of the sample reflectivity. The aberrations are thus modelled as a random phase screen of transmittance $\rev{\bm{\mathcal{T}}}$. It exhibits a Gaussian statistics of correlation length $\ell_\phi=2~\mu$m, and standard deviation $\sigma_\phi=0.2$. The aberration phase law is displayed in Supplementary Fig.~\ref{fig:corr_in_out_rpsf_aberr_SIMUS}b. 
\begin{figure}[h!]
    \centering
    \includegraphics[width=\linewidth]{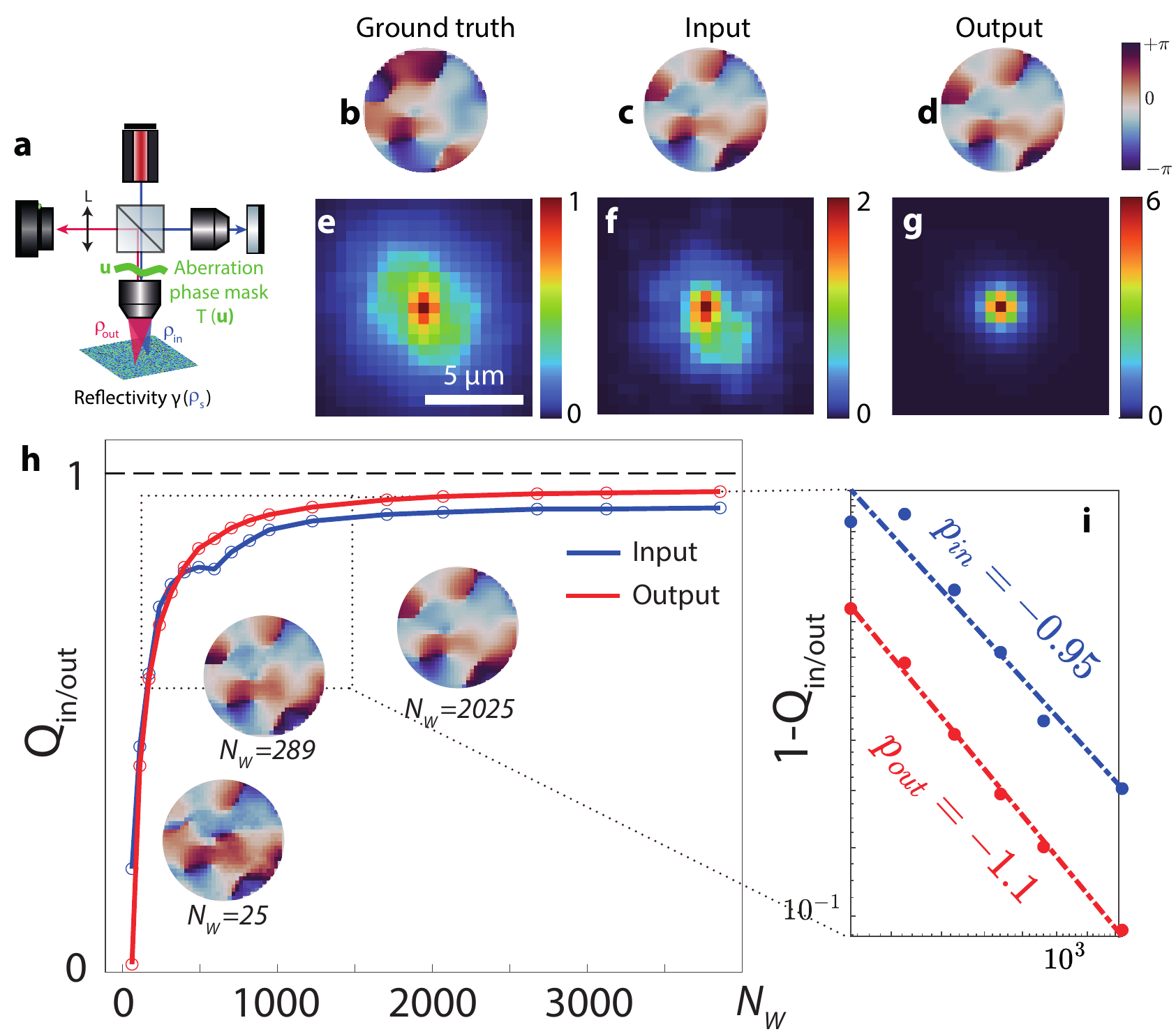}
    \caption{\textbf{Numerical validation of the iterative phase reversal algorithm}. \textbf{a.} Experimental configuration. Aberration are modelled by a random phase screen introduced in the pupil plane of the microscope objective. The object exhibits a random reflectivity. \textbf{b.} Simulated Gaussian random phase screen {($\sigma_\phi=0.2$, $\ell_\phi=2~\mu$m)}. \textbf{c}-\textbf{d.} Estimated input and output phase laws estimated by IPR {($N_\mathcal{W}=N^2=61^2$)}. \textbf{e.} Original RPSF before any correction. (scale bar: $5~\mu$m). \textbf{f.} RPSF after aberration correction at input. \textbf{g} RPSF after correction at input and output. \textbf{h.} Scalar product $Q_\textrm{in/out}$ between the estimated aberration transmittance and its ground truth as a function of $N_\mathcal{W}$ \ulysse{(averaged over 10 realizations of disorder)}. The estimated aberration phase law is displayed for different values of $N_\mathcal{W}$ as insets. \textbf{i.} ($1-Q_\textrm{in/out}$) as a function of $N_\mathcal{W}$ in log-log scale. Numerical points (disks) are fitted by linear curves (dashed lines). }
    \label{fig:corr_in_out_rpsf_aberr_SIMUS}
\end{figure}
Once the reflection matrix $\mathbf{R}_{\mathbf{uu}}$ is built in the pupil basis, a spatial Fourier transform yields the reflection matrix $\mathbf{R}_{\bm{\rho\rho}}$ in the focused basis, such that:
\begin{equation}
\mathbf{R}_{\bm{\rho\rho}} = \rev{\mathbf{T}_0^*} \times \mathbf{R}_{\mathbf{uu}}.\times \rev{\mathbf{T}_0^{\dag} }.
\end{equation}
The resulting reflection matrix yields an estimate of the reflection point-spread function  shown in Supplementary Fig.~\ref{fig:corr_in_out_rpsf_aberr_SIMUS}e. The distortion matrices $\vec{D}_\mathrm{in/out}$ in input and output are derived from the matrix $\vec{R}$, as described in Supplementary Section~\ref{sec:SM_distortion}, and the estimations of the input and output aberration transmittances $\rev{\hat{\mathcal{T}}}_\mathrm{in/out}(\vec{u}_\mathrm{in/out})$ are computed using the IPR process described in the Methods section of the accompanying paper.

Supplementary Figures~\ref{fig:corr_in_out_rpsf_aberr_SIMUS}c and d show the estimated transmittances $\rev{\hat{\bm{\mathcal{T}}}}_\mathrm{in/out}$, respectively, when the whole FOV is considered. A strong similarity is observed with the ground truth up to a phase ramp~\cite{lambert_ultrasound_2022} (Supplementary Fig.~\ref{fig:corr_in_out_rpsf_aberr_SIMUS}b). The corresponding RSPFs after each correction are displayed in Supplementary Figs.~\ref{fig:corr_in_out_rpsf_aberr_SIMUS}f and g. A diffraction-limited resolution is obtained at the end of the RMI process, which validates the IPR algorithm.

One can go further by investigating the convergence of the process as a function of $N_W$, the size of the spatial window considered for the computation of the correlation matrices $\mathbf{C}_\textrm{in/out}$. 
The similarity between the estimators $\rev{\hat{\bm{\mathcal{T}}}}_\mathrm{in/out}$ and the ground truth $\rev{\bm{\mathcal{T}}}$ is evaluated by the normalized scalar product $Q_\textrm{in/out}=N_u^{-1}\rev{\hat{\bm{\mathcal{T}}}}_\mathrm{in/out} \times \rev{{\bm{\mathcal{T}}}}^{\dag}$, or, in terms of matrix coefficients. 
\begin{equation}
   Q_\textrm{in/out}
=N_u^{-1}{\sum_{\mathbf{u}}\rev{\hat{{\mathcal{T}}}}_\mathrm{in/out}(\mathbf{u}) \rev{{{\mathcal{T}}}}^\star}(\mathbf{u}).
\end{equation}
The evolution of $Q_\textrm{in}$ and $P_\textrm{out}$ is displayed as a function of $N_W$ in Supplementary Fig.~\ref{fig:corr_in_out_rpsf_aberr_SIMUS}h. The convergence can be considered as fulfilled for $Q>0.9$, \textit{i.e $N_\mathcal{W} \sim 500$}, which is roughly the number of resolution cells contained in the final spatial windows ($L=6$ $\mu$m) in our experiment. 

This convergence rate is directly related to the bias of $\rev{\hat{\bm{\mathcal{T}}}}_\mathrm{in/out}$ (Eq.~\ref{bias2}). To show it, let us first express the intensity bias $|\delta \rev{\mathcal{T}} (\vec{u})|^2$ as a function of the phase error $\delta\phi (\vec{u})$ exhibited by the estimator $\rev{\hat{{\mathcal{T}}}}(\vec{u})$ with respect to $\rev{\mathcal{T}}(\vec{u})$, following the same formalism as in Supplementary Section~\ref{bias}:
\begin{align} \label{eq:bias3}
    |\delta \rev{\mathcal{T}}(\vec{u})|^2 = |1-\exp\left[i\delta \phi (\vec{u})\right]|^2  \underset{\delta\phi\ll1}{\sim} \left|1-(1+\delta\phi)\right|^2 \underset{\delta\phi\ll1}{\sim} |\delta\phi|^2
\end{align}
On the other hand, the scalar product as a function of the phase error $\delta\phi$ writes as such:
\begin{align}
     Q_\textrm{in/out}=N_u^{-1}\sum_{\vec{u}} \exp\left[i\delta\phi(\vec{u})\right]
\end{align}
The sum over the points in the Fourier plane $\vec{u}$ can be replaced by an ensemble average, since $N>> 1$:
\begin{align}
     Q_\textrm{in/out}=\left\langle\exp\left[i\delta\phi(\vec{u})\right] \right\rangle
\end{align}
Assuming a small phase error ($\delta\phi\ll1$),
\begin{align}
    Q_\textrm{in/out}&\sim 1+ i\left\langle \delta\phi\right\rangle - \frac{\left\langle (\delta\phi)^2\right\rangle}{2}  \\
    &\sim 1 -  \frac{\left\langle(\delta\phi)^2 \right\rangle}{2} 
\end{align}
since $\left\langle \delta\phi\right\rangle = 0$. Combining this last expression with Eq.~\ref{eq:bias3} leads to:
\begin{equation}\label{eq:bias_scalar_prod}
    Q_\textrm{in/out}\sim 1- \frac{\left\langle |\delta  {T}|^2\right\rangle}{2}  
\end{equation}
According to Eq.~\ref{bias2}, $1-Q_\textrm{in/out}$ should therefore scale as the inverse of the number of independent resolution cells contained in the spatial window: $1-P_\textrm{in/out} \propto {N_\mathcal{W}^{-1}}$.
To highlight this scaling law, $1-Q_\textrm{in/out}$ can be plotted in log-log scale as a function of $N_\mathcal{W}$ (Supplementary Fig.~\ref{fig:corr_in_out_rpsf_aberr_SIMUS}h). A slope $p$ close to 1 is obtained both at input and output: $p_{in}=-0.95$ and $p_{out}=-1.1$, confirming that the bias on the aberration estimation scales with the inverse of the number of independent resolution cells in the field of view (Eq.~\ref{bias2}). Another interesting observation is the lower bias observed at output in Supplementary Fig.~\ref{fig:corr_in_out_rpsf_aberr_SIMUS}h,i. Indeed, the first correction at input increases the coherence factor $\mathcal{C}_\textrm{in}$ and reduces the size of the virtual guide star when investigating wave distortions at output. This gain in focusing quality improves the sharpness of the estimator $\rev{\hat{\bm{\mathcal{T}}}}$, as already highlighted by the scaling of $|\delta \rev{\mathcal{T}}|^2$ as the inverse square of the coherence factor in Eq.~\ref{bias2}.  

In the present numerical simulation, the isoplanicity assumption makes the IPR algorithm converging towards an appropriate solution in one iteration at input and output. In the experiment, the situation is more complex since aberrations are spatially-distributed. In that case, an iterative compensation of wave distortions aver a multiple scale is required.  The corresponding strategy is explained in the next Section.

\clearpage

\subsection{Multi-scale compensation of wave distortions}
\begin{figure}[h!]
    \centering
    \includegraphics[width=.8\linewidth]{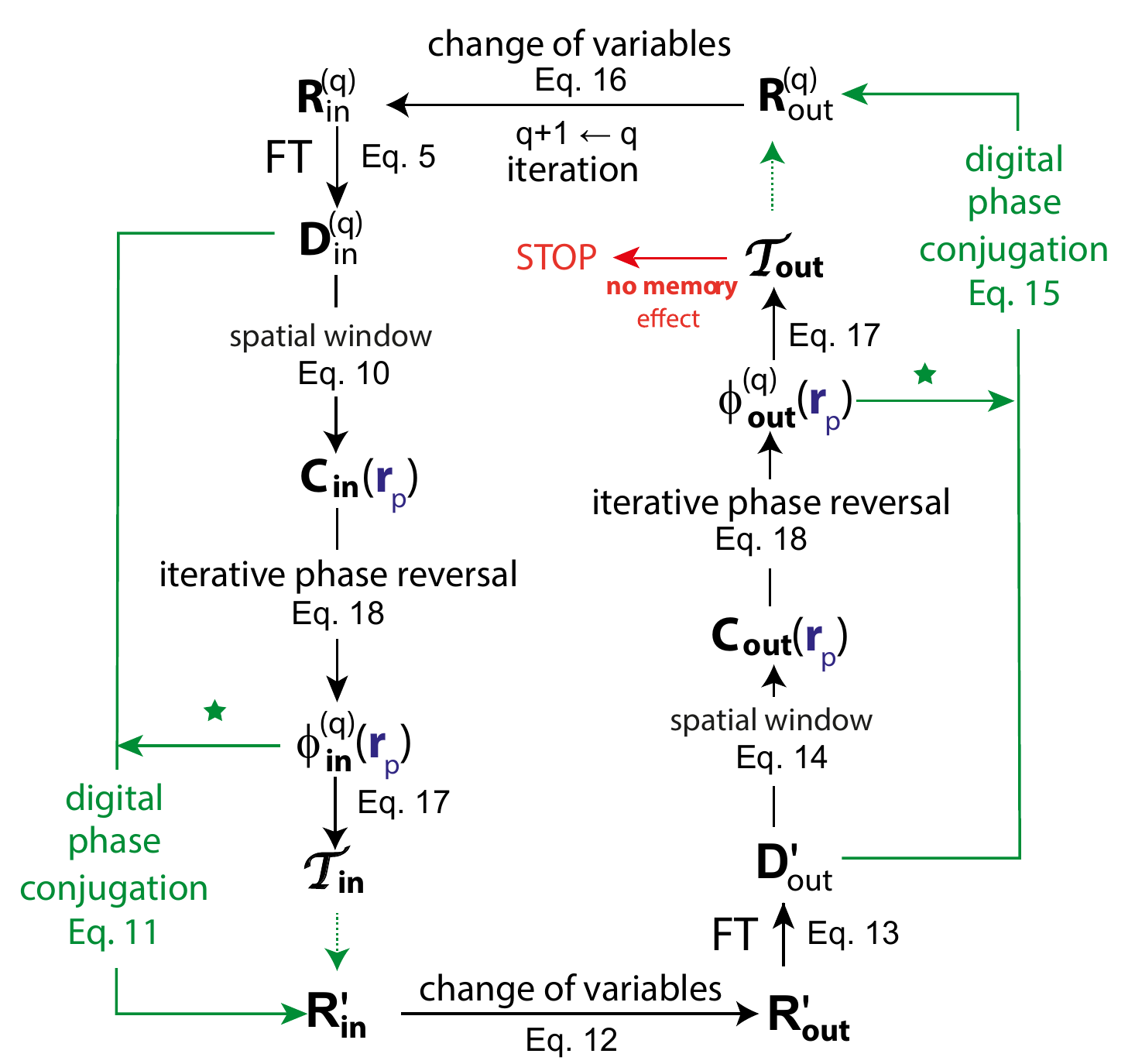}
    \caption{\textbf{Flowchart of the multi-scale matrix imaging process.}}
    \label{fig:flowchart}
\end{figure}
\begin{figure}[h!]
    \centering
    \includegraphics[width=\linewidth]{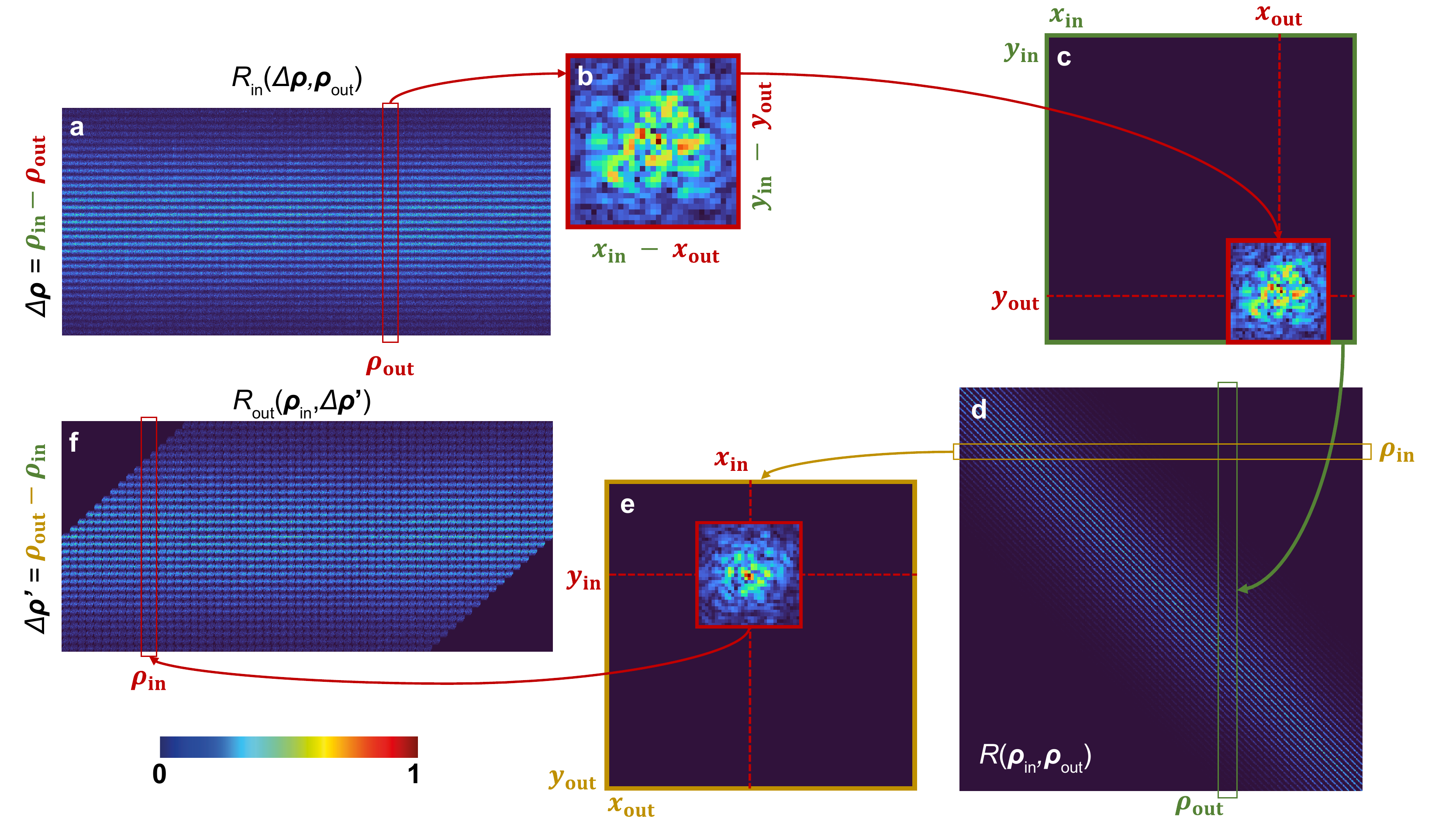}
    \caption{\textbf{Conversion of the reflection matrix from the input to the output de-scanned basis.} \textbf{a.} Reflection matrix $\mathbf{R}_\textrm{in}$ in the input de-scanned basis. \textbf{b.} Each column reshaped in 2D corresponds to the input focal spot de-scanned with respect to the output focusing point $\bm{\rho}_\textrm{out}$. \textbf{c}-\textbf{d.} Each focal spot can be re-expressed in the laboratory frame ($\bm{\rho}_\textrm{in}$) (\textbf{c}) and stored in the canonical reflection matrix $\mathbf{R}$ (\textbf{d}).  \textbf{e.} Each line of $\mathbf{R}$ can be reshaped in 2D corresponds to the output focal spot in the focused basis ($\bm{\rho}_\textrm{out}$). \textbf{f.} Each output focal spot can be de-scanned with respect to the input focusing point $\bm{\rho}_\textrm{in}$ and stored in the output de-scan matrix $\mathbf{R}_\textrm{out}$.}
    \label{fig:exch_in_out}
\end{figure}

The multi-scale compensation of wave distortions consists in dividing by two the lateral extension $L$ of the spatial windows $\mathcal{W}$ at each step. The full process is described in the Methods section of the accompanying paper and summarized in a flowchart displayed in Supplementary Fig.~\ref{fig:flowchart}. At each step, the correction process is iterated both at input and output of the reflection matrix (left and right parts of Supplementary Fig.~\ref{fig:flowchart}). Mathematically, the transfer between the input and output de-scanned bases is performed by a change of variable (Eqs.~12 and 16) illustrated by Supplementary Fig.~\ref{fig:exch_in_out}. {In particular, Supplementary Fig.~\ref{fig:exch_in_out}f shows that the output de-scan matrix $\mathbf{R}_\textrm{out}$ cannot be fully retrieved. A set of  coefficients cannot be determined in its corners and are arbitrarily fixed to zero. They correspond to de-scanned coordinates $(\bm{\rho}_\textrm{in},\Delta \bm{\rho}')$ associated with points $\bm{\rho}_\textrm{out}=\bm{\rho}_\textrm{in}+\Delta \bm{\rho}'$ outside of the initial field-of-detection. To avoid the potential detrimental impact of such zero coefficients on the estimation of the $\rev{\rev{\bm{\mathcal{T}}}}-$matrix, the output correlation matrix $\mathbf{C}_\textrm{out}$ is only computed over points $\bm{\rho}_\textrm{in}$ that are associated with a full de-scan wave-field, \textit{i.e} points $\bm{\rho}_\textrm{in}$ such that ${R}_\textrm{out}(\bm{\rho}_\textrm{in},\Delta \bm{\rho}')\neq 0$ for each de-scan position $\Delta \bm{\rho}'$.} 

In a previous work~\cite{badon_distortion_2020},  the compensation of wave-front distortions was performed in one single step and on a single side (output). The low spatial sampling of the reflection matrix at input explained this minimalist strategy. In the accompanying paper, the de-scanned measurement of the reflection matrix provides the same sampling of the wave-field at input and output. An alternate compensation of wave distortions is therefore possible and actually critical if one wants to converge towards a sharp estimator of the $\rev{\bm{\mathcal{T}}}$-matrix. Indeed, as shown by Eq.~\ref{bias2}, the bias of this estimator on one side (input/output) directly depends on the focusing quality on the other side (output/input) since it controls the blurring of the virtual guide star synthesized by a coherent combination of focal spots. By alternating aberration compensation at input and output, we can improve gradually the coherence factor $\mathcal{C}_\textrm{out/in}$ and address forward multiple scattering associated with smaller isoplanatic patches (decrease $N_\mathcal{W}$) while maintaining the bias $\delta \rev{{\mathcal{T}}}_\textrm{in/out}$ at a sufficiently low level. 
\begin{figure}[h!]
    \centering
    \includegraphics[width=\linewidth]{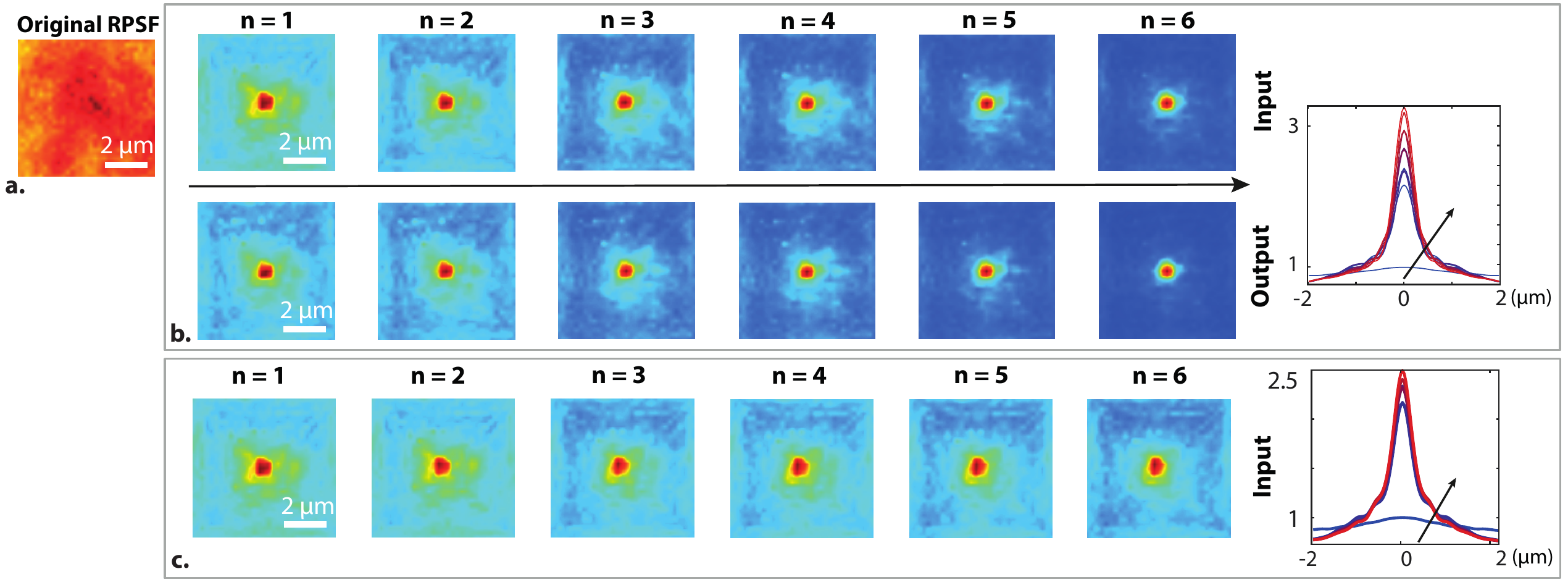}
    \caption{\textbf{On the importance of alternating the compensation of wave distortions at input and output.} \textbf{a.}  Original RPSF at the same position considered in Fig.~5 of the accompanying paper ($z=200$ $\mu$m). \textbf{b.} Local RPSF at the same position, after every step of the iterative process (multi-scale and input/output).  \textbf{c}. Local RPSF at the same position, after every step of the iterative process (multi-scale and input only). For panels \textbf{b} and \textbf{c}, the evolution of the radial profile  of the RPSF throughout the iterative process is displayed on the right. Scale bar: 2$\mu$m.}  \label{fig:alternance}
\end{figure}

Supplementary Figure \ref{fig:alternance} illustrates the importance of an alternate compensation of aberration and scattering at input and output. Supplementary Figure \ref{fig:alternance}b shows the evolution of the RPSF at each step of the algorithm when balancing between input and output. Supplementary Figure \ref{fig:alternance}c shows the evolution of the RPSF when the algorithm is only iterated at input. While a continuous balance between input and output aberration phase laws allows us to reach a diffraction-limited resolution at the end of the process (Fig.R3b), the absence of correction at output prevents from a refinement of the virtual guide star and does not allow our algorithm to converge towards a satisfying estimation of the matrix $\rev{\bm{\mathcal{T}}}_{\!\! \textrm{in}}$.

\subsection{Convergence of the multi-scale analysis process}

\begin{figure}[h!]
    \centering
    \includegraphics[width=\linewidth]{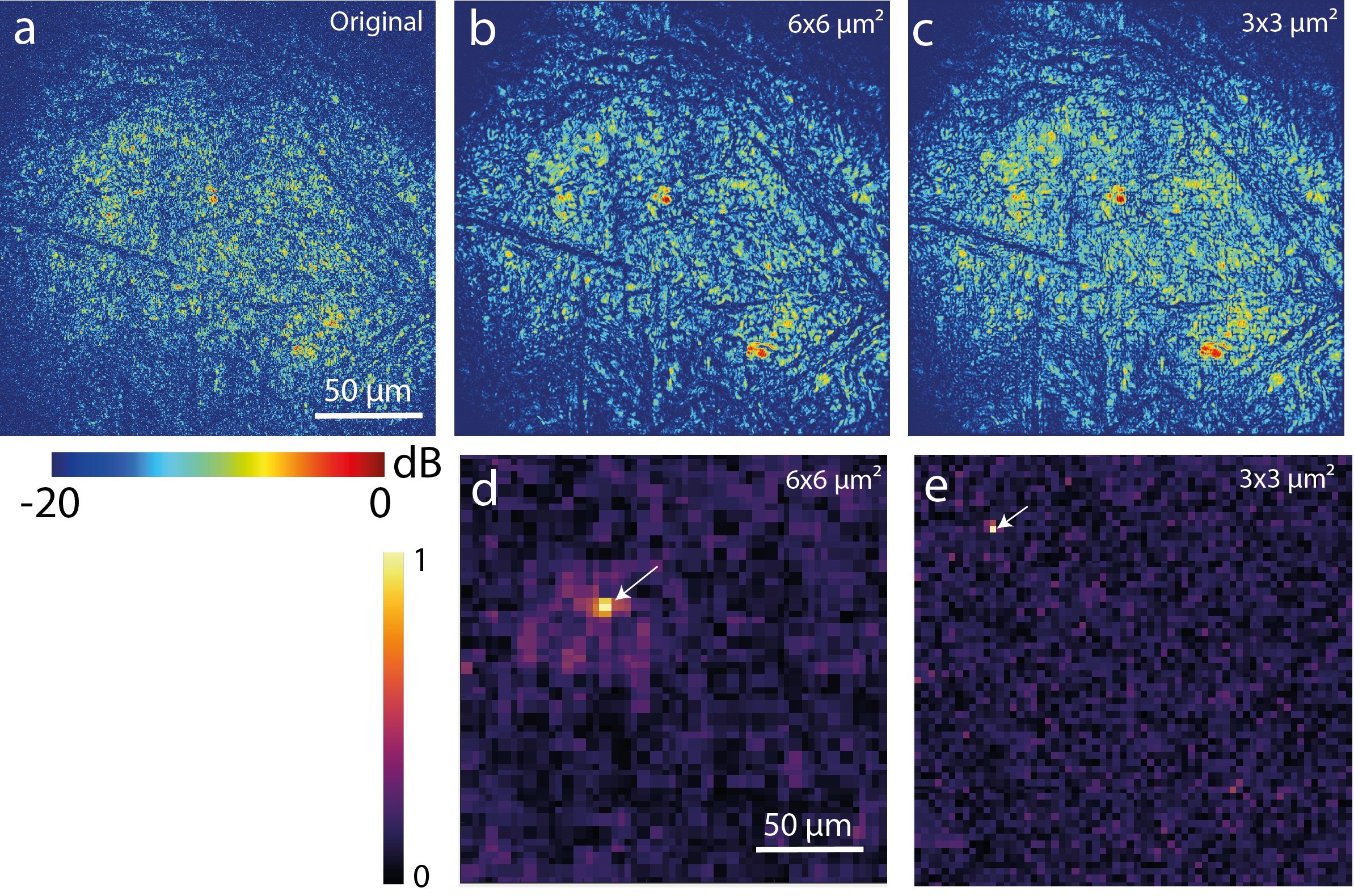}
    \caption{\textbf{Confocal images at several steps of the multi-scale analysis}. \textbf{a.} Initial en-face image of the cornea at depth $z\sim$ 100 $\mu$m. \textbf{b}-\textbf{c.} RMI images based on a $\rev{\bm{\mathcal{T}}}-$matrix estimator of spatial resolution $L=6~\mu$m and $L=3~\mu$m, respectively. \textbf{d}-\textbf{e.} Spatial correlation $\rev{C_{\mathcal{S}}}$ of $\rev{\bm{\mathcal{S}}}$ with respect to one reference location (white arrow). Scale bars: $50~\mu$m.}
    \label{fig:vignettage}
\end{figure}

The multi-scale process shown in Fig.~\ref{fig:flowchart} shall be stopped at some iterative step. Indeed, the spatial window $W_L$ cannot be reduced to a speckle grain otherwise the method would lead to a bucket image that consists in an incoherent summation of each de-scanned wave-field. 
Qualitatively, the end of the process can be determined by a careful look at the image. An incoherent compensation of aberrations induces a loss of contrast on the final image. Figure \ref{fig:vignettage} illustrates this assertion by comparing the original image (Fig. \ref{fig:vignettage}a), the RMI image obtained with a $\rev{\bm{\mathcal{T}}}-$matrix of optimal resolution ($6 \times 6$ $\mu$m$^2$, see Supplementary Fig. \ref{fig:vignettage}b) and a RMI image based on too small spatial windows $W_L$  ($3 \times 3$ $\mu$m$^2$, see Supplementary Fig.~\ref{fig:vignettage}c). The contrast of each image $I(\bm{\rho},z)$,  $\mathcal{F}(z)= \mbox{std} \left[ I(\bm{\rho},z) \right]/\left\langle I(\bm{\rho},z) \right\rangle$, tends to gradually increase when the estimator $\rev{\hat{\bm{\mathcal{T}}}}$ approaches $\rev{\bm{\mathcal{T}}}$ (see comparison between Supplementary Figs.~\ref{fig:vignettage}a and b) and decrease when the compensation of aberrations and scattering becomes bucket-like (see comparison between Supplementary Figs.~\ref{fig:vignettage}b and c). For the images displayed in Supplementary Figs. \ref{fig:vignettage}a, b and c, we find $\mathcal{F}\sim 1.48$, $\mathcal{F}\sim 1.61$ and $\mathcal{F}\sim 1.37$, respectively. Nevertheless, an optimization criterion only based on the image contrast can be misleading since the contrast also depends on the sample reflectivity distribution. 

{A more reliable observable is the spatial correlation function ${C}_S(\rp,\rp')$ of the scattering component of the $\rev{\bm{\mathcal{T}}}-$matrix between neighboring points $\rp$ and $\rp'$ (Methods). Examples of this spatial correlation function are displayed in Supplementary Figs.~\ref{fig:vignettage}d and \ref{fig:vignettage}e. 
While a spatial window of $6 \times 6$ $\mu$m$^2$ preserves a short-range correlation between neighbor windows (Supplementary Fig.~\ref{fig:vignettage}d), a spatial window of $3 \times 3$ $\mu$m$^2$ leads to a fully spatially incoherent estimator $\rev{\hat{\bm{\mathcal{T}}}}$ (Supplementary Fig.~\ref{fig:vignettage}e). This observable clearly shows whether the estimator $\rev{\hat{\bm{\mathcal{T}}}}$ leads to a coherent (i.e physical) or incoherent (i.e bucket-like) compensation of scattering. The number of iterations in the phase reversal algorithm has thus been based on this $\rev{\bm{\mathcal{T}}}-$matrix correlation criterion. }

\subsection{Benefit of a multi-scale strategy}

\begin{figure}[h!]
    \centering
    \includegraphics[width=14cm]{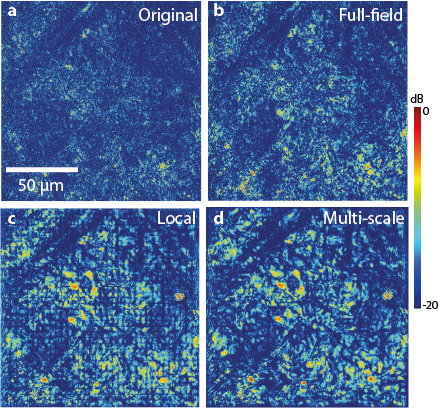}
    \caption{\textbf{On the importance of a multi-scale compensation of wave distortions.} \textbf{a.}  Original confocal image of the cornea at $z=200$ $\mu$m. \textbf{b.} Full-field correction. \textbf{c.} Direct compensation of aberrations over reduced spatial windows ($L=6$ $\mu$m). \textbf{d.} Multi-scale strategy. Scale bar: 50 $\mu$m}  \label{fig:multiscale}
\end{figure}
Supplementary Figure \ref{fig:multiscale} shows the benefit of a multi-scale compensation of wave distortions. On the one hand, a full-field correction only addresses the aberrations induced by the reference arm and does not address forward multiple scattering (Supplementary Section~\ref{CBS}). It thus only provides a blurred view of the corneal internal structure (Supplementary Fig.~\ref{fig:multiscale}b). On the other hand, a direct compensation of wave distortions leads to a strong vigneting effect (Supplementary Fig.~\ref{fig:multiscale}c). The latter phenomenon is due to the imperfect convergence of the IPR algorithm over extremely reduced spatial windows (Supplementary Sec.~\ref{bias}). Note that this detrimental effect is not limited to IPR but also exists for other algorithms such as CLASS (Supplementary Fig.~\ref{fig:CLASSb}b$_1$) or ITR (Supplementary Fig.~\ref{S3_IPRvsSVD}b$_3$). On the contrary, our multi-scale strategy limits the bias of our estimator and provides a clear view of the cornea reflectivity without being hampered by any vignetting phenomenon.

\subsection{\alex{Validation of the method with a ground-truth object}}
\label{sec:res_target}

\alex{We now provide an experimental validation of the method using a resolution target behind the scattering medium. Although such a specular object does not reproduce the reflectivity properties of tissues, this reference experiment will allow us to validate our multi-scale analysis of wave distortions and also outline its limits.}\\

\alex{This experiment is displayed in Supplementary Fig.~\ref{fig:mire}. It consists in the imaging of a resolution target placed right behind a 500-$\mu$m-thick mouse peritoneum layer (Supplementary Fig.~\ref{fig:mire}a). This tissue layer roughly displays the same scattering properties as the cornea used in the accompanying manuscript.}

\begin{figure}[h!]
    \centering
    \includegraphics[width=17cm]{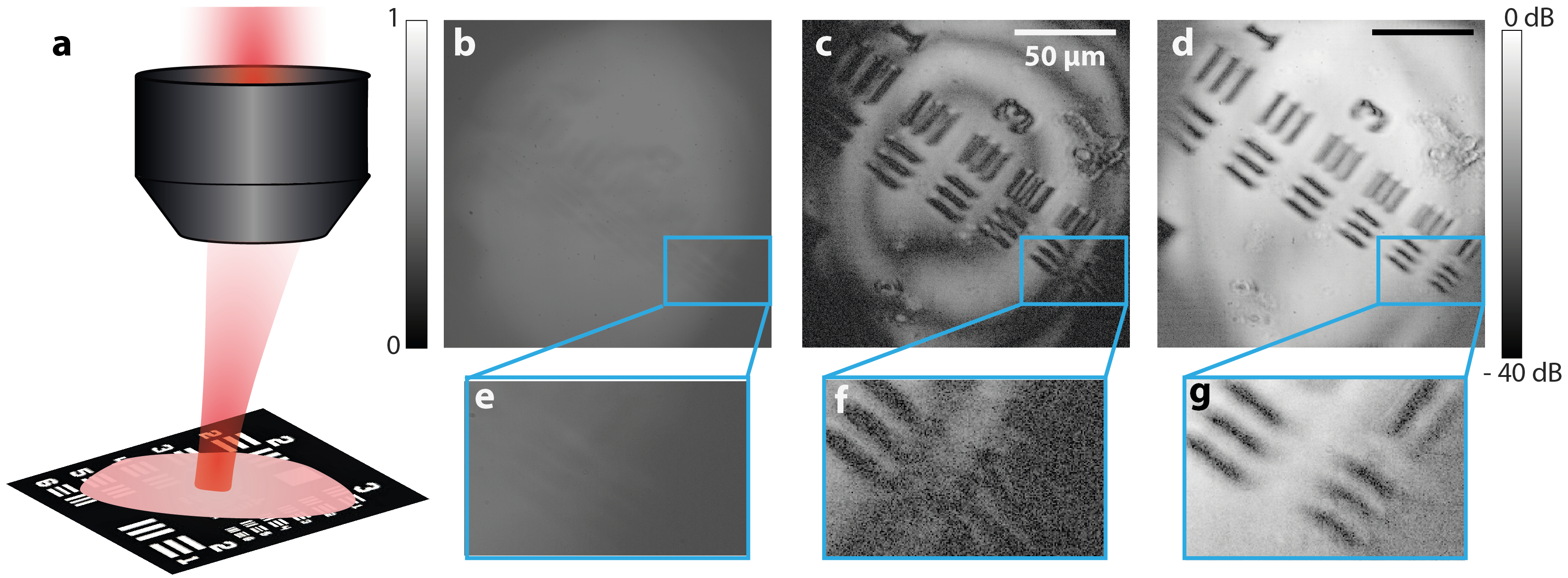}  
    \caption{\alex{\textbf{Imaging a resolution target hidden behind an opaque tissue layer.} 
    \textbf{a.} Experimental configuration.{ Image credit:  Setreset on Wikimedia Commons.} \textbf{b.} Incoherent image (colorbar in linear scale). \textbf{c.} FFOCT image (B\&W bar in log-scale). \textbf{d.} Matrix Image (B\&W bar in log-scale). \textbf{e}-\textbf{h.} Zooms on the smallest details of the resolution target images displayed in panels \textbf{b}-\textbf{d}, respectively. Scale bar: 50 $\mu$m.}}  \label{fig:mire}
\end{figure}
\alex{Supplementary Figure~\ref{fig:mire}b displays the incoherent image of the target in reflection which is obtained here by blocking the reference arm in the experimental set up. Its foggy feature highlights the strong turbidity of the scattering layer. Supplementary Figure~\ref{fig:mire}c shows the FFOCT image acquired for $\Delta \bm{\rho}_\textrm{in}=\mathbf{0}$. Its comparison with Supplementary Fig.~\ref{fig:mire}b illustrates the drastic filtering of diffuse multiple scattering operated by the time gating process in FFOCT. It also shows how the confocal filter allows to reveal the large patterns of the resolution target. The fact that FFOCT is robust with respect to aberrations for specular objects has already been noticed in a previous work~\cite{barolle_manifestation_2021}. In that case, aberrations only give rise to fluctuations of image contrast across the field-of-view. Nevertheless, this robustness to aberrations vanishes for the smallest details of the target that cannot be detected [see the corresponding zoom in Supplementary Fig.~\ref{fig:mire}g].}\\

\alex{The aberration and scattering induced by the tissue layer can be evaluated by the RPSF whose spatial evolution is displayed in Supplementary Fig.~\ref{fig:mire_RPSF}a. It shows drastic variations across the field-of-view, a manifestation of a particularly short-range memory effect across the field-of-view. Compared with the incoherent RPSFs measured inside the cornea (Fig.~4c of the accompanying paper), the coherent RPSFs here show a more contrasted feature. This is due to the specular nature of the sample. In that regime, the RPSF is a coherent function of the focusing quality since it scales as the convolution between the coherent input and output PSFs~\cite{lambert_reflection_2020}: $$I(\Delta \bm{\rho}_\textrm{in},{\bm{\rho}_{\textrm{out}},z})=|H^{\rev{(l)}}_{\textrm{in}} \stackrel{\Delta \bm{\rho}_\textrm{in}}{\circledast} H^{\rev{(l)}}_{\textrm{out}} (\Delta \rho_{\textrm{in}},\bm{\rho}_{\textrm{out}},z) |^2.$$
On the contrary, for a sample of random reflectivity (like the opaque cornea under study), the RPSF is an incoherent measure of the focusing quality. Its ensemble average can be expressed as the convolution between the incoherent input and output PSFs (Eq.~3 of the accompanying paper):  $$ \langle I(\Delta \bm{\rho}_\textrm{in},{\bm{\rho}_{\textrm{out}},z}) \rangle \propto |H^{\rev{(l)}}_\textrm{in}|^2 \stackrel{\Delta \bm{\rho}_\textrm{in}}{\circledast} |H^{\rev{(l)}}\out|^2 (\Delta \bm{\rho}_\textrm{in},{\bm{\rho}_{\textrm{out}},z}).$$}

\alex{Despite the short-range memory effect highlighted by the original RPSFs (Supplementary Fig.~\ref{fig:mire_RPSF}a), a multi-scale compensation of aberration and scattering phenomena allows us to retrieve an almost diffraction-limited RPSF across the whole surface (Supplementary Fig.~\ref{fig:mire_RPSF}b), except inside the patterns of the resolution target since there is no back-scattered wave-field there. The resulting image is displayed in Supplementary Fig.~\ref{fig:mire}d. Compared with its original counterpart (Supplementary Fig.~\ref{fig:mire}c), an homogeneous contrast is obtained throughout the field-of-view. Above all, RMI is able to retrieve the smallest details of the resolution target (Supplementary Fig.~\ref{fig:mire}g) that FFOCT initially failed to reveal (Supplementary Fig.~\ref{fig:mire}f). 
Nevertheless, the image is not perfect. A first reason comes from the limited de-scan of the RPSF. This loss of information gives rise to a residual incoherent background in the final RPSF (Supplementary Fig.~\ref{fig:mire_RPSF}c). The second reason is the limited isoplanicity. The size $L$ of spatial windows at the end of the RMI process is limited to \ulysse{10}$~\mu$m. Hence we cannot compensate for scattering phenomena giving rise to a memory effect range smaller than $L$.}
\begin{figure}[h!]
    \centering
    \includegraphics[width=17cm]{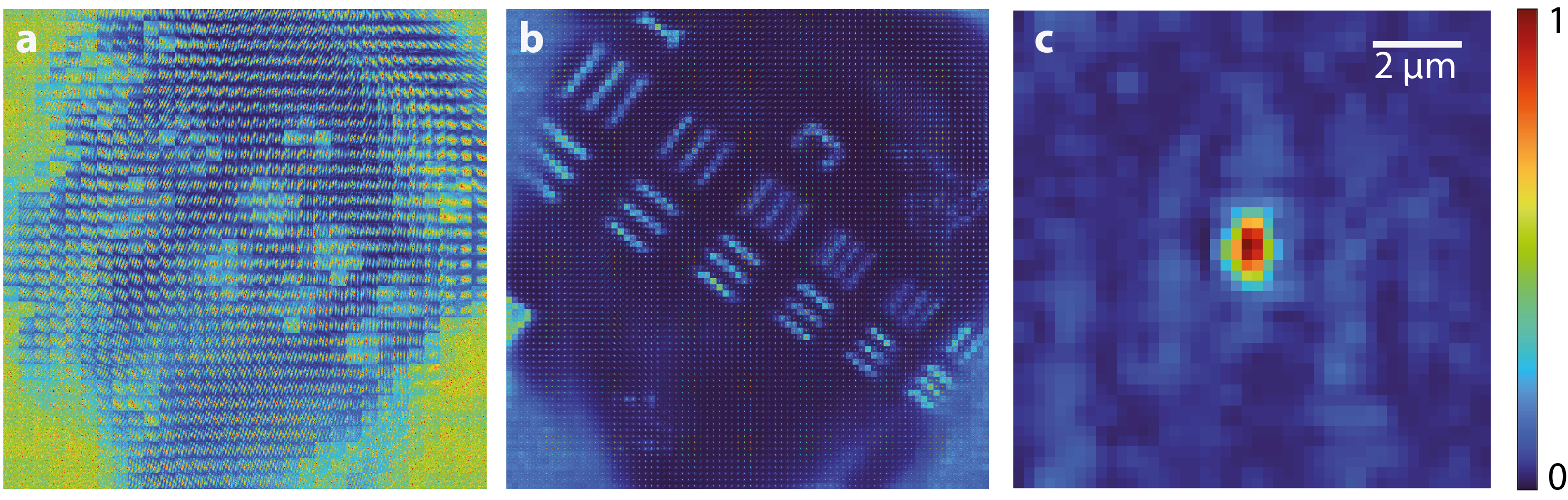} \caption{{\textbf{Reflection point spread function in the resolution target experiment.}. \textbf{a.} Map of initial RPSFs. \textbf{b.} Map of final RPSFs. \textbf{c.} Mean RPSF after aberration compensation (scale bar: $2~\mu $m).}}   \label{fig:mire_RPSF}
\end{figure}

\clearpage

\section{Analysis of the {Transmittance} Matrix}

\subsection{Discrepancy between input and output ${{\mathcal{T}}}-$matrices}
\label{sectionSR}

\begin{figure}[h!]
    \centering
    \includegraphics[width=.9\linewidth]{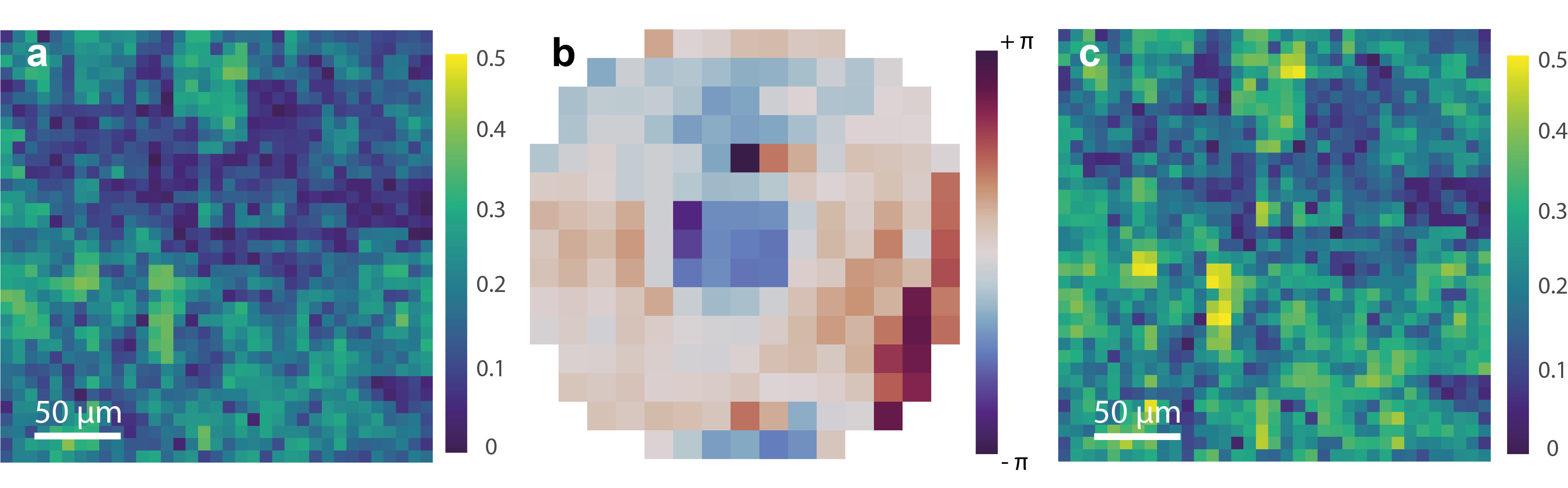}
    \caption{\textbf{Discrepancy between input and output $\mathbf{T}-$matrices.} \textbf{a.} Scalar product $P(\bm{\rho}_\textrm{p},z)$ between $\rev{\hat{\bm{\mathcal{T}}}}_\textrm{in}$ and $\rev{\hat{\bm{\mathcal{T}}}}_\textrm{out}$ at depth $z=$200 $\mu$m. \textbf{b.} Phase of the first pupil singular vector $\mathbf{U}_1$ of $\rev{\hat{\bm{\mathcal{T}}}}_\textrm{in}$. \textbf{c.} Scalar product $P'(\bm{\rho}_\textrm{p},z)$ between  $\rev{\hat{\bm{\mathcal{T}}}}'_\textrm{in}$ and $\rev{\hat{\bm{\mathcal{T}}}}_\textrm{out}$. Scale bar: 50 $\mu$m. }
    \label{fig:spatial_reciprocity}
\end{figure}
While spatial reciprocity implies a strict equality between the wave distortions undergone by the incident and reflected waves in the sample arm of our experimental set-up, the input and output estimators of the $\rev{{\bm{\mathcal{T}}}}-$matrix are far from it (see Figs.~4e and f of the accompanying paper). This discrepancy can be quantified by computing the normalized scalar product $P(\bm{\rho}_\textrm{p},z)$ between the coefficients of $\rev{\hat{\bm{\mathcal{T}}}}_\textrm{in}$ and $\rev{\hat{\bm{\mathcal{T}}}}_\textrm{out}$:
\begin{equation}
P(\bm{\rho}_\textrm{p},z)=N_u^{-1}  \sum_\mathbf{u} \rev{\hat{{\mathcal{T}}}}_\textrm{in} (\mathbf{u},\bm{\rho}_\textrm{p},z)\rev{\hat{{\mathcal{T}}}}_\textrm{out} ^* (\mathbf{u},\bm{\rho}_\textrm{p},z)
\end{equation}
Supplementary Fig.~\ref{fig:spatial_reciprocity} shows the transverse evolution of this scalar product at depth $z=200$ $\mu$m. As it could be anticipated when looking at the $\rev{\hat{\bm{\mathcal{T}}}}-$matrices in Figs.~4e and f, this scalar product is quite low: $P(\bm{\rho}_\textrm{p},z)=0.3$ in average. As we will see, this discrepancy can be, at least partially, explained by the aberrations in the reference arm. Indeed, according to Eq.~\ref{SR}, the input transmission matrix accumulates the aberrations undergone by the incident wave-field in the sample arm and the aberrations undergone by the reference wave-field,
\begin{equation}
\rev{{\bm{\mathcal{T}}}}_\textrm{in}=\rev{{\bm{\mathcal{T}}}} \circ \rev{{\bm{\mathcal{T}}}}_\textrm{ref}.  
\end{equation}
On the contrary, $\rev{\hat{\bm{\mathcal{T}}}}_\textrm{out}$ only grasps the wave distortions undergone by the reflected wave-field in the sample arm. If, in a first approximation, we assume that the aberration due to the reference arm is isoplanatic, it can be extracted by considering the first eigenstate of $\rev{\hat{\bm{\mathcal{T}}}}_\textrm{in}$ (see Supplementary Section~\ref{svdT}). The phase of the pupil singular vector $\mathbf{U}_1$ displayed in Supplementary Fig.~\ref{fig:spatial_reciprocity} is an estimator of $\rev{{\bm{\mathcal{T}}}}_\textrm{ref}$. Not surprisingly, it mainly corresponds to a spherical aberration phase law. One can subtract this reference phase to $\rev{\hat{\bm{\mathcal{T}}}}_\textrm{in}$ in order to build a matrix $\rev{\hat{\bm{\mathcal{T}}}}'_\textrm{in}=\rev{\hat{\bm{\mathcal{T}}}}_\textrm{in} \circ {\mathbf{U}}_1^*$. One can expect the scalar product $P'(\bm{\rho}_p,z)$ between $\rev{\hat{\bm{\mathcal{T}}}}'_\textrm{in}$ and $\rev{\hat{\bm{\mathcal{T}}}}_\textrm{out}$ to be increased compared to its initial value $P(\bm{\rho}_p,z)$ (see comparison between Supplementary Figs.~\ref{fig:spatial_reciprocity}a and c). This is actually what we observe even though the scalar product $P'$ remains smaller than 0.7 (Supplementary Fig.~\ref{fig:spatial_reciprocity}c). It means that the spherical aberration law induced by the reference arm account partially for the mismatch between $\rev{\hat{\bm{\mathcal{T}}}}_\textrm{in}$ and $\rev{\hat{\bm{\mathcal{T}}}}_\textrm{out}$. 

The residual mismatch can be explained by the fact that the aberration induced by the reference arm is not strictly isoplanatic. Misalignment between sample and reference arms manifests as a transverse shift of the RPSF that varies across the field-of-view as illustrated by Fig.4c.  Field curvature can also induce spatially-varying aberrations that our approach can address but they are difficult to discriminate from sample arm aberrations. Last but not least, another phenomenon that can contribute to this discrepancy between input and output aberration phase laws is the bias of our $\rev{{\bm{\mathcal{T}}}}$-matrix estimator, especially for small spatial windows $L$ as explained in Supplementary Section \ref{bias}.

\subsection{Aberration and scattering components of the ${\mathcal{T}}-$matrix}
 \label{svdT}
In a previous work, Badon \textit{et al.} showed how the singular value decomposition (SVD) of the $\mathbf{D}-$matrix provided a decomposition of the field-of-view into isoplanatic modes in the case of a specular object. In the present paper, this property does not hold since we cope with a random distribution of heterogeneities. In this regime, this is the SVD of the $\rev{{\bm{\mathcal{T}}}}$-matrix estimator that enables a mapping of isoplanatic modes. As recently shown in an ultrasound study~\cite{lambert_ultrasound_2022}, the  complexity of the associated aberration phase laws increases with the rank of the corresponding singular values while the spatial extension of the isoplanatic mode decreases. As we will show below, this complexity can  be quantified by a vorticity degree of the associated transmittance, quantity that has a direct link with the occurrence of multiple scattering paths involved in the trajectory of the wave from the focal plane to the camera sensors.

The SVD of $\rev{\hat{\bm{\mathcal{T}}}}$ writes:
\begin{equation}
\rev{\hat{\bm{\mathcal{T}}}}=\sum_k\sigma_k \mathbf{U}_k \mathbf{V}_k^\dag
\end{equation}
where $\sigma_k$ are the singular values arranged in decreasing order. $\mathbf{U}_k = [U_k(\mathbf{u})]$ and $\mathbf{V}_k = [V_k(\bm{\rho}_\textrm{p})] $ are the singular vectors of $\rev{\hat{\bm{\mathcal{T}}}}$ in the pupil and focal plane, respectively. For a physical interpretation of these vectors, we take advantage of the equivalence between the SVD of $\rev{\hat{\bm{\mathcal{T}}}}$  and the eigenvalue decomposition of the spatial correlation matrix,
\begin{equation}
\mathbf{C}_{\rev{\mathcal{T}}}=\rev{\hat{\bm{\mathcal{T}}}}\times \rev{\hat{\bm{\mathcal{T}}}}^\dag.
\end{equation}
The elements of $\mathbf{C}_{\rev{\mathcal{T}}}$ correspond to the correlation coefficients between aberration phase laws obtained for each image pixel $\bm{\rho}_p$ and $\bm{\rho}'_p$:
\begin{equation}
{C}_{\mathcal{T}}(\bm{\rho}_p,\bm{\rho}'_p)=\sum_{\mathbf{u}} \rev{\hat{{\mathcal{T}}}}(\mathbf{u},\bm{\rho}_p) \rev{\hat{{\mathcal{T}}}}^*(\mathbf{u},\bm{\rho}_p).
\end{equation}
The first eigenvector $\mathbf{V}_1$ of $\mathbf{C}_{\rev{\mathcal{T}}}$ is thus the spatial domain where the degree of correlation between aberration phase laws is maximized. This degree of correlation is quantified by the normalized eigenvalue $\overline{\sigma}_1^2$, such that 
\begin{equation}
\overline{\sigma}_k^2=\frac{{\sigma}_k^2}{\sum_l {\sigma}_l^2 }=\frac{\mathbf{V}_k^\dag \times \mathbf{C}_{\rev{\mathcal{T}}} \times \mathbf{V}_k}{\mbox{Tr}\lbrace \mathbf{C}_{\rev{\mathcal{T}}} \rbrace } 
\end{equation}
The corresponding singular vector \begin{equation}
\mathbf{U}_1=\sigma_1^{-1} \rev{\hat{\bm{\mathcal{T}}}}\times \mathbf{V}_1
\end{equation}
is the transmittance the most-spatially invariant across the field-of-view. The same process can be iterated on the matrix $\rev{\hat{\bm{\mathcal{T}}}}-\sigma_1 \mathbf{U}_1\times \mathbf{V}_1^{\dag}$ to retrieve the second eigenstate and so on. A set of orthogonal isoplanatic modes  $\mathbf{V}_k$ is finally obtained with a degree of correlation $\overline{\sigma}_k^2$ that decreases with their rank.
\begin{figure}[h!]
    \centering
    \includegraphics[width=16 cm]{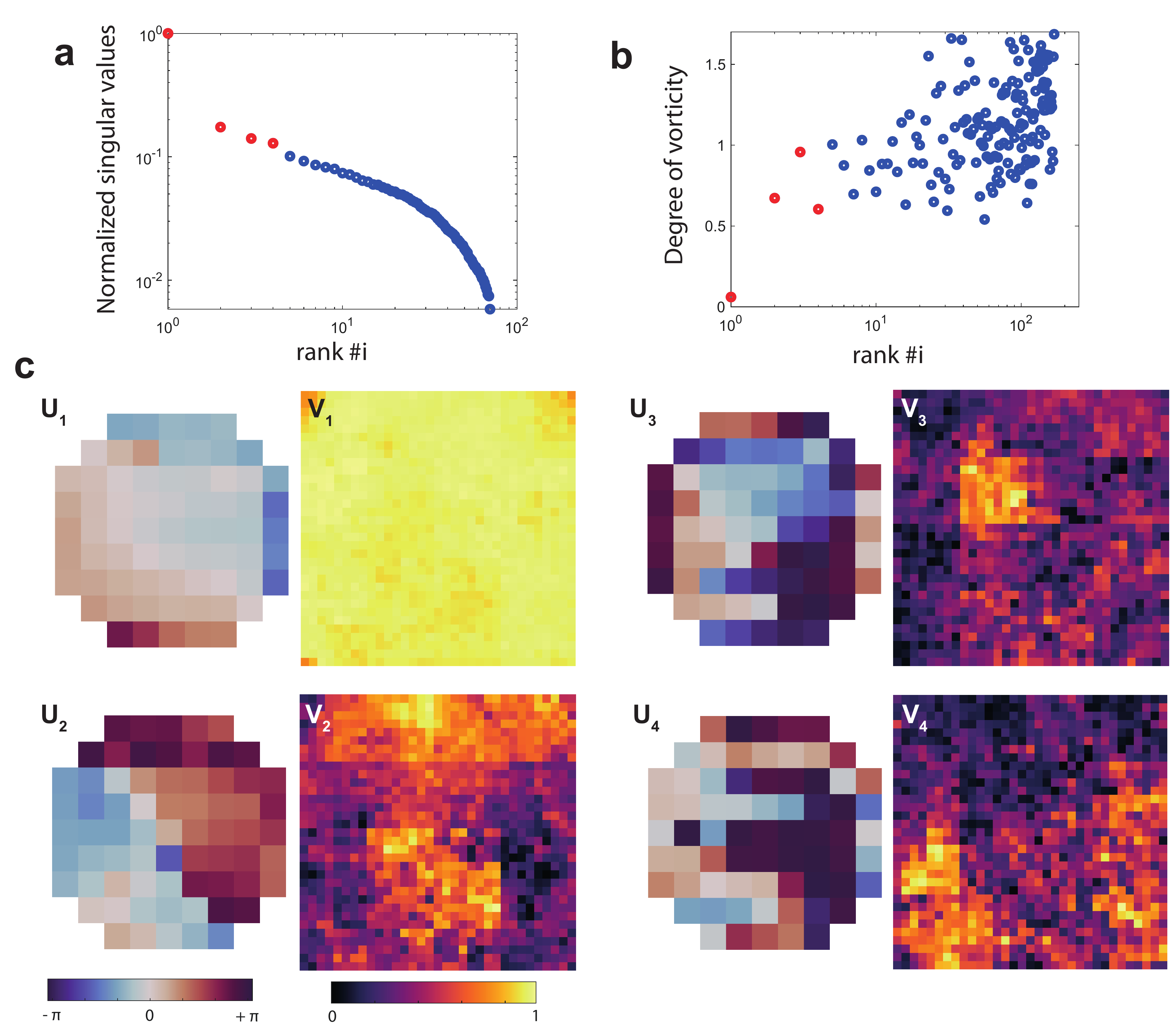}
    \caption{\textbf{Singular value decomposition of the $\hat{\rev{\bm{\mathcal{T}}}}-$matrix at depth $z=50$ $\mu$m}. \textbf{a.} Normalized singular values of $\hat{\rev{\bm{\mathcal{T}}}}_\textrm{out}$. \textbf{b.} Degree of vorticity associated with each pupil singular vector $\mathbf{U}_k$ of $\hat{\rev{\bm{\mathcal{T}}}}_\textrm{out}$. \textbf{c.} Four first eigenstates of $\hat{\rev{\bm{\mathcal{T}}}}_\textrm{out}$: Phase of the transmittance $\mathbf{U}_k$ and modulus of the corresponding isoplanatic modes $\mathbf{V}_k$.}
\label{fig:svdT}
\end{figure}

Supplementary Figure~\ref{fig:svdT} shows the result of the SVD of $\rev{\hat{\bm{\mathcal{T}}}}_\textrm{out}$ for $z=50$  $\mu$m (same depth as the one considered on top of Fig.6 of the accompanying paper). Supplementary Figure~\ref{fig:svdT}a displays its normalized singular values. A few predominant eigenvalues associated with the main isoplanatic modes seem to emerge from a continuum of lower eigenvalues associated with a multiple scattering background in each case. Supplementary Figure~\ref{fig:svdT}c shows the four first eigenstates of $\rev{\hat{\bm{\mathcal{T}}}}_\textrm{out}$. While the first eigenstate $\mathbf{V}_1$ spans over the whole field-of-view, the higher order isoplanatic modes $\mathbf{V}_k$ are associated with spatial domains whose size decrease with the rank $k$ of the eigenstate. The complexity (i.e the spatial frequency content) of the associated transmittance $\mathbf{U}_k$ also increases with this rank. 

The nature of the associated wave distortions can be investigated by considering the phase $\phi_k$ of each singular vector $\mathbf{U}_k$. More precisely, recent works~\cite{Wu2021,wu2023singleshot} showed how aberrations and scattering can be discriminated by computing the divergence and curl of the phase gradient $\nabla \phi_k$. Each phase law $\phi_k$ can be decomposed 
into: (\textit{i}) an irrotational component $\phi_k^{\textrm{(irr)}}$, such that $\nabla \wedge \nabla \phi_k^{\textrm{(irr)}}=\mathbf{0}$, associated with low-order aberrations; (\textit{ii}) a curl component $\phi_k^{\textrm{(rot)}}$, such that $\nabla \cdot \nabla \phi_k^{\textrm{(rot)}}={0}$, induced by forward multiple scattering trajectories. Indeed, this curl component is a manifestation of optical vortices that necessarily originate from, at least,
three interfering beams and thus suppose several optical
trajectories, hence multiple scattering.
A degree $\alpha$ of vorticity can be assessed by looking at the ratio between the energy of each contribution, such that:
\begin{equation}
\alpha_k=\frac{|\nabla \wedge \nabla \phi_k |^2 }{|\Delta \phi_k |^2 }
\end{equation}
This degree of vorticity is displayed for each eigenstate in Supplementary Fig.~\ref{fig:svdT}b. Although it shows some fluctuations,  $\alpha_k$ tends to increase with the rank $k$ of eigenstate. It thus seems to indicate that the higher order eigenstates associated with smaller isoplanatic patches also exhibit a higher degree of vorticity, which \alex{can be} a manifestation of forward multiple scattering paths. 

\subsection{\alex{Angular decomposition of the $\mathcal{T}$-matrix}}

To investigate the effect of forward multiple scattering, an angular decomposition of the transmitted wave-field between the cornea surface and the focal plane can be performed. Interestingly, this can be done by considering the Fourier transform of the transmission matrix estimator $\rev{\hat{\bm{\mathcal{T}}}}(z)=[\hat{\rev{\mathcal{T}}}(\mathbf{u}_\textrm{in},\bm{\rho}_\textrm{out},z)]$ \rev{along} the focused basis:
\begin{equation}
\tilde{\rev{\bm{\mathcal{T}}}}(z)=\hat{\rev{\bm{\mathcal{T}}}}(z) \times \alex{\mathbf{T}_0},
\end{equation}
which writes, in terms of matrix coefficients,
\begin{equation}
\tilde{\rev{\mathcal{T}}}(\mathbf{u},\Delta \mathbf{u},z)=\sum_{\bm{\rho}_\textrm{out}} \hat{\rev{\mathcal{T}}]}(\mathbf{u}_\textrm{in},\bm{\rho}_\textrm{out},z) \exp \left ( - i \frac{2\pi}{\lambda f } \mathbf{\Delta \mathbf{u}.\bm{\rho}_\textrm{out}} \right ).
\end{equation}

\begin{figure}[h!]
    \centering
    \includegraphics[width=16cm]{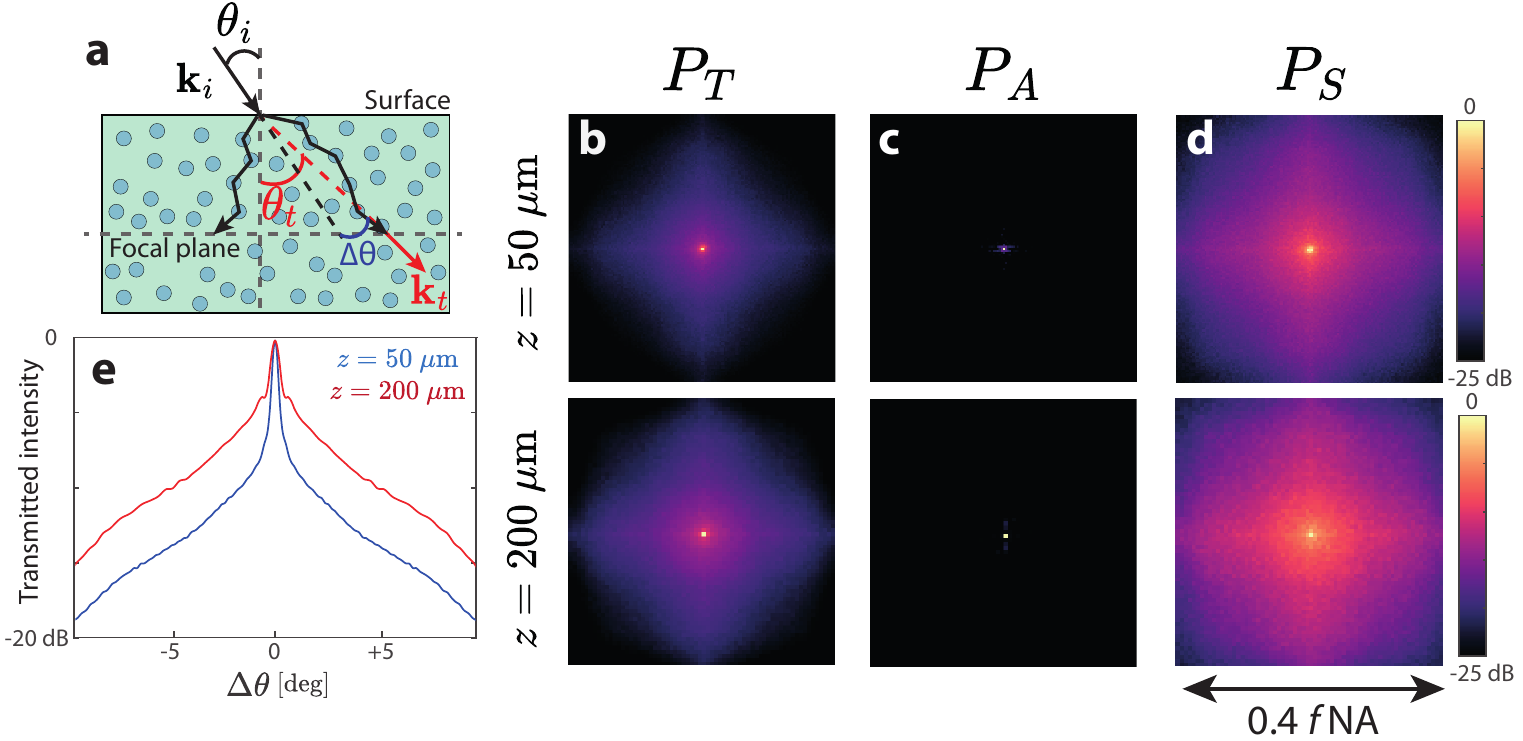}  \caption{\alex{\textbf{Plane wave decomposition of the $\bm{\mathcal{T}}-$matrix.}. \textbf{a.} Definition of the incident and transmitted wave vectors, $\mathbf{k}_{\textrm{in}}$ and $\mathbf{k}_{\textrm{out}}$, and their associated angles, $\theta_{\textrm{in}}$ and $\theta_{\textrm{out}}$, at the cornea surface and the focal plane. \textbf{b}-\textbf{d.} Wave vector deviation distribution of the transmitted wave-field $P_{\rev{\mathcal{T}}}(\Delta \mathbf{u} ,z)$ (\textbf{b}), of its aberrated component $P_{\rev{\mathcal{A}}}(\Delta \mathbf{u}  ,z)$ (\textbf{c}), and of its scattered component $P_{\rev{\mathcal{S}}}(\Delta \mathbf{u}  ,z)$ (\textbf{d}) [Top: $z=50$ $\mu$m; Bottom: $z=200$ $\mu$m]. \textbf{e.} Angular distribution $P_{\rev{\mathcal{S}}}(\Delta \theta,z)$ of the scattered wave-field at depths z= 50 $\mu$m (blue line) and  z= 200 $\mu$m (red line).}}   \label{fig:spectreT}
\end{figure}

\rev{To show the relationship between $\tilde{\mathcal{T}}(\mathbf{u}_\textrm{in},\Delta \mathbf{u},z)$ and the angular distribution of light in the focal plane, one can use the relationship between the transmission and transmittance matrices, 
\begin{equation}
\label{eq1}
\mathbf{T}=\bm{\mathcal{T}}\circ \mathbf{T}_0,
\end{equation}
which writes in terms of matrix coefficients:
\begin{equation}
{T}(\mathbf{u}_\textrm{in},\bm{\rho}_\textrm{out},z)= \mathcal{T}(\mathbf{u}_\textrm{in},\bm{\rho}_\textrm{out},z) \exp \left ( j \frac{2 \pi}{\lambda f} \mathbf{u}_\textrm{in} \cdot \bm{\rho}_\textrm{out} \right ),
\end{equation}
or, equivalently,
\begin{equation}
\mathcal{T}(\mathbf{u}_\textrm{in},\bm{\rho}_\textrm{out},z)= {T}(\mathbf{u}_\textrm{in},\bm{\rho}_\textrm{out},z) \exp \left ( -j \frac{2 \pi}{\lambda f} \mathbf{u}_\textrm{in} \cdot \bm{\rho}_\textrm{out} \right ).
\end{equation}
Injecting this last expression into Eq.~\ref{eq1} leads to the following expression for $\tilde{\mathcal{T}}$:
\begin{align}
\label{eq2}
\tilde{\mathcal{T}}(\mathbf{u}_\textrm{in},\Delta \mathbf{u},z)&=\sum_{\Delta \mathbf{u}} \exp \left [ -j\frac{2\pi}{\lambda f} (\Delta \mathbf{u}+\mathbf{u}_\textrm{in}) \cdot \bm{\rho}_\textrm{out} \right ] {{T}}(\mathbf{u}_\textrm{in},\bm{\rho}_{\textrm{out}})\\
&= \tilde{T}(\mathbf{u}_\textrm{in},\mathbf{u}_\textrm{in}+\Delta \mathbf{u}).
\end{align}
Hence, the $\tilde{\bm{\mathcal{T}}}-$matrix actually corresponds to the de-scan transmission matrix in the pupil plane. }{Its coefficients $\tilde{\rev{\mathcal{T}}}(\mathbf{u}_\textrm{in},\Delta \mathbf{u},z)$ provide the angular dispersion of the transmitted wave-field at depth $z$ with respect to the incident plane wave of transverse wave vector $\mathbf{k}_{\textrm{in}}=\mathbf{u}_\textrm{in}/f$ (Supplementary Fig.~\ref{fig:spectreT}a). For small angles, the deviation angle $ \Delta \bm{\theta}=(\Delta \theta_x,\Delta \theta_y)$ can be expressed as follows:
\begin{equation}
 \Delta \bm{\theta} \sim \sin \bm{\theta}_{\textrm{out}} - \sin \bm{\theta}_{\textrm{in}}= \Delta \mathbf{u} /f 
\end{equation}
The angular dispersion of the wave-field between the cornea surface and the focal plane can thus be obtained by averaging the intensity of the transmitted wave-field over the incident wave vector:
\begin{equation}
 P_{\rev{\mathcal{T}}}(\Delta \mathbf{u}  ,z) = \left \langle | \tilde{\rev{\mathcal{T}}}(\mathbf{u}_{\textrm{in}}, \Delta \mathbf{u} /f,z) |^2\right \rangle_{\mathbf{u}_{\textrm{in}}}.
\end{equation}}


Supplementary Figure~\ref{fig:spectreT}b shows this angular distribution at depths $z=50$ and $200$ $\mu$m, respectively. It displays the following shape: A close-to-ballistic peak around $\Delta \mathbf{u}=\mathbf{0}$ on top of a wide pedestal generated by forward multiple scattering. This statement is confirmed by investigating the angular distributions, $P_{\rev{\mathcal{A}}}(\Delta \mathbf{u}  ,z)$ (Supplementary Fig.~\ref{fig:spectreT}c) and $P_{\rev{\mathcal{S}}}(\Delta \mathbf{u}  ,z)$ (Supplementary Fig.~\ref{fig:spectreT}d), associated with the aberration and scattering matrices, ${\rev{\bm{\mathcal{A}}}}_\textrm{out}$ and ${\rev{\bm{\mathcal{S}}}}_\textrm{out}$, respectively. As anticipated, the aberration component of the ${\rev{\bm{\mathcal{T}}}}-$matrix is associated with a close-to-ballistic peak around $\Delta \mathbf{u}=\mathbf{0}$ (Supplementary Fig.~\ref{fig:spectreT}c), while its scattering component gives rise to wide distribution of deviation angles in the focal plane (Supplementary Fig.~\ref{fig:spectreT}d). As illustrated by Supplementary Fig.~\ref{fig:spectreT}h, a wider angular distribution is observed for $P_{\rev{\mathcal{S}}}$ at 200 $\mu$m than at 50 $\mu$m. The angular width of the photon distribution at -10 dB goes from 4$\deg$ at $z=$ 50$\mu$m to 10$\deg$ at $z=$ 200$\mu$m. This angular dispersion of the transmitted wave-field between the cornea surface and the focal plane is a manifestation of scattering events taking place between those two planes. It thus confirms that our ${\rev{\bm{\mathcal{T}}}}-$matrix estimator indeed contains a forward multiple scattering contribution. 

\subsection{Coherent backscattering as a manifestation of multiple scattering}
\label{CBS}
\begin{figure}[h!]
    \centering
    \includegraphics[width=16cm]{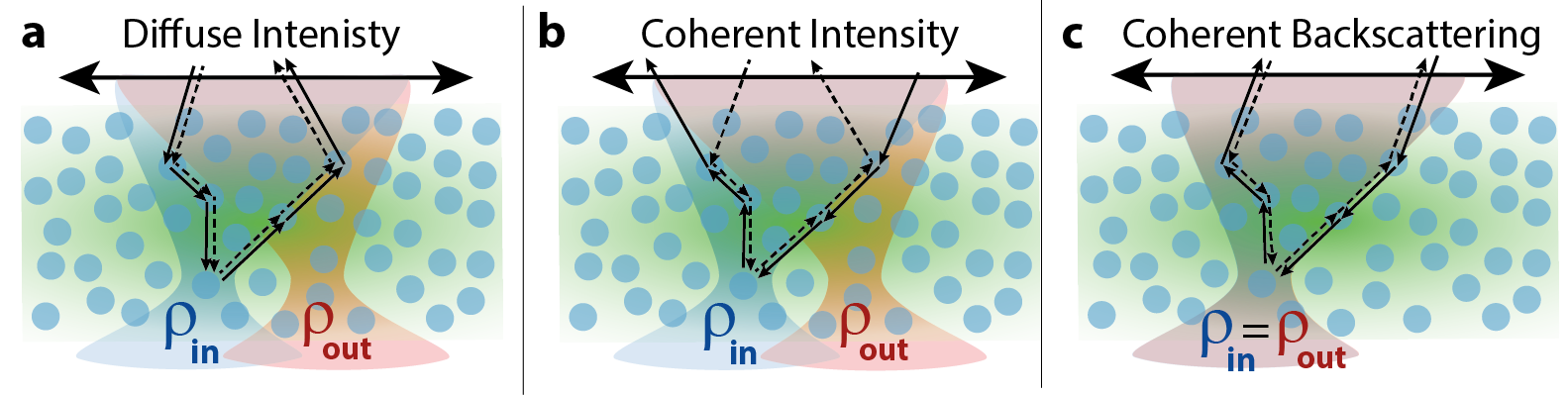}  \caption{\rev{\textbf{Coherent back-scattering phenomenon in the focused basis}. The RPSF is made of two contributions in the multiple scattering regime: \textbf{a} A diffuse component which results from the self-interference of each scattering path with itself; \textbf{b} A coherent intensity resulting from the interference between reciprocal multiple scattering paths inside the medium. Those paths undergo the same scattering sequence but in reverse order. \textbf{c} When the input and output focused beams coincide ($\bm{\rho}_\textrm{in}=\bm{\rho}_\textrm{out}$), the interference is constructive and leads to an enhancement by a factor two of the RPSF with respect to the diffuse background. This is the so-called coherent backscattering peak highlighted by Supplementary Fig.~\ref{fig:CBS}c.}}   \label{fig:CBS0}
\end{figure}

\begin{figure}[h!]
    \centering
    \includegraphics[width=16cm]{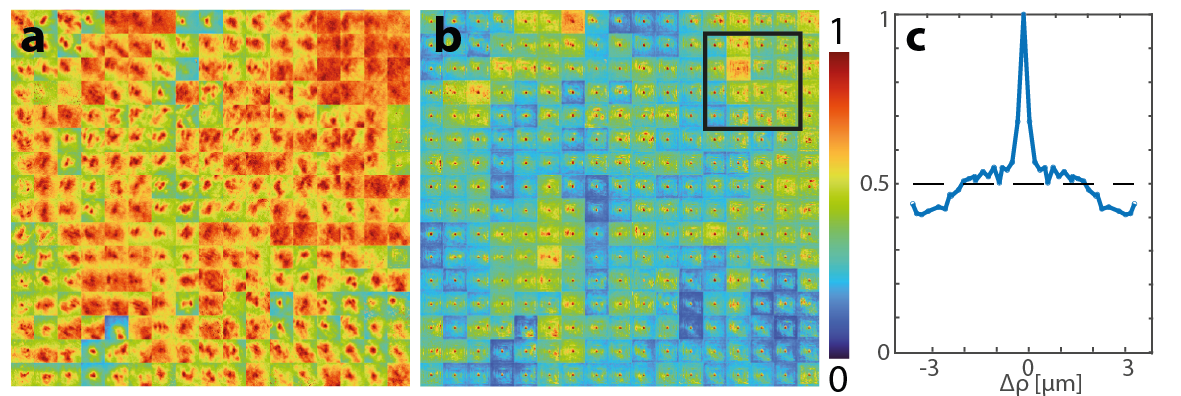}  \caption{\alex{\textbf{Reflection point spread function in the focused basis.}. \textbf{a.} Maps of initial RPSFs at $z=200$ $\mu$m. \textbf{b.} Maps of RPSFs at the same depth after compensation of the isoplanatic aberration phase law (Fig.~S10b) mainly induced by the reference arm. \textbf{c.} Radial average of the RPSF averaged over the area surrounded by a black square in panel \textbf{b}.}.}   \label{fig:CBS}
\end{figure}
\alex{Another key observable to prove the existence of multiple scattering paths is the coherent back-scattering phenomenon. This phenomenon results from the constructive interference between reciprocal multiple scattering paths. Investigated originally in a plane wave basis ($\mathbf{k}$-space)~\cite{Albada1985,Wolf1985,Akkermans1988c}, it manifests as an enhancement by a factor two of the time-gated intensity in the back-scattering direction~\cite{Tourin1997}. Studied in the focused basis (position space), it gives rise to an enhancement by a factor two of the mean intensity at $\Delta \bm{\rho}=\mathbf{0}$~\cite{Larose2004,Aubry2007,lambert_reflection_2020} (Supplementary Fig.~\ref{fig:CBS0}). This phenomenon is investigated in Supplementary Fig.~\ref{fig:CBS}. The initial maps of RPSFs provided by Fig.~4c and reproduced in Supplementary Fig.~\ref{fig:CBS}a does not exhibit a clear signature of the CBS because of the aberrations induced by our imaging system that alters its shape~\cite{Cobus2022}. }

\alex{After compensation of the reference arm aberration (Supplementary Fig.~\ref{fig:spatial_reciprocity}b) at the first step of our aberration correction process, a new map of RPSFs is obtained and shown in Fig.~\ref{fig:CBS}b. It exhibits a following shape: \rev{ A confocal peak due to the single scattering contribution and to the coherent back-scattering phenomenon that results from the construtive interference between reciprocal multiple scattering paths inside the medium (Supplementary Fig.~\ref{fig:CBS0}c); on top of a diffusve background resulting from the incoherent summation of each multiple scattering path intensity (Supplementary Fig.~\ref{fig:CBS0}a)}}. 

In absence of noise, the confocal intensity is \rev{therefore} equal to $I_C=I_S+2I_M$, with $I_S$ the single scattering intensity and $I_M$ the multiple scattering intensity. The incoherent background directly provides the multiple scattering intensity. The ratio $\beta_C=I_C/I_M$ between the confocal peak and the incoherent background can actually provide an estimate for the multiple scattering rate~\cite{lambert_reflection_2020}:
\begin{equation}
\beta_{M}=I_M/(I_S+I_M)=1/(\beta_C-1)
\end{equation}
or, equivalently, of the single scattering rate 
\begin{equation}
\label{ssrate}
\beta_{S}=I_S/(I_S+I_M)=(\beta_C-2)/(\beta_C-1)
\end{equation}
Note that, in practice, a quantitative measurement of the single/multiple scattering rates is not so easy to perform since the multiple scattering background does not exhibit a flat profile especially at shallow depth~\cite{Lambert2022a}. 

\alex{Anyway, a value of $\beta_C$ close to 2 means a predominant multiple scattering contribution. This is actually what we observe in many parts of the field-of-view, in particular in the area surrounded by a black rectangle in Supplementary Fig.~\ref{fig:CBS}b. The radial distribution of the RPSF displayed in Supplementary Fig.~\ref{fig:CBS}c shows a CBS enhancement of two, proof that multiple scattering is predominant in this region. At the end of the multi-scale aberration correction process, the maps of RPSF shows a much larger confocal ratio $\beta_C$, proof that multiple scattering trajectories have been (at least partially) compensated by our $\bm{\rev{\mathcal{T}}}-$matrix estimator.}

\subsection{Measuring the Scattering Mean Free Path}

\begin{figure}[h!]
    \centering
    \includegraphics[width=10 cm]{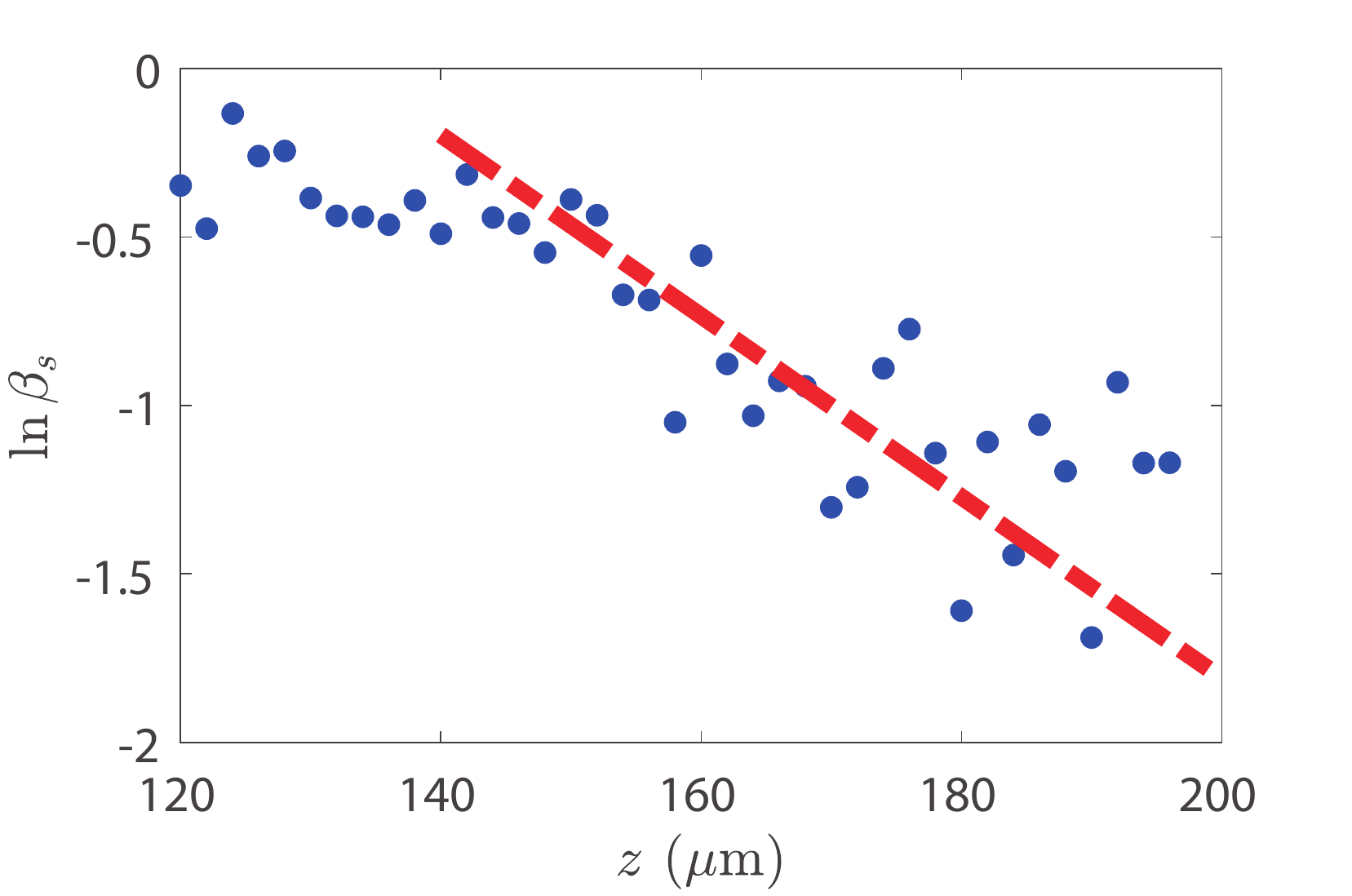}
    \caption{Logarithm of the single scattering rate $\ln \beta_S$ versus depth (blue dots) fitted with Eq.~\ref{beta} (red dashed line). }
    \label{fig:Iconfoc_MFP}
\end{figure}
In a previous work~\cite{Bocheux2019}, the scattering mean free path $\ell_s$ in the cornea was measured by investigating the depth evolution of the confocal intensity. Indeed, in the single scattering regime, under the paraxial approximation and for an homogeneous reflectivity, the time-gated confocal intensity is supposed to decrease as $\exp (-2z/\ell_s)$ if we neglect absorption losses~\cite{badon_multiple_2017,Tricoli2019}. 

Unfortunately, here, the cornea is not healthy but oedematous. The depth evolution of the confocal intensity in the stroma is thus strongly impacted by multiple scattering and cannot be used for a measurement of $\ell_s$. Moreover, in the epithelium, the different layers of cell make the cornea reflectivity too heterogeneous to provide an exponential decrease of the confocal intensity.

Recently, an alternative strategy has been proposed in presence of multiple scattering~\cite{Brutt2022}. It consists in investigating the depth evolution of the \alex{single scattering rate. In 3D, radiative transfer solution indicates the following exponential scaling for $\beta_S$~\cite{Goicoechea2023}:
\begin{equation}
\label{beta}
    \beta_S (z)\sim \exp \left (-z/\ell_s \right)
\end{equation}
To estimate $\beta_S$ (Eq.~\ref{ssrate}), the confocal ratio $\beta_c$ has been estimated as follows:
\begin{equation}
\hat{\beta}_C(z)=\frac{\underset{\Delta \bm{\rho} } {\textrm{max}} \left \lbrace I (\Delta \bm{\rho},z ) \right \rbrace}{\underset{\Delta \bm{\rho} } {\textrm{min}} \left \lbrace I (\Delta \bm{\rho},z ) \right \rbrace}
\end{equation}
This estimator $\hat{\beta}_C$} relies on the fact that the multiple scattering component of the RPSF exhibits a flat background such that it can estimated with the minimum of $I (\Delta \bm{\rho},z ) $. This hypothesis is wrong at shallow depth since the diffuse halo grows as $\sqrt{Dt}$, with $D$ the diffusion coefficient. Nevertheless, \alex{beyond a few $\ell_s$ (here 140 $\mu$m),} the multiple scattering background can be considered as flat \alex{as illustrated by Supplementary Fig.\ref{fig:CBS}c.}

\alex{Supplementary Figure \ref{fig:Iconfoc_MFP} displays the depth evolution of the single scattering rate ${\beta}_S(z)$ computed from $\hat{\beta}_C$ (Eq.~\ref{ssrate}). It exhibits an exponential decay in the stroma beyond $z=140$ $\mu$m. The single scattering rate cannot be estimated beyond $z=200$ $\mu$m because our estimator of $\beta_C(z)$ starts to be impacted by the experimental noise. Therefore, the fit of ${\beta}_S(z)$ with Eq.~\ref{beta} is performed from $z=140$ to $z=200$ $\mu$m. We find {$\ell_s \sim 35$ $\mu$m}.}

\subsection{Quantifying the contrast enhancement}

\begin{figure}[h!]
    \centering
    \includegraphics[width=10 cm]{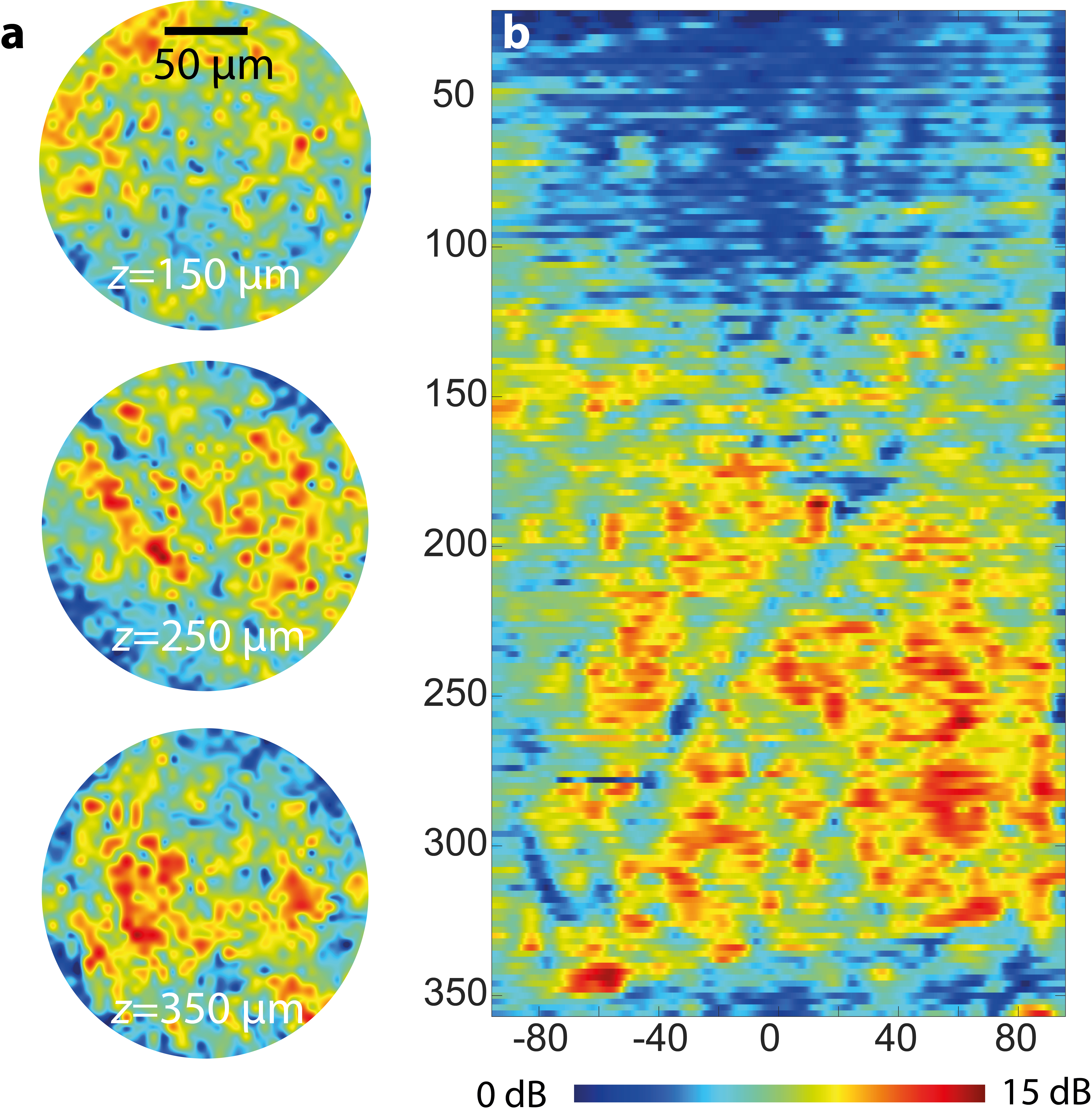}
    \caption{\textbf{Confocal gain provided by the matrix imaging process}. 
    \textbf{a}-\textbf{b.} Transverse cross-section of the confocal gain observed for the \emph{en-face} images displayed in Fig.~3b at depths $50~\mu$m, $250~\mu$m and $350~\mu$m within the cornea (scale bar: 50 $\mu$m). \textbf{c.} Longitudinal cross-section of the confocal gain observed by comparing the B-scan displayed in Fig.3f with its original version shown in Fig.~3c. In each panel, the color scale is in dB. }
    \label{fig:confocal_gain_bscan}
\end{figure}

Supplementary Figure~\ref{fig:confocal_gain_bscan} shows the enhancement of the confocal peak before and after RMI. It reaches a maximal value of 30. This gain should scale, in amplitude, as the number $P_c$ of independent coherence grains exhibited by the $\bm{\rev{\mathcal{T}}}-$matrix in the pupil plane (see, for instance, Figs.~4e and f) and that RMI tends to realign in phase by means of a digital optical phase conjugation. Supplementary Figure~\ref{fig:confocal_gain_bscan}b clearly shows that the confocal gain increases with depth $z$. Indeed, multiple scattering becomes predominant in depth and the transmission phase laws become more and more complex. Note, however, that given the complexity of phase laws displayed in Figs.~4e and f, we could have expected a larger confocal intensity enhancement. This moderate gain in contrast is explained by the fact that a part of the multiple scattering background is not addressed by RMI.

\section{Comparison with state-of-the-art methods}

\subsection{Comparison between matrix imaging and adaptive optics}

The discrimination between aberration and scattering phenomena can be also applied directly to the estimated $\bm{\rev{\mathcal{T}}}$-matrix. The interest of such a decomposition is to show the superiority of matrix imaging with respect to conventional AO. Indeed, the latter approach generally relies on Shack Hartmann sensors that only give access to the phase gradient of reflected wave-fronts. Standard numerical integration of this quantity gives access to the irrotational (\textit{i.e} aberrated) component of the wave-front but generally not to the scattering components of the wave-front that exhibits a wealth of optical vortices~\cite{wu2023singleshot}. On the contrary, the interferometric measurement of the reflected wave-field gives access to this scattering component which is crucial for deep imaging.

\begin{figure}[h!]
    \centering
    \includegraphics[width=14cm]{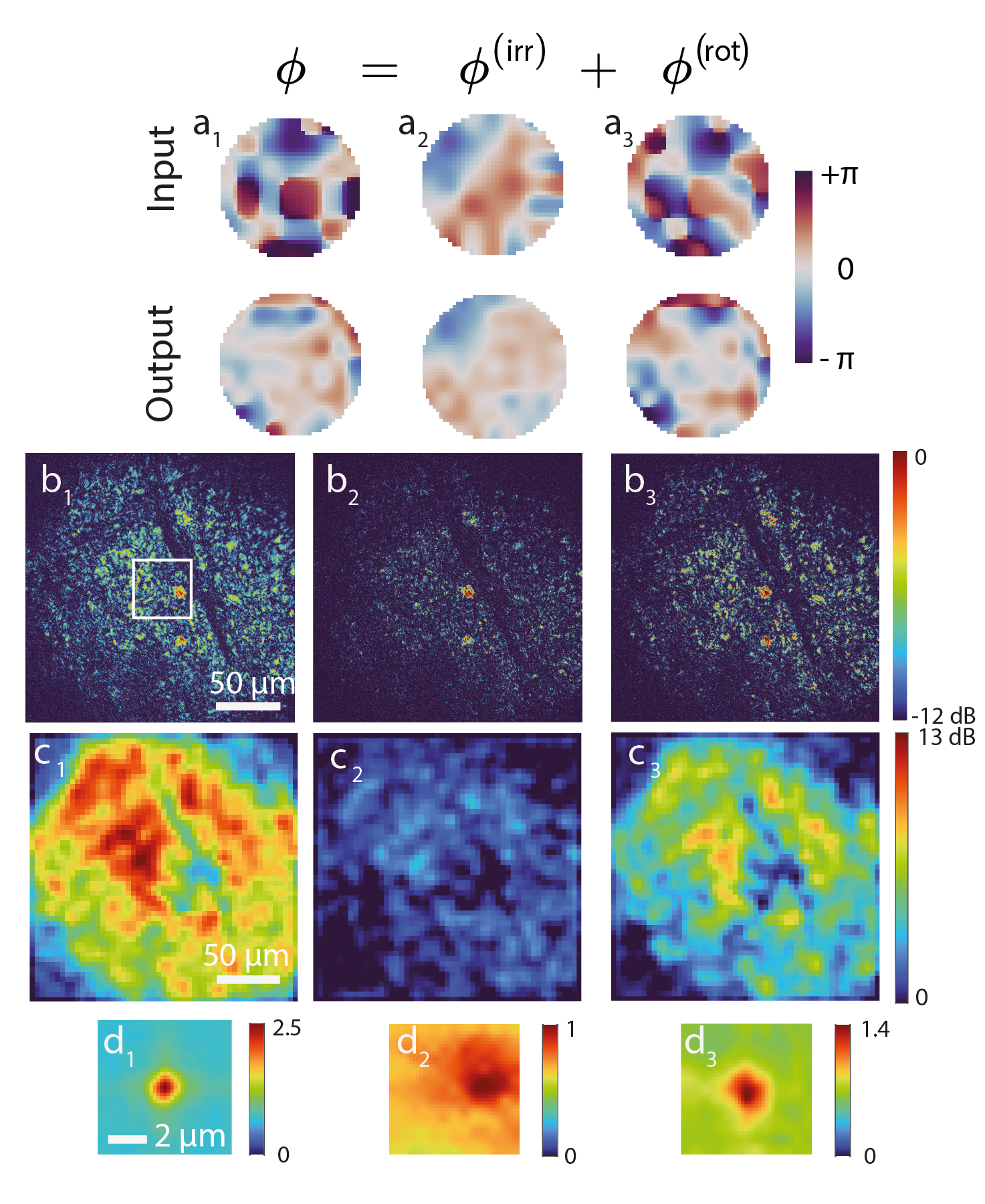}
    \caption{\textbf{Decomposition of the focusing law $\phi$ extracted by RMI at depth $z=200$ $\mu$m.} \textbf{a.} Examples of input and output phase laws ($1$) extracted by RMI and their decomposition into irrotational ($2$) and curl (\textbf{a}$3$) components. These phase laws correspond to the center of the white rectangle in \textbf{b}$_1$. \textbf{b.} Confocal image obtained using the corresponding focusing laws  (scale bar: 50 $\mu$m). \textbf{c.} Associated confocal gain (scale bar; 50 $\mu$m). \textbf{d} Example of local RPSF obtained in the white rectangle displayed in panel \textbf{b}$_1$) (scale bar: 2 $\mu$m).}
    \label{fig:correction_avec_divT1}
\end{figure}

{Supplementary Figure \ref{fig:correction_avec_divT1} illustrates this assertion by first showing the decomposition of the input and output phase laws (Supplementary Figs.~\ref{fig:correction_avec_divT1}a$_1$) into their irrotational (Supplementary Figs.~\ref{fig:correction_avec_divT1}a$_2$) and curl (Supplementary Figs.~\ref{fig:correction_avec_divT1}a$_3$) components. The access to the latter component is decisive for the compensation of wave distortions since it greatly contributes to the improvement of the confocal image (Supplementary Fig.~\ref{fig:correction_avec_divT1}b). This can be quantified by the confocal gain exhibited at the end of the matrix imaging process (Supplementary Fig.~\ref{fig:correction_avec_divT1}c) and the corresponding RPSF (Fig.~\ref{fig:correction_avec_divT1}d). While the access to the curl component of the focusing laws allows us to reach a confocal gain up to 13 dB (Supplementary Fig.~\ref{fig:correction_avec_divT1}c$_1$), conventional AO would only allow a compensation of low-order aberrations, giving rise to a weak confocal gain ($<3$ dB, Supplementary Fig.~\ref{fig:correction_avec_divT1}c$_2$).  }

{Another advantage of RMI versus AO consists in our ability of simulating any physical experiment in post-processing. If performed experimentally with an adaptive optics set up, the multi-scale compensation of wave distortions described in this paper would require: (\textit{i}) a complex adaptive optics arrangement to compensate for wave distortions both in the sample and reference arms; (\textit{ii}) an extremely long acquisition time since the focusing process would have to be repeated 12 times on each of the 10$^8$ points of the imaged volume. The performance of matrix imaging is therefore impossible to reach with conventional adaptive optics tools.}

\subsection{Comparison between iterative time reversal and iterative phase reversal }

\label{IPRvsSVD}

\begin{figure}[ht!]
\centering
\includegraphics[width=\textwidth]{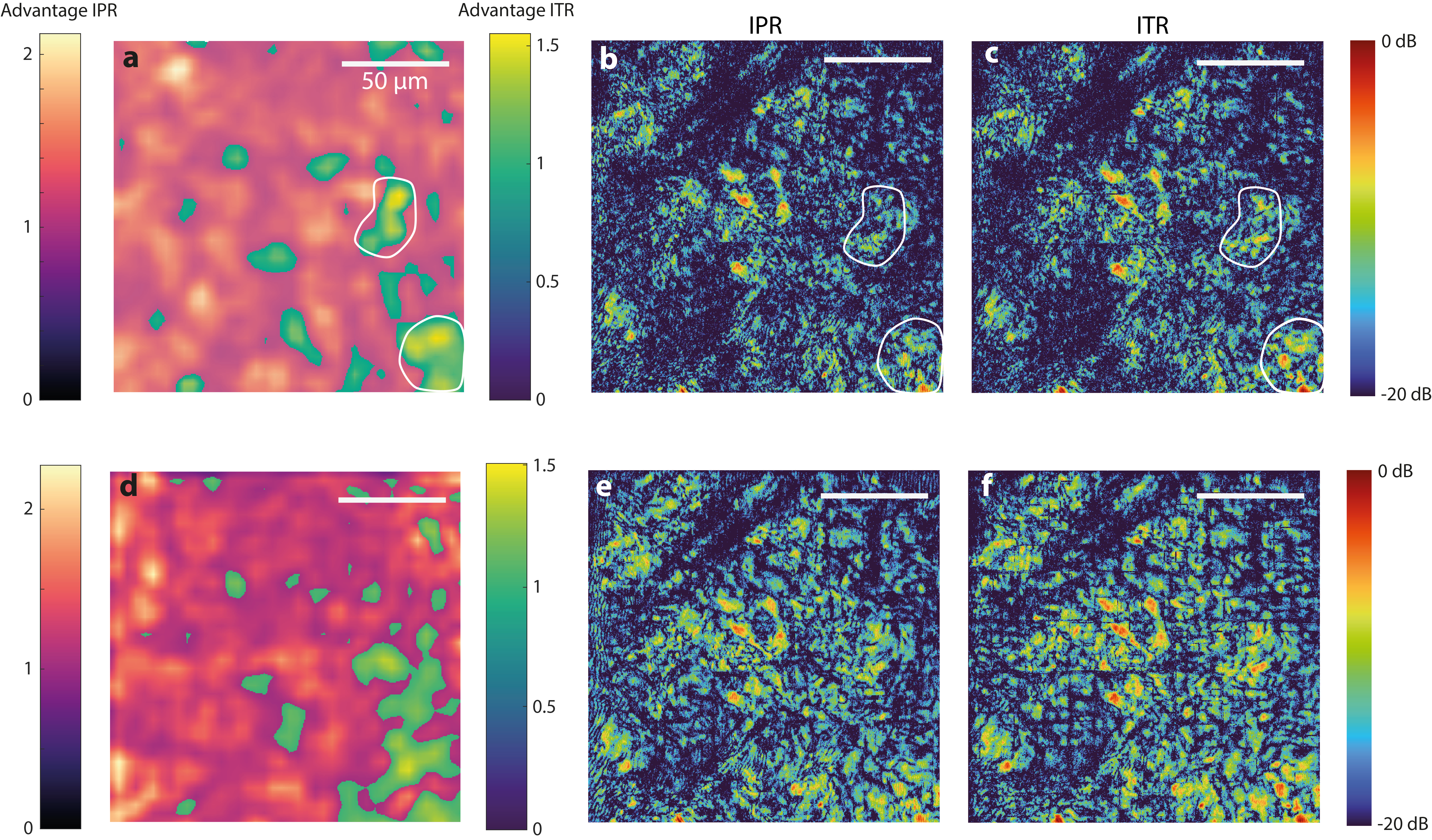}
\caption{\textbf{Multi-scale compensation of wave distortions: Iterative time reversal vs. iterative phase reversal.} \textbf{a.} Ratio between confocal gains obtained at $z=200$ $\mu$m using a ITR or IPR process. Left colorbar: Ratio between IPR and SVD confocal gains. Right colorbar: Ratio between SVD and IPR confocal gains. \textbf{b}-\textbf{c.} Confocal image obtained via IPR and ITR, respectively. The top ($1$) and bottom ($2$) figures correspond to size of spatial windows $L=25$ and 6 $\mu$m, respectively. Scale bars: 50 $\mu$m.} 
  \label{S3_IPRvsSVD}
\end{figure}  

Even though the ITR process does not provide a satisfying estimator at the first iteration of the matrix imaging process (see Supplementary Section \ref{IPC} and Supplementary Fig.~\ref{S2_IPRvsSVD}), the multi-scale analysis and continuous balance between input and output correction enable to gradually reduce the size of the virtual scatterer and improves the ITR estimator at depth $z=$200 $\mu$m. In Supplementary Figs.~\ref{S3_IPRvsSVD}a$_2$ and a$_3$, we compare the confocal images provided by ITR and IPR for $L=25$ $\mu$m. Qualitatively, a careful look at the images shows that the ITR correction leads to a smoother image of the sample reflectivity than IPR. This is in agreement with Supplementary Fig.~\ref{S2_IPRvsSVD} that showed that ITR tends to concentrate on a low spatial frequency spectrum while IPR addresses the whole spatial frequency components of the wave-field. In terms of image resolution, IPR thus shows a better performance than ITR. 

{To understand why, let us first express the confocal image as a function of the reflection matrix coefficients expressed in the plane wave basis:
\begin{equation}
\label{Ic}
I(\Delta \bm{\rho}=\mathbf{0},\bm{\rho}_\textrm{out},z)=\sum_{\mathbf{u}_{\textrm{in}}} \sum_{\mathbf{u}_{\textrm{out}}} R(\mathbf{u}_{\textrm{out}},\mathbf{u}_{\textrm{in}},z) \exp \left [  i\frac{2\pi}{\lambda f} (\mathbf{u}_{\textrm{in}}+\mathbf{u}_{\textrm{out}}).\bm{\rho}_\textrm{out}\right ]
\end{equation}
Under an isoplanatic assumption, the coefficients of the reflection matrix in the plane wave basis can be expressed as follows~\cite{lambert_distortion_2020}:
\begin{equation}
\label{Rkk}
R(\mathbf{u}_\textrm{in},\mathbf{u}_\textrm{out},z)=\rev{\mathcal{T}}_{\textrm{out}}(\mathbf{u}_{\textrm{out}},z) \tilde{\gamma}(\mathbf{u}_{\textrm{in}}+\mathbf{u}_{\textrm{out}},z) \rev{\mathcal{T}}_{\textrm{in}}(\mathbf{u}_{\textrm{in}},z) 
\end{equation}
with $\tilde{\gamma}(\mathbf{u},z)=\int d\bm{\rho} \gamma(\bm{\rho},z) \exp(-i 2\pi \mathbf{u}.\bm{\rho}/(\lambda f))$, the spatial frequency spectrum of the sample  reflectivity at depth $z$. In absence of aberrations, each antidiagonal ($\mathbf{u}_\textrm{in} +\mathbf{u}_\textrm{out}=$ constant) encodes one spatial frequency of the sample reflectivity: $R(\mathbf{u}_\textrm{out},\mathbf{u}_\textrm{in},z)=\tilde{\gamma}(\mathbf{u}_{\textrm{in}}+\mathbf{u}_{\textrm{out}},z) $ in the pupil support. $I(\Delta \bm{\rho}=\mathbf{0},\bm{\rho}_\textrm{out},z)$ is then a satisfying estimator of $\gamma(\bm{\rho},z)$ with a transverse resolution only limited by diffraction. In presence of aberrations, phase fluctuations of $\rev{\mathcal{T}}_{\textrm{in/out}}$
implies phase distortions between each spatial frequency component of the object while amplitude variations of $\rev{\mathcal{T}}_{\textrm{in/out}}$ implies the attenuation of some spatial frequency components of the object. }\\

\alex{The compensation of aberrations consists in applying the phase conjugate of the aberration transmittance estimators, $\hat{\rev{\mathcal{T}}}_{\textrm{out}}$ and $\hat{\rev{\mathcal{T}}}_{\textrm{in}}$, at input and output of the reflection matrix:
\begin{equation}
\label{Rkkcorr}
R''(\mathbf{u}_\textrm{out},\mathbf{u}_\textrm{in},z)=\hat{\rev{\mathcal{T}}}_{\textrm{out}}^*(\mathbf{u}_\textrm{out},z)R(\mathbf{u}_\textrm{out},\mathbf{u}_\textrm{in},z)\hat{\rev{\mathcal{T}}}_{\textrm{in}}^*(\mathbf{u}_\textrm{in},z) 
\end{equation}
Assuming ITR provides correct estimators of the aberration transmittance ($\hat{\rev{\mathcal{T}}}_{\textrm{in/out}} \equiv \rev{\mathcal{T}}_{\textrm{in/out}} $) and injecting Eq.~\ref{Rkk} into the last equation leads to the following expression for the reflection matrix corrected by ITR processing:
\begin{equation}
R''_{ITR}(\mathbf{u}_\textrm{out},\mathbf{u}_\textrm{in},z)=|\rev{\mathcal{T}}_{\textrm{out}}(\mathbf{u}_\textrm{out},z)|^2 \tilde{\gamma}(\mathbf{u}_{\textrm{in}}+\mathbf{u}_{\textrm{out}},z) |\rev{\mathcal{T}}_{\textrm{in}}(\mathbf{u}_\textrm{in},z) |^2
\end{equation}
Because the amplitude of $\rev{\mathcal{T}}_{\textrm{in/out}}$ vanishes for high spatial frequency $|\mathbf{u}_{\textrm{in/out}}|$ (Supplementary Fig.~\ref{S2_IPRvsSVD}b), ITR tends to filter the high spatial frequency component of the object.} 

\alex{On the contrary, IPR converges towards a normalized version of the transmittances ${\bm{\rev{\mathcal{T}}}}_{\textrm{in/out}}$, such that: $\hat{\rev{\mathcal{T}}}_{\textrm{in/out}} \equiv \exp \left ( j \mbox{arg} \left \lbrace {{\rev{\mathcal{T}}}}_{\textrm{in/out}} \right \rbrace \right)$. Using this last expression of $\hat{\bm{\rev{\mathcal{T}}}}_{\textrm{in/out}} $ and injecting Eq.~\ref{Rkk} into \ref{Rkkcorr} leads to the following expression for the reflection matrix corrected by IPR processing: 
\begin{equation}
R''_{IPR}(\mathbf{u}_\textrm{out},\mathbf{u}_\textrm{in},z)=|\rev{\mathcal{T}}_{\textrm{out}}(\mathbf{u}_\textrm{out},z)| \tilde{\gamma}(\mathbf{u}_{\textrm{in}}+\mathbf{u}_{\textrm{out}},z) |\rev{\mathcal{T}}_{\textrm{in}}(\mathbf{u}_\textrm{in},z) |
\end{equation}
IPR leaves the spatial frequency spectrum of the reflection matrix unchanged and does not filter the high spatial frequency components of the reflection matrix. Of course, a better compensation of aberration would consist in an inverse filter where the amplitude decrease of $\rev{\mathcal{T}}_{\textrm{in/out}}$ could be compensated on top of phase distortions. Nevertheless, this operation is extremely sensitive to noise and suppose a perfect match between the aberration transmittance $\rev{\mathcal{T}}_{\textrm{in/out}}$ and their estimators $\hat{\rev{\mathcal{T}}}_{\textrm{in/out}}$. This is wrong especially for the phase of $\rev{\mathcal{T}}_{\textrm{in/out}}$ at high spatial frequencies where the amplitude of $\hat{\rev{\mathcal{T}}}_{\textrm{in/out}}$ vanishes. Therefore, IPR is an adequate compromise between the matched filter operated by ITR that affect the high spatial frequencies of the object and an inverse filter that is extremely sensitive to noise.}\\

In terms of contrast, the relative performance between IPR and ITR can be assessed by investigating the confocal energy at the end of the whole process. Supplementary Fig.~\ref{S3_IPRvsSVD}a$_1$ shows the ratio between their confocal gains at depth $z=$200 $\mu$m. The IPR process exhibits a better performance on a major part of the field-of-view. The ITR process is only better when it can hang on a highly reflecting structure. In that case, the maximization of backscattered energy on which the ITR process is based can lead to a better result than a criterion based on coherence as done by IPR. In practice, one could apply both IPR and ITR and keep the best option. Nevertheless, for sake of clarity and image continuity, the IPR process has been considered for all figures shown in the accompanying paper. Indeed, for small spatial windows ($L=6$ $\mu$m), the ITR process leads to an image with strong vignetting effects (Supplementary Fig.~\ref{S3_IPRvsSVD}b$_3$), which is a manifestation of a lack of correlations of the estimator $\hat{\mathbf{T}}$ between adjacent windows. On the contrary, the IPR leads to an estimator whose scattered component $\hat{\mathbf{S}}$ exhibits the expected memory effect for this size of spatial windows (Supplementary Fig.~\ref{fig:vignettage}d). Supplementary Figure~\ref{S3_IPRvsSVD} thus shows the superiority of IPR in terms of spatial resolution for the $\mathbf{T}-$matrix estimation.  

\begin{figure}
    \centering
    \includegraphics[width=\linewidth]{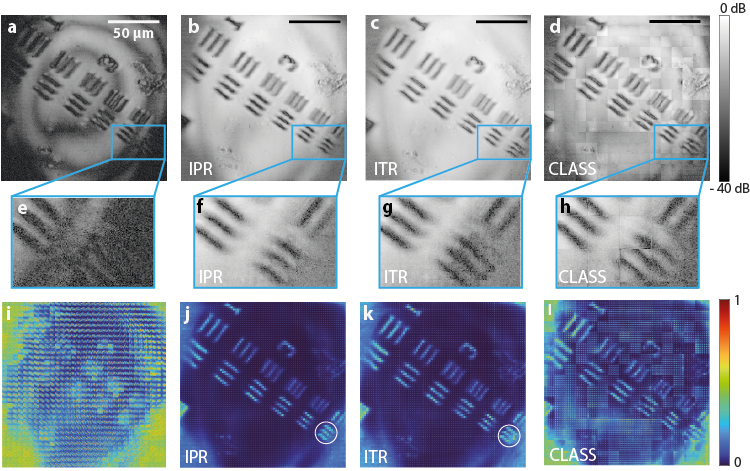}
    \caption{\alex{\textbf{Comparison of IPR with other state-of-the-art methods for the resolution target experiment.}
    \textbf{a}-\textbf{d.} Confocal images (scale bar: 50 $\mu$m): Original image (\textbf{a}), corrected image using the IPR multi-scale process (\textbf{b}), corrected image using the ITR multi-scale process (\textbf{c}) and corrected image using the local CLASS algorithm (\textbf{d}).
    \textbf{e}-\textbf{h.} Corresponding blowups of the area contained inside the blue rectangle displayed in panels a-d.
    \textbf{i}-\textbf{k.} Corresponding maps of RPSFs (computed over spatial windows of size $L=2~\mu$m)}.
    }
    \label{fig:res_target_comp}
\end{figure}

\alex{The superiority of IPR compared with ITR is also confirmed by the resolution target experiment presented in Supplementary Section \ref{sec:res_target}. Supplementary Figure~\ref{fig:res_target_comp} compares the images obtained via IPR and ITR using the multi-scale process described above. The full-field images show a slightly better contrast with IPR (Supplementary Fig.~\ref{fig:res_target_comp}b) than ITR (Supplementary Fig.~\ref{fig:res_target_comp}c). The zoom on the smallest patterns of the target also shows the benefit of IPR (Supplementary Fig.~\ref{fig:res_target_comp}f) compared with ITR (Supplementary Fig.~\ref{fig:res_target_comp}g) in terms of resolution. Indeed, the three strips are clearly resolved with IPR, while this pattern remains quite blurred with ITR . This superiority is also confirmed by the corresponding RPSFs displayed in Supplementary Figs.~\ref{fig:res_target_comp}j and k. In the area of the smallest pattern surrounded by a white circle, the final RPSFs obtained with IPR display a weaker incoherent background than with ITR. This reference experiment thus confirms the overall superiority of IPR compared with ITR that we have already noticed in the cornea experiment.}

\subsection{Comparison between CLASS and RMI}\label{sec:CLASS_vs_RMI}

For the last five years, an alternative matrix approach has been developed by W. Choi and colleagues, this is the so-called CLASS algorithm~\cite{kang_high-resolution_2017,yoon_laser_2020,Kwon2023}. Based on the recording of a time-gated reflection matrix, it exploits its input-output correlations in the plane wave basis to estimate input and output aberration phase laws. The CLASS algorithm amounts to find the aberration phase laws that maximize the confocal intensity in the focused basis. The input and output aberration laws are computed simultaneously and the whole process is then iterated several times to converge towards satisfying phase laws.
On the contrary, our algorithm maximizes sequentially the coherence of input and output wave-fields generated by the virtual guide star. It allows us to decrease gradually the size of this guide star and improve the estimation of the $\mathbf{T}$-matrix, while improving its resolution by gradually reducing the size of spatial windows.
\begin{figure}[h!]
    \centering  \includegraphics[width=\textwidth]{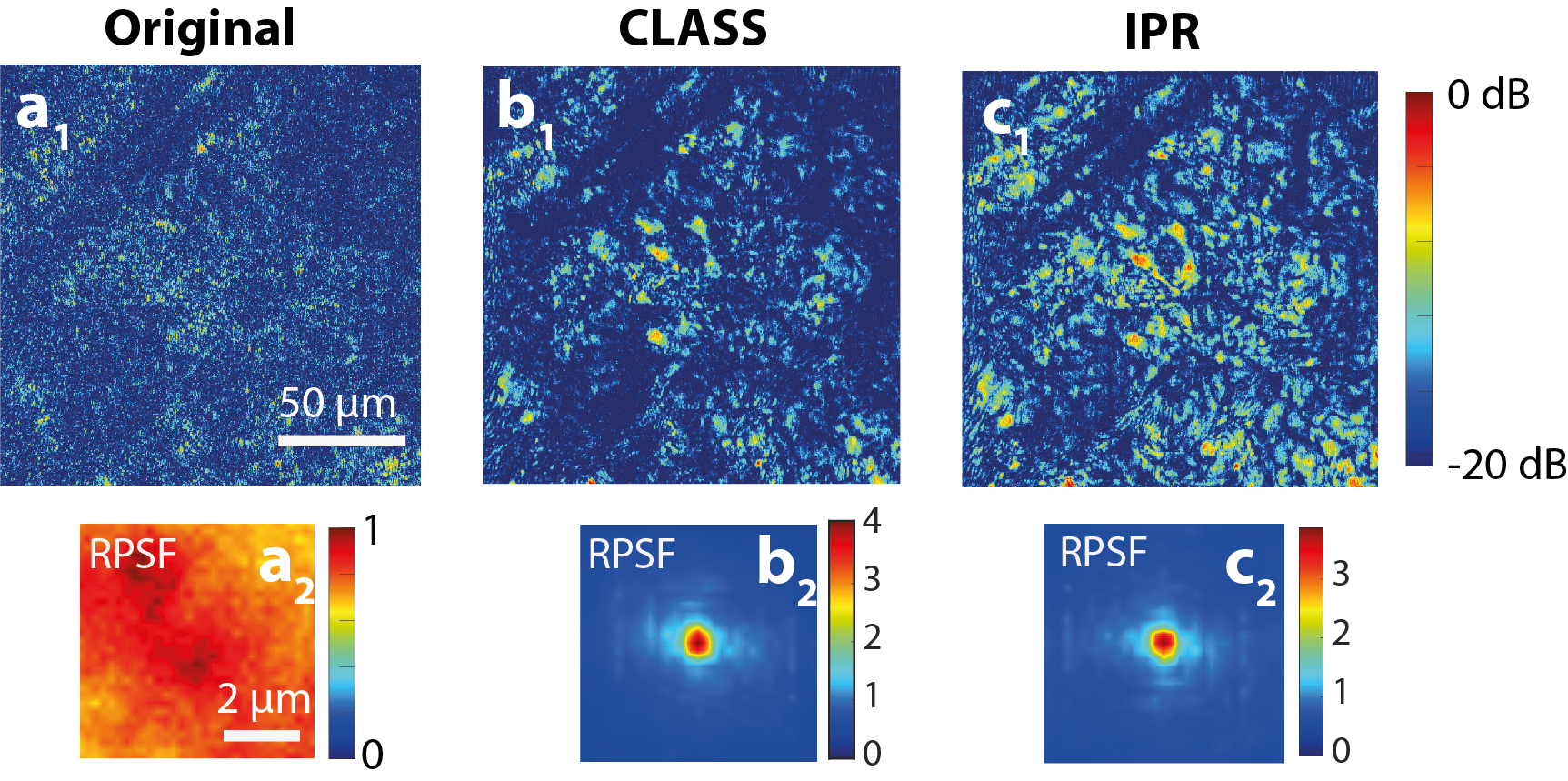}
    \caption{\textbf{Comparison between CLASS and RMI upon convergence}. \textbf{a.} Original confocal image (scale bar: 50 $\mu$m) (\textbf{a}$_1$) and mean RPSF (scale bar: 2 $\mu$m) (\textbf{a}$_2$) at z=200 $\mu$m. \textbf{b}-\textbf{c.} Corresponding CLASS  (\textbf{b}$_1$)  and IPR (\textbf{c}$_1$) images with their mean RPSFs (\textbf{b}$_2$ and \textbf{c}$_2$, respectively). The size of spatial windows is $L=13$ $\mu$m.}
\label{fig:CLASSa}
\end{figure}

To compare the performance of both approaches, we have applied CLASS on our experimental data at z=200 $\mu$m. Supplementary Figure~\ref{fig:CLASSa} shows the results obtained by CLASS and IPR when the extension $L$ of spatial windows $W_L$ is sufficiently large to allow the convergence of each method ($L=14$ $\mu$m). Although the corrected RPSF display similar properties in average for each method [Supplementary Figs.~\ref{fig:CLASSa}b$_2$ and c$_2$], the obtained images show significant differences [Supplementary Figs.~\ref{fig:CLASSa}b$_1$ and c$_1$]. On the one hand, the CLASS image seems more contrasted because it tends to focus on the main scatterers of the field-of-view. On the other hand, the IPR image displays a more homogeneous reflectivity across the field-of-view. This difference can be understood by the maximization of the confocal energy operated by CLASS which gives more weight to the most echogenic scatterers to the detriment of weaker reflectivity regions.  
\begin{figure}[h!]
    \centering\includegraphics[width=\textwidth]{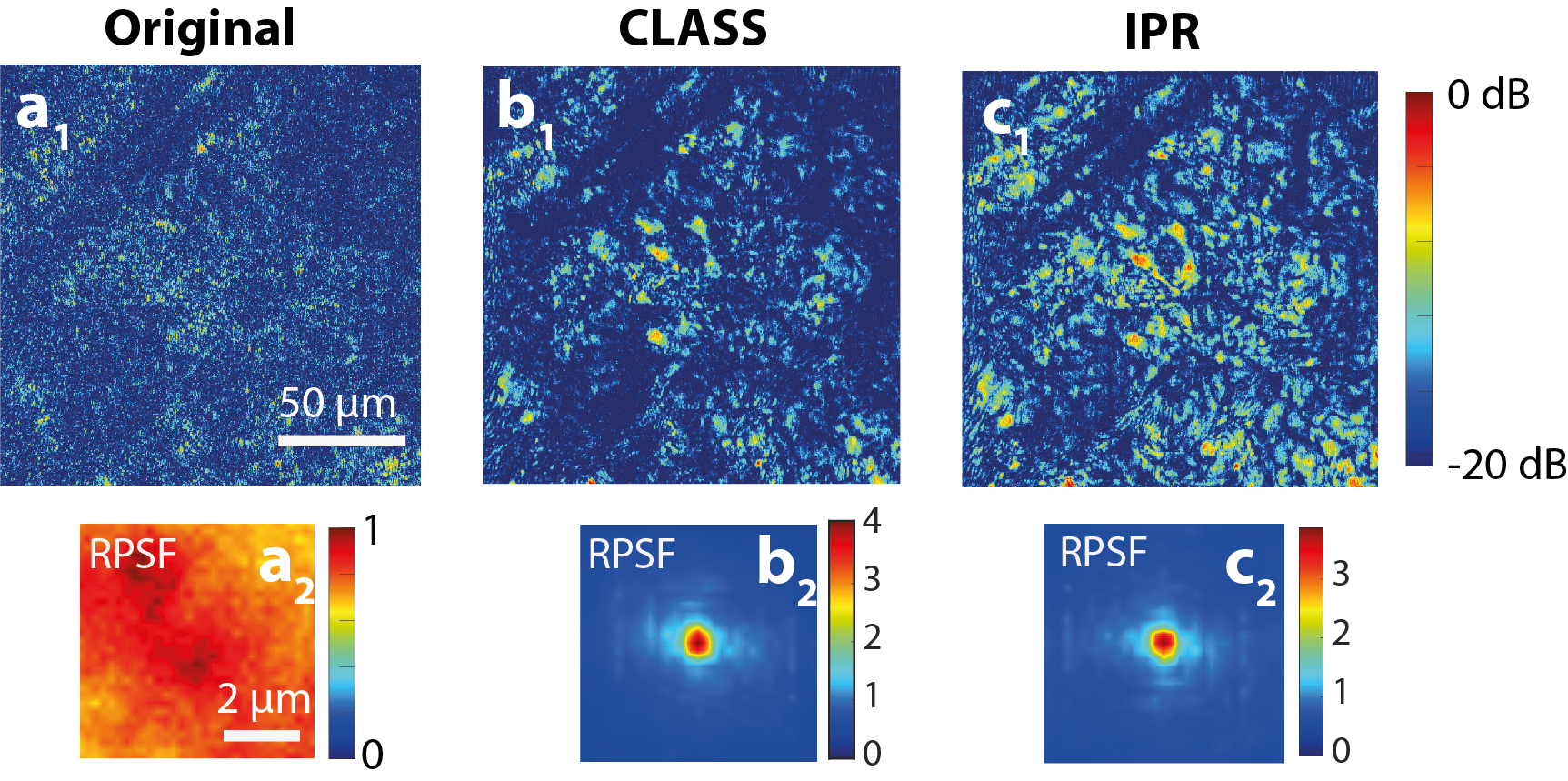}
\caption{\textbf{Comparison between CLASS and RMI for small isoplanatic patches}. \textbf{a.} Original confocal image (scale bar: 50 $\mu$m) (\textbf{a}$_1$) and mean RPSF (scale bar: 2 $\mu$m) (\textbf{a}$_2$) at $z=$200 $\mu$m. \textbf{b}-\textbf{c.} Corresponding CLASS  (\textbf{b}$_1$) and IPR (\textbf{c}$_1$) images with their mean RPSFs (\textbf{b}$_2$ and \textbf{c}$_2$, respectively). The size of spatial windows is $L=6$ $\mu$m.}
\label{fig:CLASSb}
\end{figure}

Another difference between CLASS and IPR approach lies in the higher resolution capability of the latter method. This assertion is supported by Supplementary Fig.~\ref{fig:CLASSb} that compares the results of CLASS and IPR for a smaller spatial window ($L=5.75$ $\mu$m). A local RPSF obtained via CLASS is displayed in Supplementary Fig.~\ref{fig:CLASSb}b$_2$. It shows an important background, a manifestation of an imperfect compensation of forward multiple scattering. On the contrary, the RPSF obtained at the same location via IPR is close to be ideal (Supplementary Fig.~\ref{fig:CLASSb}c$_2$). When looking into details to the CLASS image (Supplementary Fig.~\ref{fig:CLASSb}b$_1$), strong vignetting effects can be observed while the IPR image exhibits a continuous reflectivity (Supplementary Fig.~\ref{fig:CLASSb}c$_1$). 

\begin{figure}[h!]
    \centering\includegraphics[width=0.8\textwidth]{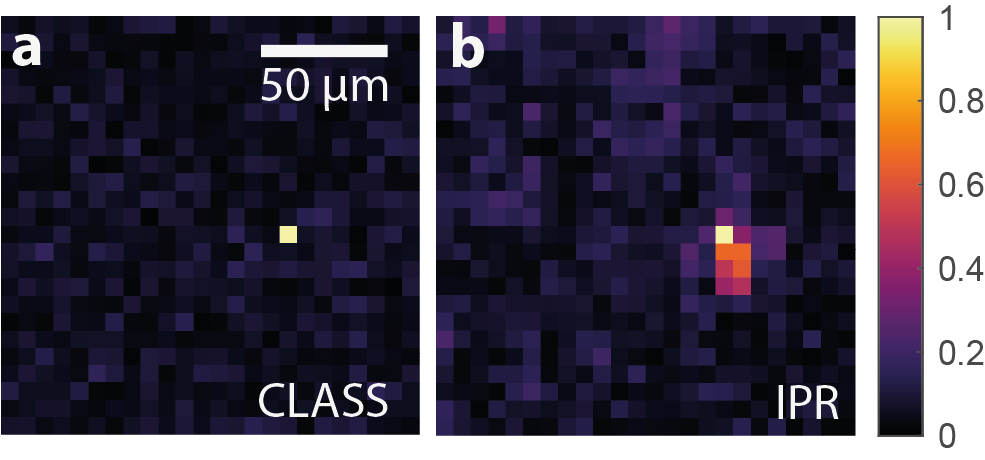}
\caption{\textbf{Memory effect exhibited the forward multiple scattering estimator: Comparison between CLASS and RMI}. Spatial correlation function $C_S(\rp,\rp')$ of the scattering component $\hat{\mathbf{S}}$ for a given point $\rp'$. \textbf{a.} CLASS. \textbf{b.} RMI. The depth is $z=200$ $\mu$m and the size of spatial windows is $L=6$ $\mu$m. Scale bar: 50 $\mu$m.}
\label{fig:CLASSc}
\end{figure}

This result can be understood by comparing the spatial correlation properties of the scattering component ${\bm{\rev{\mathcal{S}}}}$ of the $\bm{\rev{\mathcal{T}}}-$matrix obtained by CLASS and IPR (see Methods of the accompanying paper). Supplementary Figure~\ref{fig:CLASSc} shows this correlation map for the mid-point $\rp$ of the area displayed with {a white square in Supplementary Fig.~\ref{fig:CLASSb}}.  While the $\bm{\rev{\mathcal{S}}}$-matrix derived by IPR preserves a short-range correlation between neighbouring windows (see Supplementary Fig.~\ref{fig:CLASSc}c), the CLASS algorithm leads to a fully spatially incoherent estimator $\hat{\bm{\rev{\mathcal{S}}}}$ (see Supplementary Fig.~\ref{fig:CLASSc}a). This observable clearly shows that the IPR estimator leads to a coherent (i.e physical) compensation of multiple scattering while CLASS leads to an incoherent correction (i.e bucket-like). The result of Supplementary Fig.~\ref{fig:CLASSc} accounts for the vignetting effects observed for the CLASS image in Supplementary Fig.~\ref{fig:CLASSb}b$_1$.

\alex{This vignetting effect is also highlighted by the resolution target experiment in Supplementary Fig.~\ref{fig:res_target_comp}. The corresponding images obtained with the multi-scale IPR process and the local CLASS algorithm are displayed in Supplementary Figs.~\ref{fig:res_target_comp}b and d, respectively. For each method, the size of spatial windows over which aberration phase laws are estimated is $L=10$ $\mu$m.  As in the cornea experiment (Supplementary Fig.~\ref{fig:CLASSc}), the CLASS image lacks of continuity between adjacent spatial windows (Supplementary Fig.~\ref{fig:res_target_comp}d). The map of RPSFs also shows the imperfect convergence of the CLASS process (Supplementary Fig.~\ref{fig:res_target_comp}l). Unlike the RPSFs obtained at the end of the IPR process (Supplementary Figs.~\ref{fig:res_target_comp}j), the CLASS RPSFs are far from displaying a diffraction-limited feature. They also show a strong variability across the field-of-view, which is another manifestation of the vignetting effect observed in the full-field CLASS image (Supplementary Figs.~\ref{fig:res_target_comp}d). The resolution target experiment thus confirms the superiority of a multi-scale IPR process with respect to the local CLASS algorithm.}

\clearpage

\section{Glossary}

\begin{table}[h!]
    \centering
    \begin{tabular}{l|l}
      operator  &  definition \\
      \hline\hline
      $\vec{R} = [R(\rin,\rout,z)]$   &  reflection matrix in the focused basis \\
      $\vec{R}_\textrm{in/out} = [{R}_\textrm{in}(\Delta \bm{\rho}_\textrm{in/out},\bm{\rho}_\textrm{out/in},z)]$   &  de-scanned matrix at input/output \\
      $\vec{D}_\textrm{in/out} = [D(\vec{u}_\textrm{in/out},\bm{\rho}_\textrm{out/in}),z]$   &  input/output  distortion matrix  \\
      $\vec{H}_\mathrm{in/out} = [H_\mathrm{in}(\vec{\rho}_s,\bm{\rho}_\textrm{in/out},z)]$ & input/output focusing matrix \\
      \rev{$\mathbf{T}_0=[T_0(\vec{u},\vec{\rho})]$} & Fourier transform operator \\
      $\vec{C}_\mathrm{in/out}=[C_\mathrm{in/out}(\vec{u}_\textrm{in/out},\vec{u}'_\textrm{in/out},\mathbf{r}_p)]$ &  input/output pupil correlation matrix of $\vec{D}_\mathbf{in}$  \\
         $\vec{C}_H = [C_H (\vec{u},\vec{u'})]$ & correlation matrix associated with  \\
         & a virtual reflector of reflectivity $|H(\bm{\rho_s})|^2$ \\
      $\vec{T}_\mathrm{in/out} = [T_\mathrm{in/out}(\vec{u}_\textrm{in/out},\vec{r}_p)]$ & input/output transmission matrix \\
      \rev{$\vec{\mathcal{T}}_\mathrm{in/out} = [\mathcal{T}_\mathrm{in/out}(\vec{u}_\textrm{in/out},\vec{r}_p)]$ }& \rev{input/output transmittance matrix }\\
       $\rev{\hat{\vec{\mathcal{T}}}_\mathrm{in/out} = [\mathcal{T}_\mathrm{in/out}(\vec{u}_\textrm{in/out},\vec{r}_p)]}$ & \rev{estimator of the input/output transmittance matrix} \\
        ${\bm{\phi}}_\mathrm{in/out} = [\phi_\mathrm{in/out}(\vec{u}_\textrm{in/out},\vec{r}_p)]$ & phase of the input/output transmission matrix estimator \\
        $\rev{\vec{\mathcal{A}}_\mathrm{in/out}} = [\rev{\mathcal{A}}_\mathrm{in/out}(\vec{u}_\textrm{in/out},\vec{r}_p)]$ & input/output aberration matrix \\
        $\rev{\vec{\mathcal{S}}}_\mathrm{in/out} = [\rev{\mathcal{S}}_\mathrm{in/out}(\vec{u}_\textrm{in/out},\vec{r}_p)]$ & input/output forward multiple scattering matrix \\
    \end{tabular}
    \caption{\textbf{Glossary of the operators used in this study}}
    \label{tab:glossary_matrices}
\end{table}

\clearpage

%

\end{document}